%% file: top-renormalons-arXiv.tex
\documentclass[a4paper,11pt]{article}
\usepackage[english]{babel}
\usepackage{jheppub}
\usepackage{amsmath,amssymb,graphicx,textcomp,enumerate,alltt,xspace,
  multirow,mathtools,slashed,subcaption}
\usepackage{soul}
\usepackage[utf8]{inputenc}
\input{texmacros}

\title{All-orders behaviour and renormalons\\ in top-mass observables}

\author[a]{Silvia Ferrario Ravasio,}
\author[b]{Paolo Nason,}
\author[a]{Carlo Oleari}

\emailAdd{silvia.ferrario@mib.infn.it}
\emailAdd{paolo.nason@mib.infn.it}
\emailAdd{carlo.oleari@mib.infn.it}

\affiliation[a] {Universit\`a di Milano\,-\,Bicocca and INFN, Sezione di
  Milano\,-\,Bicocca, Piazza della Scienza 3, 20126 Milano, Italy}

\affiliation[b]{CERN, CH-1211 Geneve 23, Switzerland, and INFN, Sezione di
  Milano\,-\,Bicocca, Piazza della Scienza 3, 20126 Milano, Italy}

\abstract{We study a simplified model of top production and decay, consisting
  in a virtual vector boson $W^*$ decaying into a massive-massless
  $t$-$\bar{b}$ quark-antiquark pair. The top has a finite width and further
  decays into a stable vector boson $W$ and a $b$ quark. We then consider the
  emission or the virtual exchange of one gluon, with all possible
  light-quark loop insertions. These are the dominant diagrams in the limit
  of an infinite number of light flavours.  We devise a procedure to compute
  this process fully, by analytic and numerical methods, and for any
  infrared-safe final-state observables. We examine the results at arbitrary
  orders in perturbation theory, and assess the factorial growth associated
  with renormalons.  We look for renormalon effects leading to corrections of
  order $\LambdaQCD$, that we dub ``linear'' renormalons, in the inclusive cross
  section (with and without selection cuts), in the mass of the
  reconstructed-top system, and in the average energy of the final-state $W$
  boson, considering both the pole and the \MSB{} scheme for the top mass. We
  find that the total cross section without cuts, if expressed in terms of
  the \MSB{} mass, does not exhibit linear renormalons, but, as soon as
  selection cuts are introduced, jets-related linear renormalons arise in any
  mass scheme. In addition, we show that the reconstructed mass is affected
  by linear renormalons in any scheme and that the average energy of the $W$
  boson (that we consider as a simplified example of leptonic observable), in
  any mass scheme, has a renormalon in the narrow-width limit, that is
  however screened at large orders for finite top widths, provided the top
  mass is in the \MSB{} scheme.  }


\begin{document}

\maketitle

\section{Introduction}
\label{sec:intro}

The top mass is measured quite precisely at the LHC by both the
ATLAS~\cite{Pearson:2017jck} and the CMS~\cite{Castro:2017yxe}
Collaborations. Up to now, the methods that yield the most accurate results
are the so called ``direct'' methods, that are based upon the reconstruction
of the top-decay products. The measurement is performed by fitting kinematic
distributions that are closely related to the top mass with those obtained
using an event generator, and by extracting the fitted value of the top mass.

Current uncertainties are now near 500~MeV~\cite{Aaboud:2016igd,
  Khachatryan:2015hba}, so that one can worry whether QCD non-perturbative
effects may substantially affect the result. In fact, the experimental
collaborations estimate these and other effects by varying parameters in the
generators, and eventually comparing different generators. This method has
been traditionally used in collider physics to estimate theoretical
uncertainties due to the modeling of hadronization and underlying events, and
also to estimate uncertainties related to higher perturbative orders, as
produced by the shower algorithms~\cite{Ravasio:2018lzi}.  As such, it is a
valuable method, but it should not be forgotten that it may only provide a
lower bound on the associated errors. It is thus important, at the same time,
to investigate the associated uncertainties from a purely theoretical point
of view. In consideration of our poor knowledge of non-perturbative QCD,
these investigations can at most have a qualitative value, but may help us to
understand sources of uncertainties that we might have missed.  One such work
is presented in Ref.~\cite{Hoang:2018zrp}, where the authors attempt to
relate a theoretically well-defined mass parameter with a corresponding
shower Monte Carlo one, using as observable the jet mass of a highly boosted
top.

In the present work, we consider the interplay of non-perturbative effects
with the behaviour of perturbative QCD at large orders in the coupling
constant, focusing in particular upon observables that, although quite
simple, may be considered of the kind used in ``direct measurements''.

It is known that in renormalizable field theories, the renormalization group
flow of the couplings leads to the so called renormalons, i.e.~to the
factorial growth of the coefficients of the perturbative expansion as a
function of the order~\cite{Gross:1974jv, Lautrup:1977hs, tHooft:1979gj,
  Parisi:1978az, Mueller:1984vh,
  Mueller:1992xz, Altarelli:1995kz, Beneke:1998ui}. Renormalons lead to a
divergence of the perturbative expansion, that thus becomes asymptotic. In
particular, in the case of \emph{infrared} renormalons in asymptotically-free
field theories, the ambiguity in the summation of the series corresponds to a
power suppressed effect.

For top-mass observables, ambiguities of order $\Lambda/m_t$ (where $\Lambda$
is some hadronic scale and $m_t$ is the top mass) are particularly important,
since they affect the top-mass measurements by an amount close to the level
of the current accuracy.

In the following we will refer to renormalons leading to linear power
suppressions as ``linear renormalons'' (or, unless explicitly specified
differently, simply as ``renormalons'').

The full renormalon structure of QCD is not known. There is however a fully
consistent simplified model where higher order corrections are accessible up
to all orders in the coupling, namely the large-$n_f$ limit of QCD, where the
number of flavours $n_f$ is taken large and negative (see, for example,
Ref.~\cite{Beneke:1994qe}).  Very often, estimates of non-perturbative
effects are performed starting with the large-$n_f$ result, where, at the end
of the calculation, one makes the replacement
\begin{equation}
n_f \to -\frac{11 \CA}{4\TR} + n_l\,,
\end{equation}
where $\CA=3$, $\TR=1/2$ and $n_l$ is the number of light flavors.  This
approach is called ``large-$b_0$ approximation".

With such replacement, the $\beta$ function of the large (negative) $n_f$
theory becomes the $\beta$ function of the full QCD with $n_l$ massless
flavours.

In the present work we consider a fictitious process $W^* \rightarrow
t\bar{b} \to W b \bar{b}$, where the $W$ boson has only a vector coupling to
quarks, and examine the behaviour of the cross section, of the
reconstructed-top mass and of the energy of the $W$ boson, order by order in
the strong coupling expansion, taking the large-$n_f$ limit. We consider up
to one gluon exchange (or emission), and dress this gluon with an arbitrary
number of fermion vacuum-polarization insertions. Furthermore, we also
consider final states where the gluon has undergone a splitting into a
fermion-antifermion pair, corresponding to a cut vacuum polarization
diagram. We assume a finite width for the top quark.

We have devised a method that allows us to compute in principle any
observable in our process, without further approximations, making use of
simple numerical techniques. We can thus compute the perturbative expansion
at any finite order and infer its asymptotic nature for any observable, with
the only limitation of the numerical precision.

We focus for simplicity upon simple top-mass observables, such as the inclusive
cross section with or without cuts, the reconstructed top mass, defined as
the mass of a system comprising the $W$ and a $b$ (not $\bar{b}$) jet, and,
as a simplified example of leptonic observable, the average value of the
energy of the final-state $W$ boson. As discussed earlier, we consider our
reconstructed top mass as an oversimplified representation of observables of
the kind used in the so called ``direct'' measurements. We also stress that
we consider the kinematic region where the top energy is not much larger than
its mass, that is the region typically used in direct measurements.

\section{Generalities on renormalons}
\label{sec:renorm_intro}
Infrared renormalons~\cite{tHooft:1979gj, Parisi:1978az} provide a connection
between the behaviour of the perturbative expansion at large orders in the
coupling constant and non-perturbative effects.  They arise when the the last
loop integration in the $(n+1)$-loop order of the perturbative expansion
acquires the form (see e.g.~\cite{Altarelli:1995kz, Beneke:1998ui})
\begin{equation}
\label{eq:factgrowth}
 \as^{n+1}(Q) \frac{1}{Q^k} \int^Q \!\mathd l\, l^{k-1}\, b_0^n\,
\(\log \frac{Q^2}{l^2}\)^n
= n!\, \left(\frac{2b_0}{k} \right)^n \as^{n+1}(Q)\equiv c_{n+1}\,\as^{n+1}(Q)\,,
\end{equation}
where $Q$ is the typical scale involved in the process and $b_0$ is the first
coefficient of the QCD beta function
\begin{equation}
b_0 = \frac{11\, \CA}{12\pi}  - \frac{n_f \,\TR }{3\pi}\,.
\end{equation}
The coefficient $b_0$ arises because the running coupling is the source of
the logarithms in eq.~(\ref{eq:factgrowth}). A naive justification of the
behaviour illustrated in eq.~(\ref{eq:factgrowth}) can be given by
considering the calculation of an arbitrary dimensionless observable,
characterized by a scale $Q$, including the effect of the exchange or
emission of a single gluon with momentum $l$, leading to a correction that,
for small $l$, takes the form
\begin{equation}
  \label{eq:irsensitivity}
 \frac{1}{Q^k} \int^Q \!\mathd l\, l^{k-1}  \as,
\end{equation}
where $k$ is an integer greater than zero for the result to be
infrared finite.  Assuming that higher order corrections will lead to the
replacement of $\as$ with the running coupling evaluated at the scale
$l$, given by the geometric expansion
\begin{equation}
  \label{eq:alphsrunning}
  \as(l)=\frac{1}{b_0 \log \frac{l^2}{\LambdaQCD^2}}=\frac{\as(Q)}{1-\as(Q)\, b_0
    \log \frac{Q^2}{l^2}} 
  = \sum_0^\infty \as^{n+1}(Q)\, b_0^n \,\log^n \frac{Q^2}{l^2}\,,
\end{equation}
substituting eq.~(\ref{eq:alphsrunning}) into eq.~(\ref{eq:irsensitivity}),
we obtain the behaviour of eq.~(\ref{eq:factgrowth}).

The coefficients of the perturbative expansion display a factorial growth.
The series is not convergent and can at most be interpreted as an asymptotic
series. In general, the terms of the series decrease for low values of $n$,
until they reach a minimum, and then they start to increase with the order.
The minimum is reached when
\begin{equation}
  c_n\, \as^n(Q) \approx c_{n+1}\, \as^{n+1}(Q)\,,
\end{equation}
that corresponds to $n \approx k/(2 b_0 \as(Q))$, and the size of the minimal
term is
\begin{eqnarray}
 n!  \left(\frac{2 b_0}{k}\right)^{n} \as^{n+1}(Q) &\approx&
 Q^k\,\as(Q)\,n^{-n}  \(n^{n+1/2} e^{-n}\)
 \nonumber \\
  &\approx& \as(Q) \, n^{\frac{1}{2}}
\exp\left(-\frac{k}{2\, b_0\, \as(Q)}\right) \approx  \sqrt{\frac{k\,
    \as(Q)}{2b_0}}\left(\frac{\LambdaQCD}{Q}\right)^k. 
\end{eqnarray}
The value of $k$ depends upon the process under consideration.  In this
paper, we are interested in \emph{linear} IR renormalons, corresponding to
$k=1$, that can lead to ambiguities in the measured mass of the top quark of
relative order $\LambdaQCD/m_t$, i.e.~ambiguities of order $\LambdaQCD$ in
the top mass. Larger values of $k$ lead to corrections of relative order
$\LambdaQCD^{k}/m_t^k$, that are totally negligible.

It is in general not possible to compute the normalization of the tower of
factorially growing terms in non-trivial field theories. There is, however, a
context where this calculation simplifies to such an extent that it can be
carried out exactly. This is the leading number of flavors approximation, in
which one considers the corrections given by the exchange of a single gluon,
including all possible vacuum-polarization diagrams given by a single fermion
loop. Each vacuum polarization diagram yields a factor of $\as n_f$, where we
denote by $n_f$ a fictitious number of light flavours, so that, in the large
$n_f$ limit, these contributions are dominant. In order to obtain an estimate
of the renormalon effects in the full non-Abelian theory, at the end of the
calculation one performs the replacement $n_f\to -11\CA/(4\TR)+n_l$, where
$n_l$ is the true number of light flavours in the theory. This leads to the
correct, non-Abelian running of the coupling constant.  This procedure, known
as the ``large-$b_0$ approximation'', has been used in several
contexts~\cite{Beneke:1998ui}, and it leads to reasonable results.

In this work, we study renormalon effects on top-mass related observables in
the large-$b_0$ approximation. We know that, in this framework, there are
renormalons arising in the computation of the position of the pole in the top
propagators, and we also know that there must be renormalons associated to
jets requirements. We will also be able to compute the
perturbative expansion order by order in perturbation theory, and thus
determine explicitly the effects of renormalons in the perturbative
expansion.

Our results can be given in terms of the top mass expressed either in the
pole or in the \MSB{} mass scheme. We know that the expression of the pole
mass in terms of the \MSB{} mass has a linear renormalon.  If the \MSB{} mass
is considered a fundamental parameter of the theory, this is to be interpreted
as an uncertainty of the order of a typical hadronic scale associated to the
position of the pole in the top propagator. One may wonder whether the pole
mass could instead be used as a fundamental parameter of the theory, which
would imply that the \MSB{} mass has an uncertainty of the order of a
hadronic scale. In fact, it is well known and clear (but nevertheless we wish
to stress it again) that this last point of view is incorrect. QCD is
characterized by a short distance Lagrangian, and its defining parameters are
short distance parameters.  Thus, if we compute an observable in terms of the
\MSB{} mass, and we find that it has no linear renormalons, we can conclude
that the observable has no \emph{physical} linear renormalons, since its
perturbative expansion in terms of the parameters of the short distance
Lagrangian has no linear renormalons. On the other end, in the opposite case
of an observable that has no linear renormalons if expressed in terms of the
pole mass, we must conclude that this observable has a physical renormalon,
that is precisely the one that is contained in the pole mass.  We also stress
that it is the \MSB{} mass that should enter more naturally in the
electroweak fits~\cite{Patrignani:2016xqp-EWreview, deBlas:2016ojx,
  Baak:2014ora} and in the calculations relative to the stability of the
vacuum~\cite{Degrassi:2012ry, Buttazzo:2013uya, Andreassen:2017rzq,
  Chigusa:2017dux}, although in practice the pole mass if often used also in
these contexts.

\section{Description of the calculation}
\label{sec:description_calc}
\begin{figure}[htb]
 \centering
 \begin{subfigure}{0.45\textwidth}
 \includegraphics[width=\textwidth]{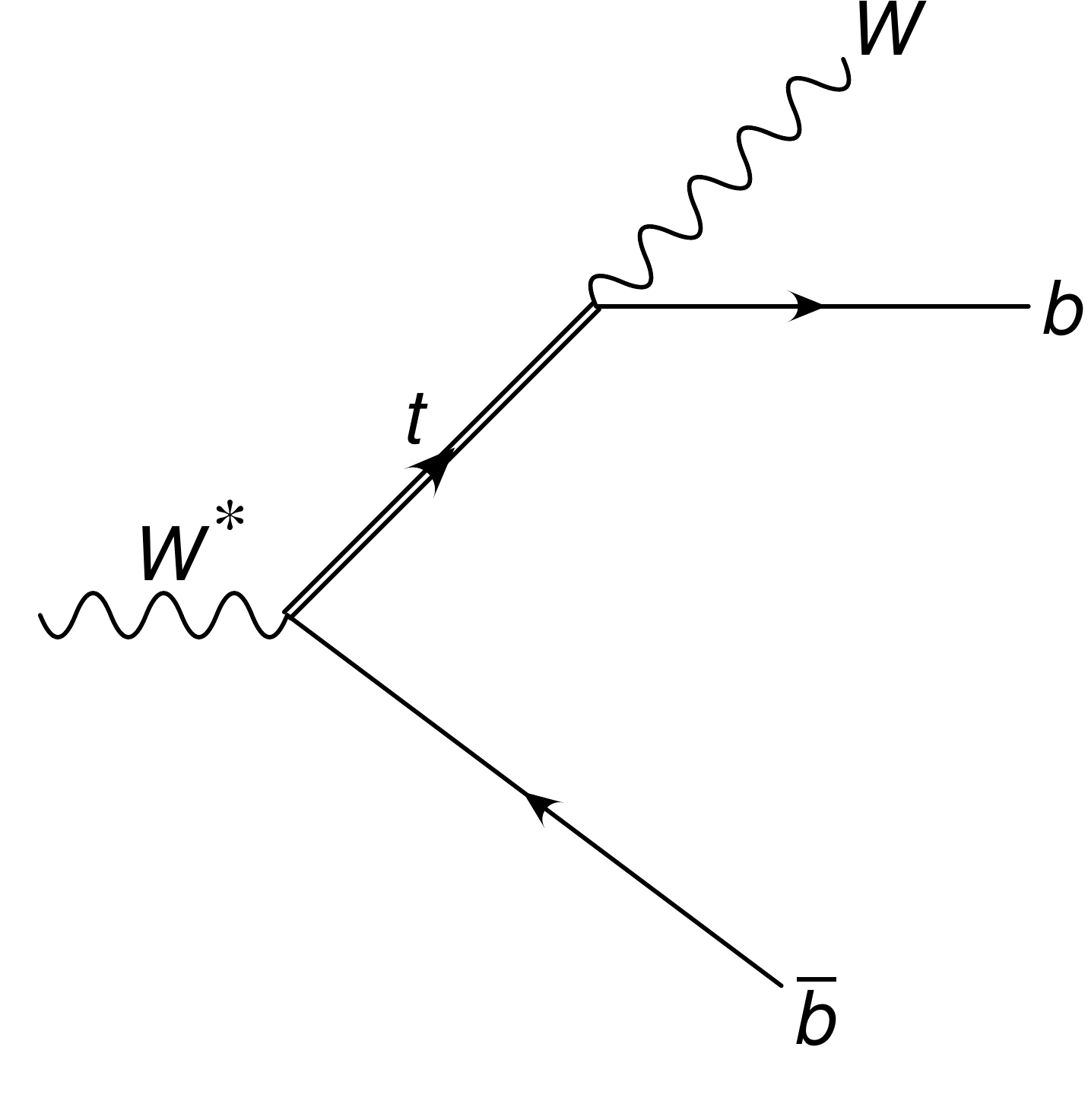}
 \caption{}
 \end{subfigure}
 \begin{subfigure}{0.45\textwidth}
 \includegraphics[width=\textwidth]{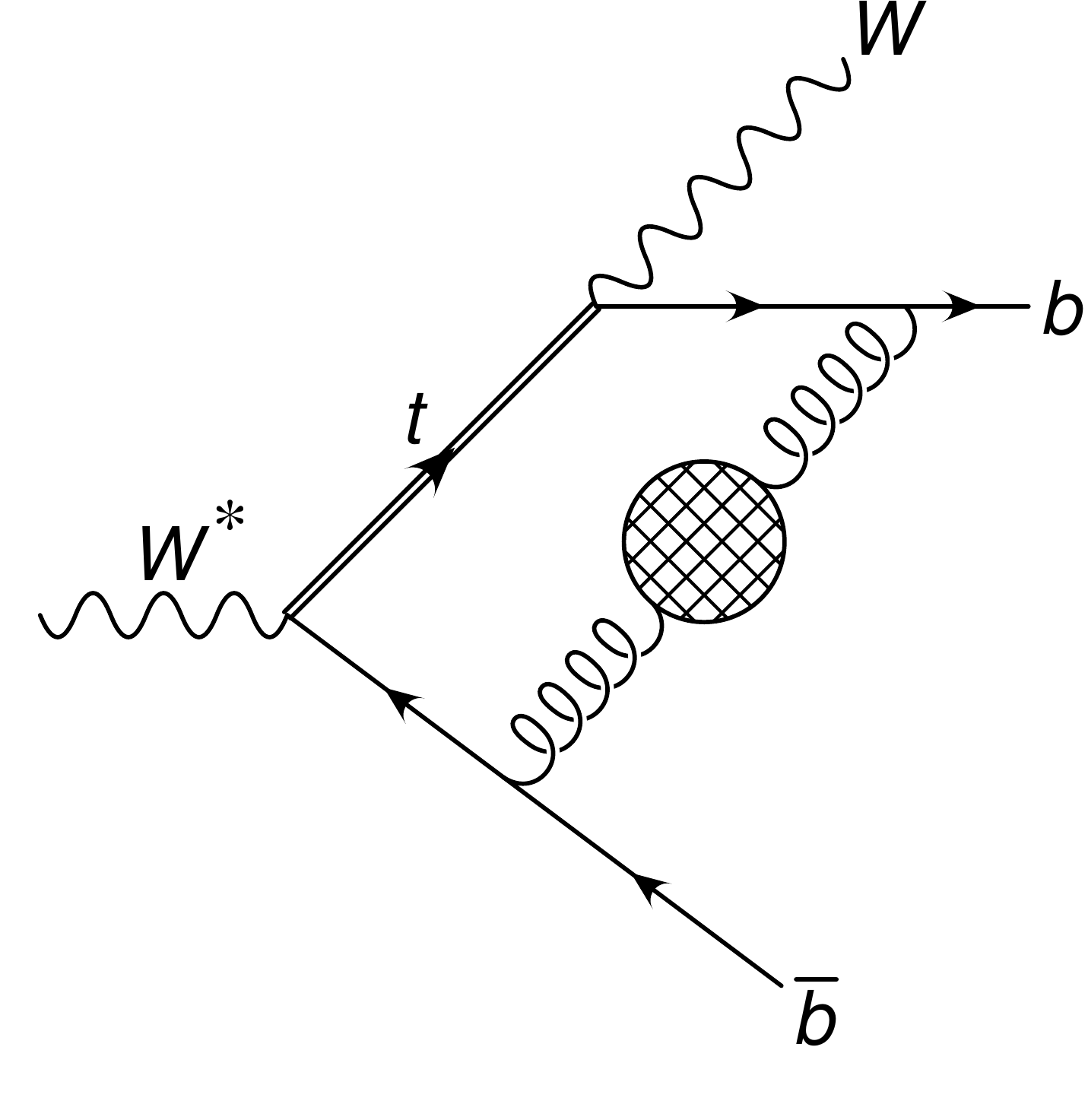}
 \caption{}
 \end{subfigure}

 \begin{subfigure}{0.45\textwidth}
 \includegraphics[width=\textwidth]{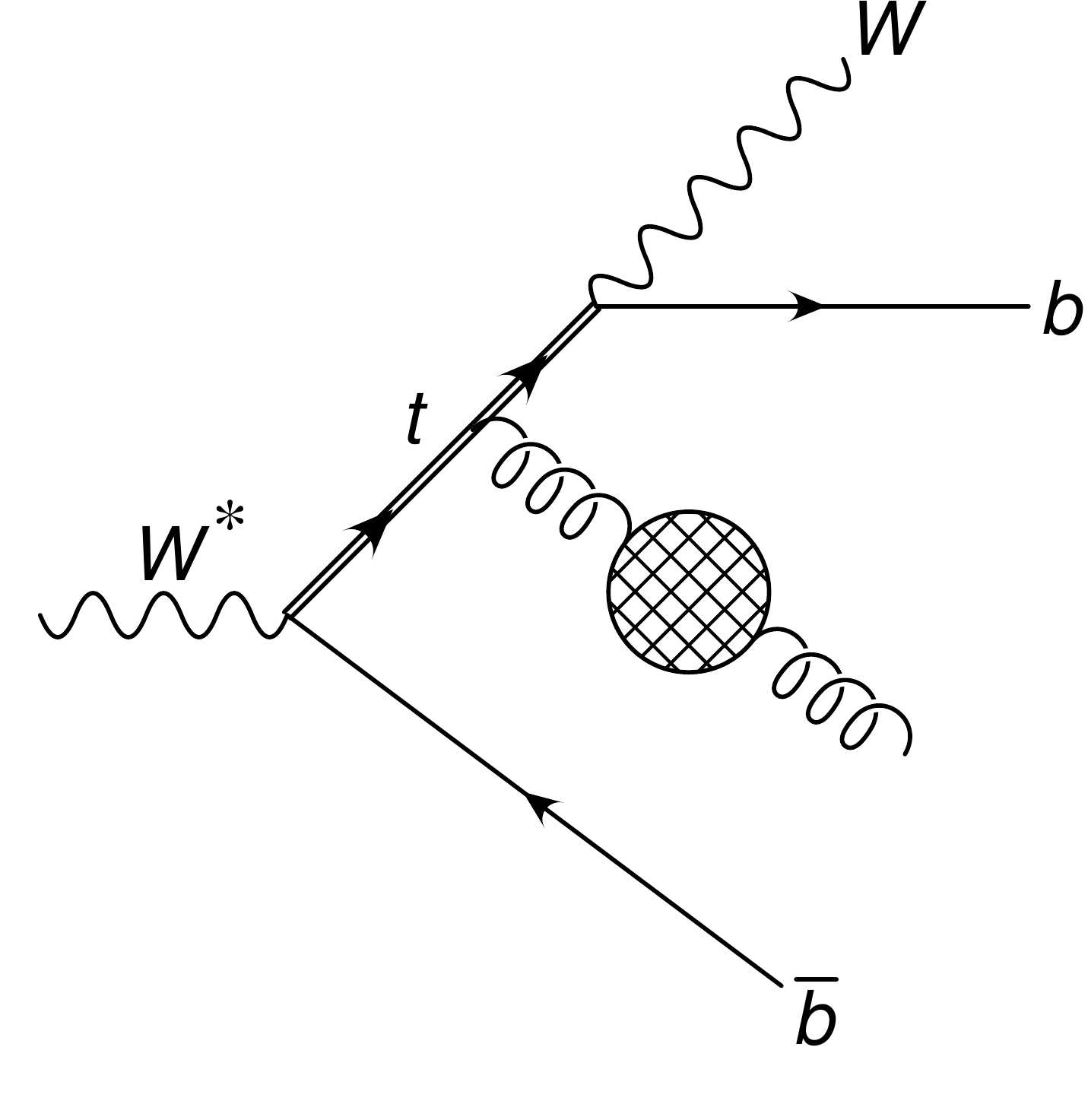}
 \caption{}
 \end{subfigure}
 \begin{subfigure}{0.45\textwidth}
 \includegraphics[width=\textwidth]{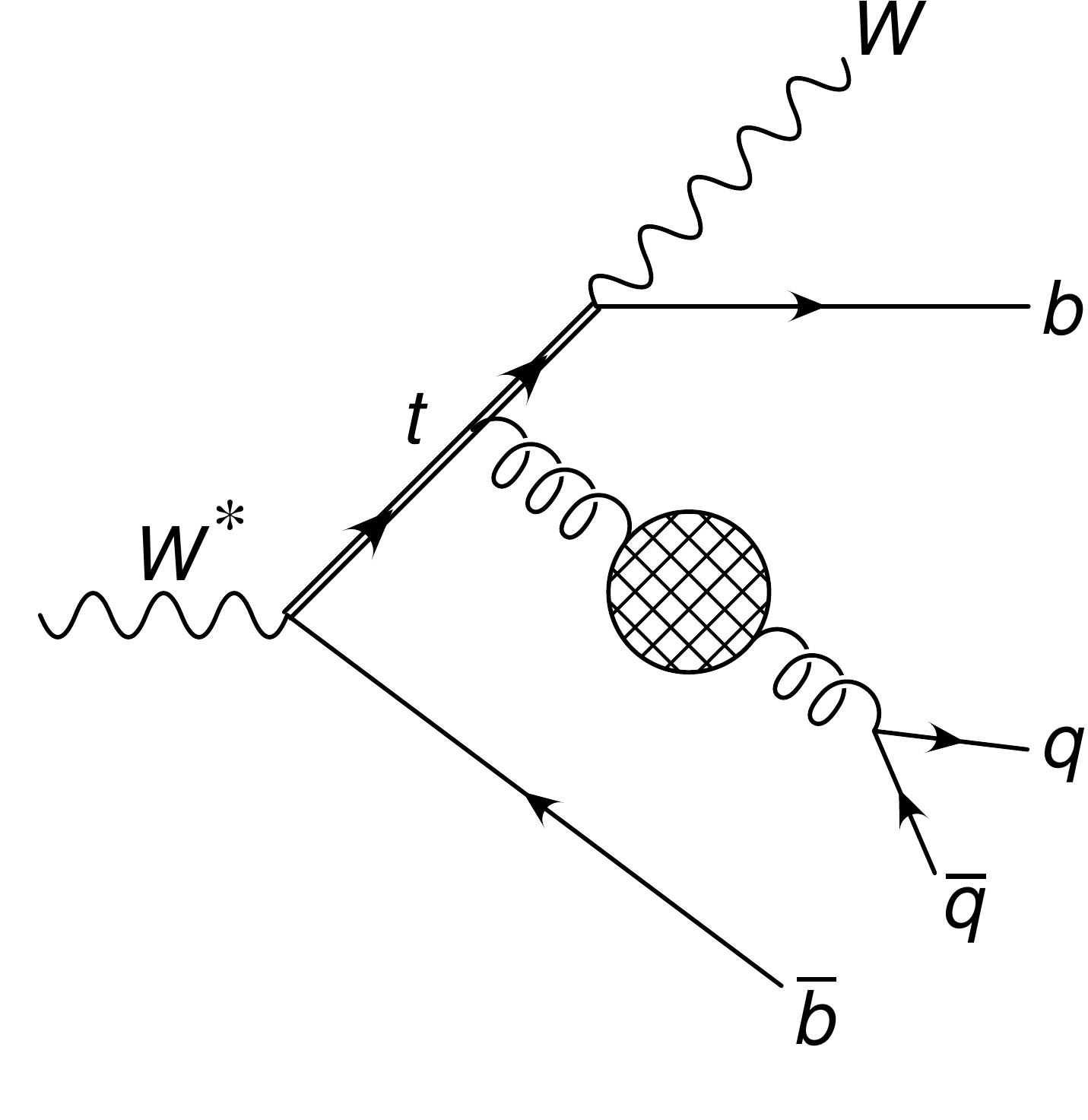}
 \caption{}
 \end{subfigure}

 \vspace{1cm}
 
 \begin{subfigure}{\textwidth}
\centerline{\includegraphics[width=0.9\textwidth]{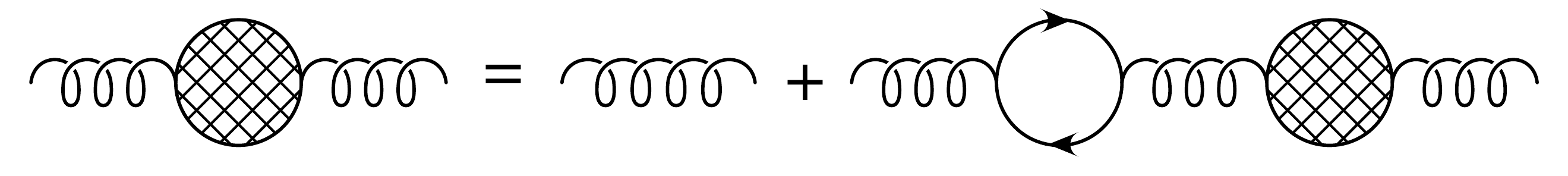}}
   \caption{}
 \end{subfigure}
 \caption{Feynman diagram for the Born $W^*\to W b\bar{b}$ process~(a),
 and samples of Feynman diagrams for the virtual contribution~(b), for the
 real-emission contribution~(c) and for $W^*\to W b\bar{b}\, q\bar{q}$
 production~(d).
 \label{fig:wbbbar}}
\end{figure}

A sample of Feynman diagrams contributing to the process $W^* \rightarrow
t\bar{b} \to W b \bar{b}$ is depicted in Fig.~\ref{fig:wbbbar}.  The dashed
blob represents the summation of all self-energy insertion in the large-$n_f$
limit.

We want to compute a generic observable, function of the final-state
kinematics $\Phi$, that we denote with $O(\Phi)$.  We assume the eventual
presence of a set of cuts $\Theta(\Phi)$, also function of the final-state
kinematics, and define
\begin{equation}
  O_\Theta(\Phi)=O(\Phi) \times \Theta(\Phi) .
\end{equation}
The average value of $O$ can be written as
\begin{eqnarray}
  \label{eq:def_aveM}
  \langle O \rangle & = & N_\Theta \lg
  \int \mathd \Phi_{\rm b} \,\sigma_{\rm b} (\Phi_{\rm b}) \, O_\Theta (\Phi_{\rm b})
  +  \int \mathd \Phi_{\rm b} \,\sigma_{\rm v} (\Phi_{\rm b}) \, O_\Theta (\Phi_{\rm b})
  +   \int \mathd \Phi_g \, \sigma_g (\Phi_g) \, O_\Theta (\Phi_g) \right. \nonumber \\
  && \hspace{0.7cm} \left . +   \int \mathd \Phi_{ q \bar{q}} \,\sigma_{q
    \bar{q}} (\Phi_{q \bar{q}}) \, O_\Theta (\Phi_{q \bar{q}}) \rg, 
\end{eqnarray}
where the first term represents the Born contribution, the second the virtual
one, the third the one due to the emission of a real gluon and the fourth
represents the contribution of the real production of $n_f$ $q \bar{q}$
pairs.  Equation~(\ref{eq:def_aveM}) implicitly defines our notation for the
different phase space integration volumes.

We always imply that the gluon propagator, in the last three contributions,
includes the sum of all vacuum-polarization insertions of light-quark loops.

$N_\Theta$ is a normalization factor, given by
\begin{eqnarray}
  \label{eq:def_N}
  N_\Theta^{- 1} &=& \int \mathd \Phi_{\rm b} \,\sigma_{\rm b}
  (\Phi_{\rm b}) \, \Theta (\Phi_{\rm b})
  + \int \mathd \Phi_{\rm b} \,\sigma_{\rm v}  (\Phi_{\rm b}) \, \Theta (\Phi_{\rm b})
  \nonumber \\
 &&  + \int \mathd \Phi_g \, \sigma_g (\Phi_g) \, \Theta (\Phi_g)
   + \int \mathd \Phi_{ q \bar{q}}
  \,\sigma_{q \bar{q}} (\Phi_{q \bar{q}}) \, \Theta (\Phi_{q \bar{q}}) \,.
\end{eqnarray}
 We can then rewrite eq.~(\ref{eq:def_aveM}) as
\begin{eqnarray}
  \label{eq:Mave}
  \langle O \rangle & = & N_\Theta \lg \int \mathd \Phi_{\rm b} \, \sigma_{\rm b}
  (\Phi_{\rm b}) \, O_\Theta (\Phi_{\rm b})
  +   \int \mathd \Phi_{\rm b} \, \sigma_{\rm v}(\Phi_{\rm b}) \, O_\Theta
  (\Phi_{\rm b}) \right.
   +   \int \mathd \Phi_g \, \sigma_g(\Phi_g)
                          \, O_\Theta (\Phi_g)
  \nonumber  \\
  &  & 
  +\int  \mathd  \Phi_{q \bar{q}} \,
   \sigma_{q \bar{q}} (\Phi_{q \bar{q}}) \, O_\Theta\(\Phi_{g^{*}}\) 
 + \left. \int \mathd  \Phi_{q \bar{q}} \, \sigma_{q \bar{q}}
  (\Phi_{q \bar{q}}) \, \lq O_\Theta (\Phi_{q \bar{q}}) - O_\Theta  (\Phi_{g^{*}})\rq \rg,
\end{eqnarray}
where we have subtracted and added the same quantity to the $q\bar{q}$
contribution.  In the last two lines, $O_\Theta(\Phi_{g^{*}})$ is defined in
terms of the $\Phi_{g^{*}}$ phase space, that is obtained from the $\Phi_{q \bar{q}}$
phase space by clustering the $q\bar{q}$ pair into a single pseudoparticle,
that we denote with $g^*$.  We also define
\begin{eqnarray}
  \label{eq:Mv}
  \langle O \rangle_{\rm v} & \equiv & N_\Theta \int \mathd \Phi_{\rm b} \,
  \sigma_{\rm v} (\Phi_{\rm b}) \, O_\Theta (\Phi_{\rm b})\,,
  \\
    \label{eq:Mg}
  \langle O \rangle_g &  \equiv  & N_\Theta \int \mathd \Phi_g  \, \sigma_g (\Phi_g) \, O_\Theta
  (\Phi_g) + N_\Theta \int \mathd
  \Phi_{q \bar{q}}  \,\sigma_{q \bar{q}} (\Phi_{q \bar{q}}) \, O_\Theta
  (\Phi_{g^{*}})\,,
  \\
    \label{eq:Mqq}
  \langle O \rangle_{q \bar{q}} & \equiv & N_\Theta \int \mathd \Phi_{q \bar{q}}\, \sigma_{q \bar{q}}
  (\Phi_{q \bar{q}}) \, \lq O_\Theta (\Phi_{q \bar{q}}) - O_\Theta
    (\Phi_{g^{*}})\rq .
\end{eqnarray}

\subsection{The  normalization factor}
\label{sec:normalization_factor}
The factor $N_\Theta$ appearing in eqs.~(\ref{eq:Mv})--(\ref{eq:Mqq}) is in
fact simply the inverse of the Born cross section, since the quantities it
multiplies are already at NLO level.  Thus, in these cases,
\begin{equation}
  \label{eq:N0}
 N_\Theta \rightarrow N^{(0)}_\Theta = \lg \int\mathd \Phi_{\rm b}\,
 \sigma_{\rm b} (\Phi_{\rm b})\,\Theta\(\Phi_{\rm b}\)\rg^{-1}. 
\end{equation}
The factor of $N_\Theta$ in front of the Born term in eq.~(\ref{eq:Mave}), on
the other hand, generates extra contributions of the form
\begin{eqnarray}
  N_\Theta &=&
  \lg \int \mathd \Phi_{\rm b}\,\sigma_{\rm b}\(\Phi_{\rm b}\)\Theta\(\Phi_{\rm b}\)
 + \int \mathd \Phi_{\rm b}\,\sigma_{\rm v}\(\Phi_{\rm b}\)\Theta\(\Phi_{\rm b}\)
 + \int  \mathd \Phi_g  \,\sigma_g (\Phi_g)\Theta\(\Phi_g\) \right.
 \nonumber \\
 && \left.  + \int \mathd \Phi_{q \bar{q}} \,\sigma_{q \bar{q}} (\Phi_{q
     \bar{q}})\Theta\(\Phi_{q\bar{q}}\) \rg ^{-1}
   \nonumber\\
   & = & N^{(0)}_\Theta  \lg  1 - N^{(0)}_\Theta
   \lq
\int \mathd \Phi_{\rm b}\,\sigma_{\rm v}\(\Phi_{\rm b}\)\Theta\(\Phi_{\rm b}\)
+ \int  \mathd \Phi_g  \,\sigma_g (\Phi_g)\Theta\(\Phi_g\) \right.\right.
\nonumber \\
 && \hspace{2.5cm}\left.\left.  + \int \mathd \Phi_{q \bar{q}} \,\sigma_{q \bar{q}} (\Phi_{q
     \bar{q}})\Theta\(\Phi_{q\bar{q}}\) 
\rq \rg +  \mathcal{O}\(\as^2 \(\as \TF \)^n\),  \phantom{aaaa}
\end{eqnarray}
where
  \begin{equation}
    \TF = n_f \, \TR.
  \end{equation}
This gives rise to a Born term of the form
\begin{equation}
 \obsb \equiv N^{(0)}_\Theta  \int \mathd
 \Phi_{\rm b}\, \sigma_{\rm b} 
   (\Phi_{\rm b}) \, O_\Theta (\Phi_{\rm b})\, , 
\end{equation}
plus an NLO correction equal to
\begin{equation}
  - N^{(0)}_\Theta \, \obsb  \lq \int \mathd \Phi_{\rm b} \, \sigma_{\rm
      v}(\Phi_{\rm b})\,\Theta(\Phi_{\rm b}) 
  +  \int \mathd \Phi_g \, \sigma_g (\Phi_g)\,\Theta(\Phi_g) \,
  + \int \mathd
   \Phi_{q \bar{q}}\,  \sigma_{q \bar{q}} (\Phi_{q \bar{q}})\,\Theta(\Phi_{q \bar{q}}) \rq\! . 
\end{equation}
In summary, eq.~(\ref{eq:Mave}) becomes
\begin{eqnarray}
  \label{eq:Mave_final}
  \langle O \rangle =   \obsb\!\!\!\! &&{} +  N^{(0)}_\Theta  \int \mathd
  \Phi_{\rm b} \, \sigma_{\rm v} 
  (\Phi_{\rm b}) \lq O_\Theta  (\Phi_{\rm b}) -  \obsb\Theta(\Phi_{\rm b}) \rq
  \nonumber\\
  && {}+  N^{(0)}_\Theta  
  \int \mathd \Phi_g \, \sigma_g(\Phi_g) \lq O_\Theta  (\Phi_g) -  \obsb \Theta(\Phi_{g})\rq
  \nonumber \\
   && {}+  N^{(0)}_\Theta  \int
  \mathd  \Phi_{q \bar{q}} \, \sigma_{q \bar{q}} (\Phi_{q \bar{q}}) \lq
    O_\Theta (\Phi_{g^{*}})  -  \obsb  \Theta(\Phi_{g^{*}})\rq
  \nonumber  \\
 && {}+  N^{(0)}_\Theta  \int \mathd  \Phi_{q \bar{q}} \, \sigma_{q \bar{q}}
  (\Phi_{q \bar{q}}) \,
 \nonumber \\
 && \hspace{1cm}\times \lg \lq  O_\Theta(\Phi_{q \bar{q}})  -  \obsb
   \Theta(\Phi_{q \bar{q}})\rq
 - \lq  O_\Theta(\Phi_{g^{*}})
   - \obsb  \Theta(\Phi_{g^{*}}) \rq\rg\!.\phantom{aaaa}
\end{eqnarray}

\subsection{Final results}

In App.~\ref{app:details_calc} we prove that the full result with the gluon
propagator dressed with all fermionic self-energy corrections can be computed
in terms of the matrix elements for the process $W^* \to W b \bar{b}$ with
the real emission or virtual exchange of one massive gluon of mass $\lambda$, and
the matrix element for the $W^* \to W b \bar{b} q\bar{q}$ tree-level process.

The general result for the average value of a generic observable $O$, in the
presence of final-state cuts $\Theta$, obtained by combining the results of
Sec.~\ref{sec:normalization_factor} and App.~\ref{sec:dressed_gluon}
and~\ref{app:details_calc}, is
\begin{equation}
  \label{eq:final_M_wcuts}
  \langle O \rangle = \obsb -
  \frac{1}{b_0\,\as}
   \int_0^{\infty}
  \!\frac{\mathd \lambda}{\pi} \, \frac{\mathd
    \widetilde{T}(\lambda)}{\mathd \lambda}
  \,\atanpilam,
\end{equation}
where
\begin{equation}
   \as=\as(\mu),\quad\quad \mu_{\sss C} = \mu\, e^{\frac{C}{2}},\quad\quad
   C=\frac{5}{3},\quad\quad b_0=-\frac{\TF}{3\pi}, 
\end{equation}
\begin{eqnarray} \label{eq:Mbdef}
   \obsb &=&  N^{(0)}_\Theta  \int \mathd \Phi_{\rm b}\, \sigma_{\rm b}
   (\Phi_{\rm b}) \, O (\Phi_{\rm b}) \, \Theta(\Phi_{\rm b})  \,,
   \\[1.5mm]
 \label{eq:Ttdef}
 \widetilde{T}\!\(\lambda\) & = &
 \widetilde{V}\!\(\lambda\) + \widetilde{R}\!\(\lambda\) + 
 \widetilde{\Delta}\(\lambda\)\,,
 \\[1.5mm]
 \label{eq:final_Vtilde}
 \widetilde{V}\!\(\lambda\) &=& N^{(0)}_\Theta \int \mathd \Phi_{\rm b} \, \sigma_{\rm v}^{(1)}
 (\lambda,\Phi_{\rm b})\, \lq O (\Phi_{\rm b}) -  \obsb \rq \Theta(\Phi_{\rm b})\, ,
 \\[1.5mm]
  \label{eq:final_Rtilde}
 \widetilde{R}\!\(\lambda\) &=& N^{(0)}_\Theta \int \mathd \Phi_{g^{*}}\, \sigma^{(1)}_{g^{*} } (\lambda,
 \Phi_{g^{*}})   \lq
    O(\Phi_{g^{*}})  -  \obsb \rq \Theta(\Phi_{g^{*}})\,,
 \\[1.5mm]
   \widetilde{\Delta}\!\(\lambda\) &=&  \, \frac{3\pi}{\as  \TF} \,
   N^{(0)}_\Theta  \lambda^2  \int  \mathd 
  \Phi_{q \bar{q}} \, \delta\!\(\lambda^2-k^2\) \sigma^{(2)}_{q \bar{q}} (\Phi_{q \bar{q}})
  \nonumber\\[1.5mm]
  \label{eq:final_Deltaqqtilde}
&& \hspace{1cm} \times \lg
  \lq  O(\Phi_{q \bar{q}})  -  \obsb \rq \Theta(\Phi_{q \bar{q}})
  - \lq  O(\Phi_{g^{*}})  -  \obsb \rq \Theta(\Phi_{g^{*}})\rg \,,
\end{eqnarray}
and $\sigma^{(1)}_{g^{*}}(\lambda, \Phi_{g^{*}})$ and $\sigma_{\rm
  v}^{(1)}(\lambda,\Phi_{\rm b})$ are the real/virtual corrections to the
process $W^* \to W b \bar{b}$ for the emission/exchange of a single gluon
with mass $\lambda$, and $\sigma^{(2)}_{q \bar{q}}(\Phi_{q \bar{q}})$ is the
tree-level cross section for the process $W^* \to W b \bar{b} q\bar{q}$.
We denote with $k$ the four-momentum of the $q\bar{q}$ pair.
Notice that, in eq.~(\ref{eq:final_M_wcuts}), $\as$ in $\widetilde{T}(\lambda)$
cancels against the $1/\as$ in front of the integral, and
\begin{equation}
\frac{\as}{1+b_0\,\as \log\displaystyle{\frac{\lambda^2}{\mu_{\sss
        C}^2}}}=\as\!\left(\lambda\, e^{-\frac{C}{2}}\right), 
\end{equation}
so that the resummed result for $\langle O \rangle$ does not depend upon the
value of $\mu$.  As discussed in App.~\ref{app:details_calc},
$\widetilde{T}(\lambda)$ vanishes for large $\lambda$, so that the integral in
eq.~(\ref{eq:final_M_wcuts}) is convergent.

In order to use the above formulae, we computed analytically the cross
sections $\sigma^{(2)}_{q \bar{q}} (\Phi_{q \bar{q}})$,
$\sigma^{(1)}_{g^{*}}(\lambda,\Phi_{g^{*}})$ and $\sigma_{\rm v}^{(1)}
(\lambda,\Phi_{\rm b})$.  Due to the finite gluon mass, only ultraviolet
divergences arise in the intermediate steps of the calculation. These
divergences were dealt with in dimensional regularization. After the mass
renormalization has been carried out (adopting a complex
mass~\cite{Denner:2005fg, Denner:2006ic} in order to account for the finite
top width), the UV divergences cancel in the virtual contribution because of
the vector nature for the incoming $W^*$ current.  For reasons that will
become clear later, we have also computed the same cross sections for
$\lambda=0$. In this case, also soft and collinear divergences are treated in
dimensional regularization, and the full result is obtained applying a
subtraction method. Notice that, for a finite gluon mass, large logarithms of
the mass arise in the real and virtual contributions, that cancel in the
sum. In the massless limit, these large cancellations are handled by the
subtraction method, and do not affect the accuracy of the result.

We evaluated the scalar integrals using \Collier~\cite{Denner:2016kdg}.  The
final numerical implementation has been built using the \RES{}
framework~\cite{Jezo:2015aia}.

We performed the phase-space integral for $\widetilde{V}$, $\widetilde{R}$,
$\widetilde{\Delta}$ numerically for several values of $\lambda$. For small $\lambda$
both $\widetilde{V}$ and $\widetilde{R}$ have logs of $\lambda$ that cancel in
the sum, so that one recovers the result corresponding to the NLO corrections
to the $W^*\to W b {\bar b}$ process involving the exchange or emission of a
single massless gluon. The $\widetilde{\Delta}$ term is instead finite by
itself, and vanishes for small $\lambda$.  We then combine these
results, for each observable, in our function $\widetilde{T}$ that we fit as
a function of $\lambda$. This allows us to compute the
coefficients of the perturbative expansion of our observable at any order in
perturbation theory, and also to determine its asymptotic behaviour.

We find that, in general, the behaviour of $\widetilde{T}$ for small $\lambda$ is
given by a constant plus a linear term in $\lambda$. It is this linear term that is
associated with linear renormalons.  As shown in Sec.~\ref{sec:renorm_intro},
these correspond to power suppressed contributions of order $\Lambda^p$ with
$p=1$, where $\Lambda$ is a typical hadronic scale.  Higher values of $p$
arise from higher powers of $\lambda$ in the expansion of $\widetilde{T}$.  In the
present work, we are interested only in $p=1$, since, because of the size of
the top mass, higher values are suppressed by a further $\Lambda/m_t$ factor.

The inclusive cross section, with or without cuts, is given by formulae
similar to the ones from (\ref{eq:Mbdef}) to (\ref{eq:final_Deltaqqtilde}),
setting $O=1$ and omitting the normalization factor $N^{(0)}_\Theta$. We
then write
\begin{equation}
  \label{eq:final_sigma_wcuts}
  \sigma = \sigma_{\rm b}
  - \frac{1}{b_0\, \as}
  \int_0^{\infty}
  \!\frac{\mathd \lambda}{\pi} \, \frac{\mathd T(\lambda) }{\mathd \lambda}
        \atanpilam,
\end{equation}
where
\begin{eqnarray} \label{eq:sbdef}
   \sigma_{\rm b}  &=&  \int \mathd \Phi_{\rm b}\, \sigma_{\rm b}
   (\Phi_{\rm b}) \, \Theta(\Phi_{\rm b})  \,,
   \\[1.5mm]
 \label{eq:Tdef}
 T(\lambda) & = & V(\lambda) + {R}\!\(\lambda\) +  \Delta(\lambda),
 \\[1.5mm]
 \label{eq:final_V}
 V(\lambda) &=&  \int \mathd \Phi_{\rm b} \, \sigma_{\rm v}^{(1)}
 (\lambda,\Phi_{\rm b}) \, \Theta(\Phi_{\rm b})\, ,
 \\[1.5mm]
  \label{eq:final_R}
 R(\lambda) &=& \int \mathd \Phi_{g^{*}}\, \sigma^{(1)}_{g^{*} } (\lambda,
 \Phi_{g^{*}})  \,  \Theta(\Phi_{g^{*}})\,,
 \\[1.5mm]
   \label{eq:final_Deltaqq}
 \Delta(\lambda) &=&   \frac{3\pi}{\as  \TF} \,
   \lambda^2 \int  \mathd \Phi_{q \bar{q}} \; \delta\!\(\lambda^2-k^2\) \sigma_{q\bar{q}}
  (\Phi_{q \bar{q}})  \lq \Theta(\Phi_{q \bar{q}})  -  \Theta(\Phi_{g^{*}})\rq .
\end{eqnarray}
We notice that when computing inclusive quantities or quantities that do not
depend upon the jet kinematics, the $\widetilde{\Delta}(\lambda)$ and
$\Delta(\lambda)$ terms of eqs.~(\ref{eq:final_Deltaqqtilde})
and~(\ref{eq:final_Deltaqq}) are zero. In these cases, our results can just
be expressed as functions of the NLO differential cross section computed
with a non-zero gluon mass.  In general, however, the
$\widetilde{\Delta}(\lambda)$ and $\Delta(\lambda)$ contributions cannot be
neglected, since observables built with the full kinematics may differ from
those obtained by clustering the $q\bar{q}$ pair into a massive gluon.  This
was first discussed in Ref.~\cite{Nason:1995np}, in the context of $e^+ e^-$
annihilation into jets.\footnote{In Refs.~\cite{Dokshitzer:1997iz,
    Dokshitzer:1998pt} it was shown that, for a large set of jet-shape
  observables, in order to account for the effect of the $\Delta$ term, the
  naive predictions computed considering only the $V+R$ contributions must be
  rescaled by a factor, dubbed the ``Milan factor'', to get the correct
  coefficient for the $1/Q$ non-perturbative effects.}

\subsection{Changing the mass scheme}
\label{sec:changescheme}
The relation between the pole mass $m$ and the \MSB{} mass $\mMSB$ in the
large-$n_f$ limit is discussed in App.~\ref{app:PoleMSB}. We have
\begin{equation}
  \label{eq:msbpole}
  \mMSB(\mu) = m \left\{ 1 - \as \left[ r_f(\mpole,\mu,\as) +
    r_d^{\rm (f)}(\mpole,\mu,\as) \right] + \mathcal{O}(\as^2(\as b_0)^n)
  \right\}\,,
\end{equation}
where $r_f$ and $r_d$ are defined in eqs.~(\ref{eq:rtilde_f})
and~(\ref{eq:rtilde_d}) respectively, and $r_d^{\rm (f)}$ is the finite part
of $r_d$, that does not receive any contribution from
linear terms in $\lambda$.  The $r_f$ contribution can be written in the form
\begin{equation}
  r_f(\mpole,\mu,\as) =
 -  \frac{1}{b_0\,\as}
\int_{0 }^{\infty}
\frac{\mathd \lambda}{\pi} \, \frac{\mathd\ }{\mathd \lambda}
\lq r_{\lambda,f}\!\(\mpole,\mu\) \rq
\atanpilam,
\end{equation}
where (see eq.~(\ref{eq:rf_small_lambda}))
\begin{equation}
  \label{eq:rlamf_small_lambda}
  r_{\lambda,f}\!\(\mpole,\mu\) = - \frac{\CF}{2} \frac{\lambda}{m}
  + {\cal O}\(\lambda^2\)\,.
\end{equation}
Note that the $\mu$ dependence disappears in the leading term.  The ${\cal
  O}(\lambda)$ term in eq.~(\ref{eq:rlamf_small_lambda}) is responsible for
the presence of a linear renormalon in the relation between the pole mass and
the \MSB{} one.\footnote{The relation between the pole and the \MSB{} mass in
  the large-$n_f$ limit is well-known (see e.g.~Refs.~\cite{Beneke:1994rs,
    Beneke:1994qe, Beneke:1994sw}).  Here we have re-derived it so as to put
  it in a form similar to eqs.~(\ref{eq:final_M_wcuts})
  and~(\ref{eq:final_sigma_wcuts}).}

In the present work we deal with the finite width of the top quark by using
the complex mass scheme~\cite{Denner:2005fg, Denner:2006ic}. Thus, in our
mass relation, both $m$ and $\mMSB$ are complex, and also $r_f$ and $r_d$.

Given a result for a quantity $\langle O \rangle$ expressed in terms of the
pole mass, representing the average value of some kinematic quantity
(possibly including cuts and possibly normalized to the total cross section),
in order to find its expression in terms of the \MSB{} mass we need to
Taylor-expand its mass dependence in its leading order expression, and
multiply it by the appropriate mass correction. In order to do so, we express
$O$ in terms of the pole mass and its complex conjugate, as if they were
independent variables (one can think of $m$ appearing in the amplitude, and
$m^*$ appearing in its complex conjugate). Denoting with $\obsb$ the LO
prediction, we can write
\begin{eqnarray}
  \label{eq:changescheme}
  \obsb(m,m^*)\!&=&\!\obsb(\mMSB,\mMSB^*) +\lg
  \frac{\partial
      \obsb(\mMSB,\mMSB^*) }{\partial \mMSB}\(m-\mMSB\)+{\rm cc}\rg
                  \nonumber \\
  &\approx & \!\obsb(\mMSB,\mMSB^*) +\lg \frac{\partial
             \obsb(\mpole,\mpole^*) }{\partial \mpole}\(m-\mMSB\)+{\rm cc}\rg
              \nonumber \\
  &= &\! \obsb(\mMSB,\mMSB^*)
  +\as\!\lg \frac{\partial \obsb(\mpole,\mpole^*) }{\partial
      \mpole}\, \mpole\! \lq r_f(\mpole,\mu,\as) +
      r^{\rm (f)}_d(\mpole,\mu,\as) \rq \! +{\rm
    cc}\rg\!,
  \nonumber\\
\end{eqnarray}
where we have neglected $\as^2(\as b_0)^n$ terms and we have dropped the
$\mu$ dependence in $\mMSB$ for ease of notation.  Notice that, as far as the
linear term in $\lambda$ is concerned, we get the simplified form
\begin{eqnarray}
\obsb(m,m^*)\!&=& \! \obsb(\mMSB,\mMSB^*)+\left[\frac{\partial
    \obsb(\mpole, \mpole^*) }{\partial \mpole}+{\rm cc}\right] 
\nonumber \\
&\times&\!\! 
\( -\frac{1}{b_0\, \as} \) \! \int_0^{\infty} \!\frac{\mathd
  \lambda}{\pi} \, \frac{\mathd\ }{\mathd \lambda} \!\lq -
\as\frac{\CF}{2} \lambda \rq
\atanpilam \,. 
\end{eqnarray}
Furthermore, we have
\begin{equation}
  \frac{\partial \obsb(\mpole,\mpole^*) }{\partial
    \mpole}+{\rm cc} =\frac{\partial \obsb(\mpole,\mpole^*) }{\partial
  \,  {\rm Re}(\mpole)}.
\end{equation}
Thus, when going from the pole to the \MSB{} mass scheme, the definition for
$\widetilde{T}$ is modified for small $\lambda$ into
\begin{equation}
  \label{eq:Ttchangescheme}
  \widetilde{T}(\lambda)  \to  \widetilde{T}(\lambda) - \frac{\partial
    \obsb(\mpole,\mpole^*) }{\partial\, {\rm Re}(\mpole)} \, \frac{\CF \as}{2}
  \, \lambda  + {\cal O}\!\(\lambda^2\)\,.
\end{equation}
One may wonder where the $\obsb$ subtraction term, that is present by
definition in $\widetilde{T}$, is hiding here.  In fact, in the case of
normalized observables, it should be kept in mind that $\obsb$ includes a
division by the total cross section. When taking the derivative, the
denominator is also derived, yielding the $\obsb$ subtraction term.

This procedure is still valid for a generic observable $O$ that does not
involve the normalization factor $N_\Theta$, like the total cross section, so
also in this case we need to replace $T$ with
\begin{equation}            
   \label{eq:Tchangescheme}
  T(\lambda)  \to  T(\lambda) -  \frac{\partial O_{\rm b}(\mpole,\mpole^*)}{\partial
   \, {\rm Re}(\mpole)} \,\frac{\CF \as}{2} \,\lambda + {\cal O}\!\(\lambda^2\)\,.
\end{equation}
Notice that the same expression holds for $\widetilde{T}$ and $T$. 

We also stress that eqs.~(\ref{eq:Ttchangescheme})
and~(\ref{eq:Tchangescheme}) also apply to any so called ``short distance''
mass schemes~\cite{Czarnecki:1997sz, Beneke:1998rk, Hoang:1998ng,
  Pineda:2001zq, Fleming:2007qr, Jain:2008gb, Hoang:2008yj}.  These schemes
are such that no mass renormalon affects their definition, and of course in
order for this to be the case, their small $\lambda$ behaviour should be the same
one of the \MSB{} scheme.

\section{Physical objects}
The numerical values of the parameters used in our study are given by
\begin{eqnarray}
  \label{eq:m0}
  m_{\sss 0} & = & 172.5\,{\rm GeV}, \\
  \Gamma_t & = & 1.3279 \,{\rm GeV}, \\
  m & = & \sqrt{m_{\sss 0}^2 - i m_{\sss 0} \Gamma_t}, \\
  m_{\sss W} & = & 80.419\,{\rm GeV}, \\
  E_{\sss\rm CM} & = & 300 \,{\rm GeV}, \\
  \mu & = & m_{\sss 0}\,.
\end{eqnarray}
Furthermore we have set  $n_l=5$ and, from $\as(M_{\sss Z})=0.1181$, we have
\begin{equation}
  \as(\mu) = 0.108\,.
\end{equation}

\subsection{Selection cuts}
\label{sec:cuts}
In order to better mimic realistic experimental analyses adopted at hadron
colliders, at times we introduce selection cuts for our cross sections,
requiring the presence of a $b$ jet and a (separated) $\bar{b}$ jet, both
having energy greater than 30~GeV. Jets are reconstructed using the {\tt
  Fastjet}~\cite{Cacciari:2011ma} implementation of the anti-$k_t$
algorithm~\cite{Cacciari:2008gp} for $e^+ e^-$ collisions, for various values
of the radius parameter $R$.

\section{Inclusive cross section}
The formula for the inclusive cross section is given in
eq.~(\ref{eq:final_sigma_wcuts}), that will be applied both without and with
cuts.

\subsection{Inclusive cross section without cuts}
\label{sec:Xtot_wocuts}
\begin{figure}[tb!]
  \centering
  \includegraphics[width=0.65\textwidth]{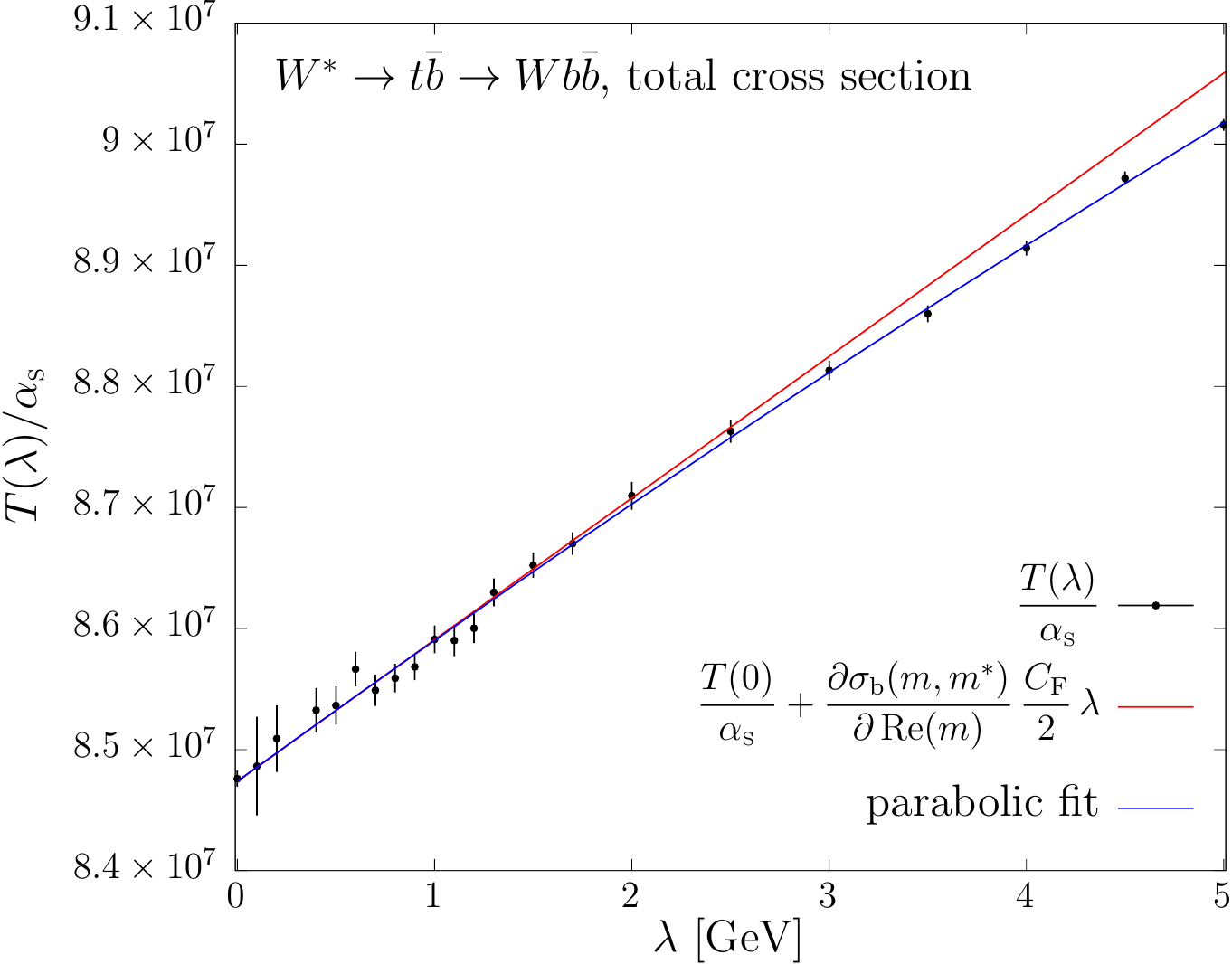}
  \caption{Small $\lambda$ behaviour of $T(\lambda)$ for the total cross
    section as function of the gluon mass $\lambda$.  In black the data
    points computed with our numerical calculations, in red the linear
    $\lambda$ dependence and in blue the parabolic fit of the points.  The
    $\lambda=0$ point has been obtained by performing the standard NLO
    computation in dimensional regularization.}
  \label{fig:sigtot_smallk}
\end{figure}

In the absence of cuts, the expression for $T(\lambda)$ in
eq.~(\ref{eq:Tdef}) simplifies, since $\Delta(\lambda)$, given by
eq.~(\ref{eq:final_Deltaqq}), is identically zero.
Its small
$\lambda$ behaviour is shown in Fig.~\ref{fig:sigtot_smallk}.
From the figure we can see that the error on $T(\lambda)$ increases for
small $\lambda$. However, the point at $\lambda=0$ is directly computed with
a massless gluon, by dealing with the soft and collinear singularities with
the usual dimensional regularization techniques, and has negligible error.
As discussed in Sec.~\ref{sec:changescheme}, the same calculation performed
in the \MSB{} mass scheme would yield, for the total cross section, to the
replacement given in eq.~(\ref{eq:Tchangescheme})
\begin{equation}
T(\lambda)\to T(\lambda)-\frac{\partial\sigma_b(\mpole,\mpole^*)}{\partial \,{\rm Re}(m)}
\,\frac{\CF \as}{2}\, \lambda+{\cal O}\!\(\lambda^2\).
\end{equation}
So, in the same figure, we also plot (in red) the expression
\begin{equation}
T(0)+\frac{\partial \sigma_b(\mpole,\mpole^*)}{\partial \,{\rm Re}(m)} \,\frac{\CF \as}{2}
\, \lambda\,.
\end{equation}
Since this coincides with $T(\lambda)$ for small $\lambda$, we infer that the \MSB{}
result has no linear term in $\lambda$, so that no linear renormalons arise for the
total cross section in the \MSB{} scheme.  From the figure it is also clear
that this holds for both $\lambda\lesssim \Gamma_t$ and for $\lambda\gg \Gamma_t$, where
$\Gamma_t$ is the top width. The $\lambda\lesssim \Gamma_t$ behaviour is justified
by the fact that, because of the finite width, phase-space points where the
top is on shell are never reached (see App.~\ref{app:KLN}). Thus, no linear
renormalon is present \emph{unless} one uses the pole-mass scheme, that has a
linear renormalon in the counterterm.

\begin{figure}[tb!]
  \centering
  \includegraphics[width=0.65\textwidth]{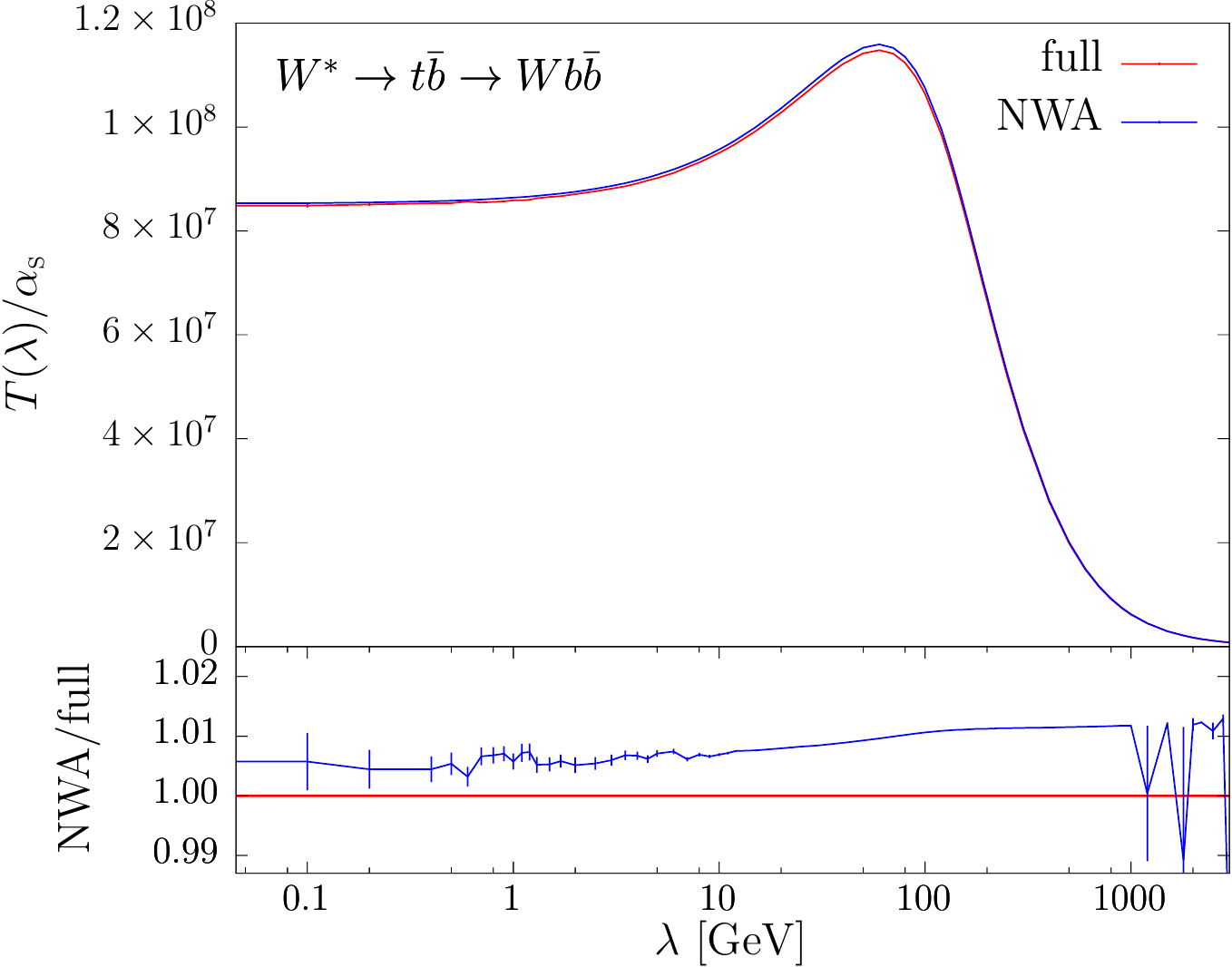}
  \caption{$T(\lambda)$ for the NLO total cross section, as function of the
    gluon mass $\lambda$, computed in the pole-mass scheme using the exact full
    matrix elements, in red, and the narrow-width approximation, NWA, in
    blue.}
  \label{fig:sigNWA}
\end{figure}

As far as the $\lambda\gg \Gamma_t$ limit is concerned, we notice that the $\lambda$
behaviour should be the same as that of the narrow width approximation~(NWA),
where the cross section factorizes in terms of the on-shell top-production
cross section, and its decay partial width
\begin{equation}
  \sigma\!\(W^* \rightarrow W b\bar{b}\) = \sigma\!\(W^* \rightarrow
  t\bar{b}\) \frac{\Gamma(t\rightarrow W b)}{\Gamma_t} +
  \mathcal{O}\left(\frac{\Gamma_t}{\mpole}\right).
\end{equation}
The behaviour of $T(\lambda)$, computed either exactly or in the NWA, is shown in
Fig.~\ref{fig:sigNWA}. 

The factor $\sigma(W^* \rightarrow t\bar{b})$ is clearly free of linear
renormalons, since it is a totally inclusive decay of a colour-neutral
system. Although less obvious, this is also the case for the factor
$\Gamma(t\rightarrow W b)$ (see Refs.~\cite{Bigi:1994em, Beneke:1994bc,
  Beneke:1994qe}).

\subsection{Inclusive cross section with cuts}
\label{sec:xsec_cuts}
When the selection cuts discussed in Sec.~\ref{sec:cuts} are imposed, the
cross section depends explicitly upon the jet radius~$R$. We expect that jets
requirements will induce the presence of linear renormalons, and thus linear
small-$\lambda$ behaviour of $T$, with a slope that goes like $1/R$ for small
$R$~\cite{Dasgupta:2007wa}.
\begin{figure}[tb!]
  \centering
  \includegraphics[width=0.65\textwidth]{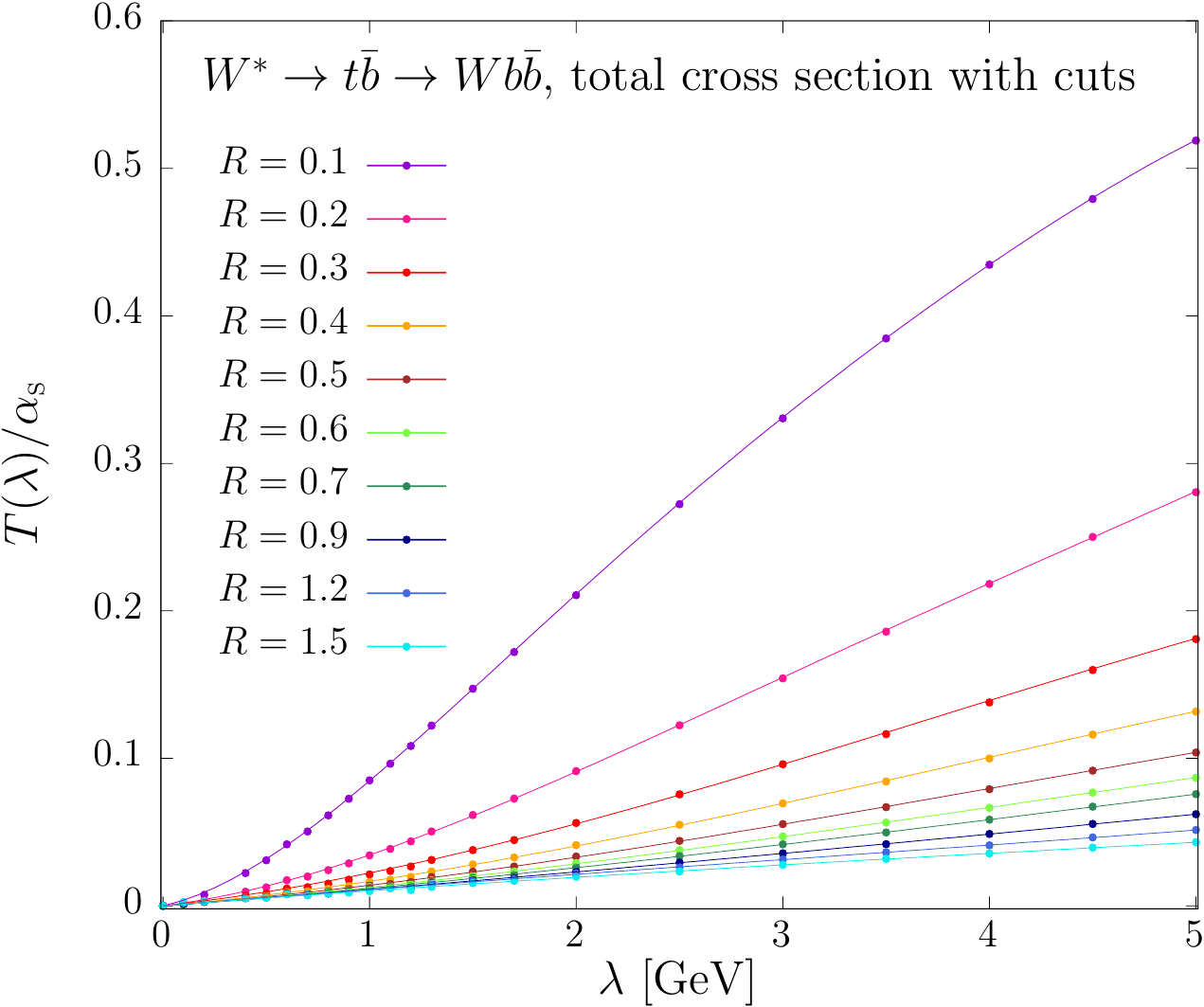}
  \caption{Small $\lambda$ behaviour for $T(\lambda)$ for the inclusive cross section
    with cuts, for several jet radii.  The points are obtained with our
    numerical calculations, while the solid lines represent their polynomial
    fit.  The fitting functions are order 5, 4 and 3 polynomials for $R=0.1$,
    $R=0.2$ and $R \ge 0.3$ radii respectively.}
  \label{fig:sigvrqq_cut2}
\end{figure}
In Fig.~\ref{fig:sigvrqq_cut2} we display the small $\lambda$ behaviour for
$T(\lambda)$ for the inclusive cross section with cuts, for several jet radii.
Together with the results of our calculation, we also plot, for each value of
$R$, a polynomial fit to the data.

When changing from the pole to the \MSB{}-mass scheme, we only expect a mild
$R$ dependent correction\footnote{The change of scheme is governed by
  formula~(\ref{eq:Tchangescheme}), where the only radius dependence comes
  from the derivative of the LO value of the observable, and this is mild for
  small $R$.} to the slope of $T(\lambda)$ at $\lambda=0$, and thus we cannot expect
the same benefit that we observed for the cross section without cuts.
\begin{figure}[tb!]
  \centering
  \includegraphics[width=0.65\textwidth]{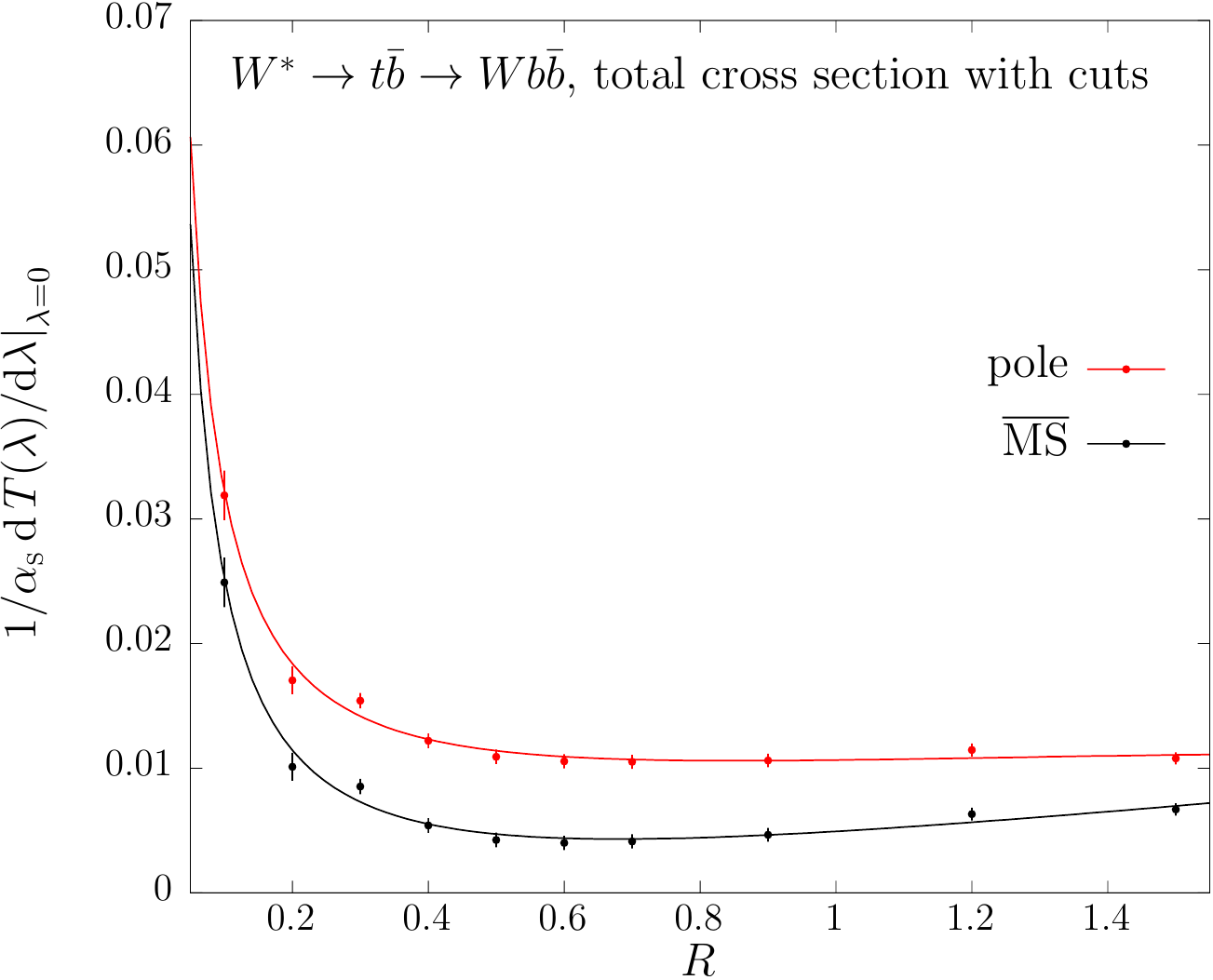}
  \caption{$R$ dependence of the slope of $T(\lambda)$ for the inclusive cross
    section, at $\lambda=0$, using the pole~(red) or the \MSB{} mass
    scheme~(black). The solid lines represent fits of parametric form
    $a/R+b+c\,R+d\,R^2$.}
  \label{fig:sigtot-linear-allR}
\end{figure}
This is illustrated in Fig.~\ref{fig:sigtot-linear-allR} for several jet
radii. The $1/R$ behaviour is clearly visible. In addition, for relatively
large-$R$ values, the use of the \MSB{} scheme brings about some reduction to
the slope of the linear term. This may be due to the fact that the cross
section with cuts captures a good part of the cross section without cuts, and
thus it partially inherits its benefits when changing scheme. However, it is
also clear that linear non-perturbative ambiguities remain important also in
the \MSB{} scheme when cuts are involved.

\section{Reconstructed-top mass}
\label{sec:rec-top-mass}
In this section we consider the average value $\langle M \rangle$, where $M$
is the mass of the system comprising the $W$ boson and the $b$ jet. Such
observable is closely related to the top mass, and, on the other hand, is
simple enough to be easily computed in our framework. We use the same
selection cuts described previously.


We computed $\langle M \rangle$ also in the narrow width limit, by simply
setting the top width to $10^{-3}$~GeV.  In this limit, top production and decay
factorize, so that we have an unambiguous assignment of the final state
partons to the top decay products. We first compute $\langle M \rangle$ in
the narrow width limit, using only the top decay products, and without
applying any cuts.  We then compute it again, still using only the top decay
products, but introducing our standard cuts. Finally we compute it again
using all decay products and our standard cuts.
\begin{figure}[tb]
  \centering
  \includegraphics[width=0.48\textwidth]{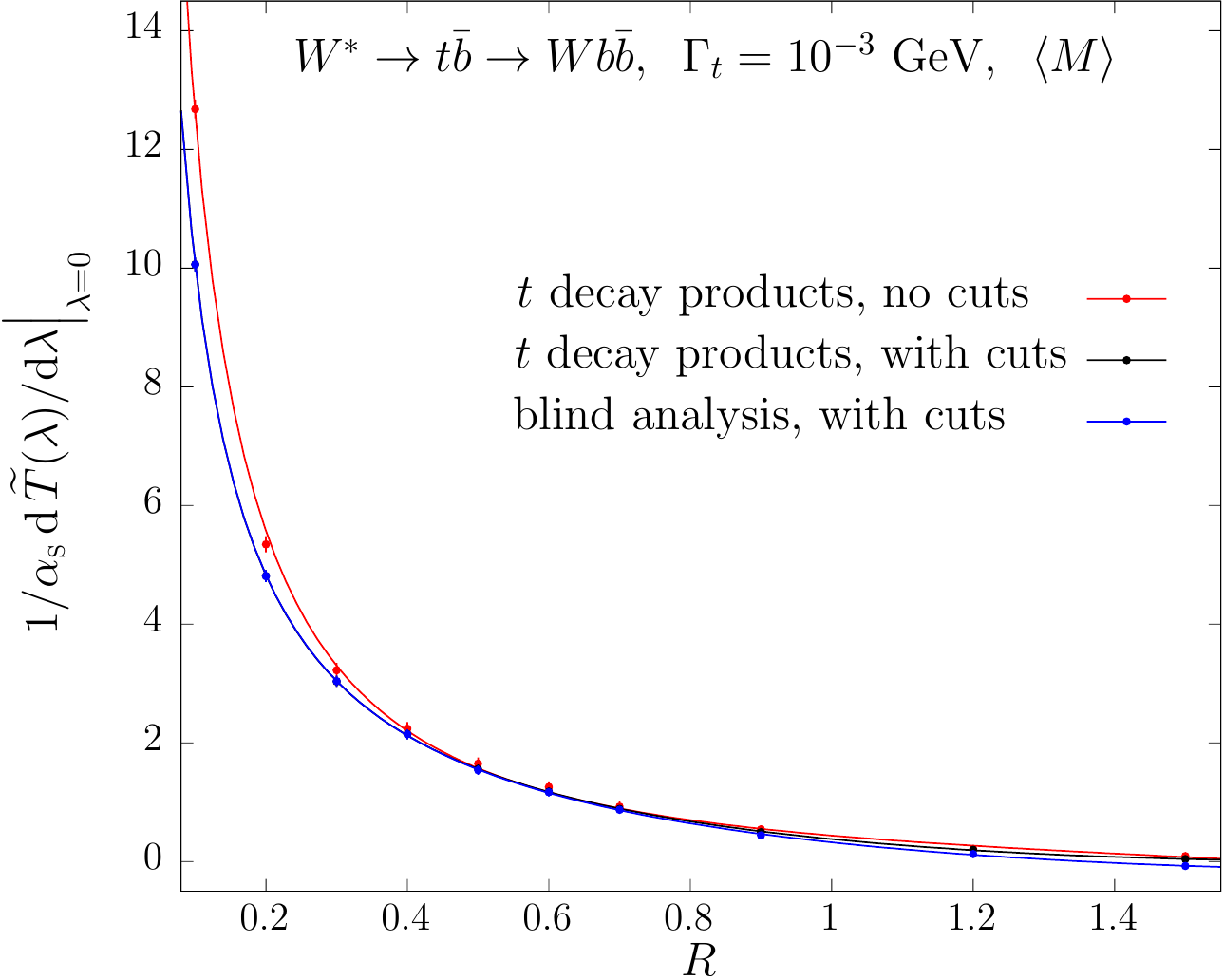}
  \includegraphics[width=0.48\textwidth]{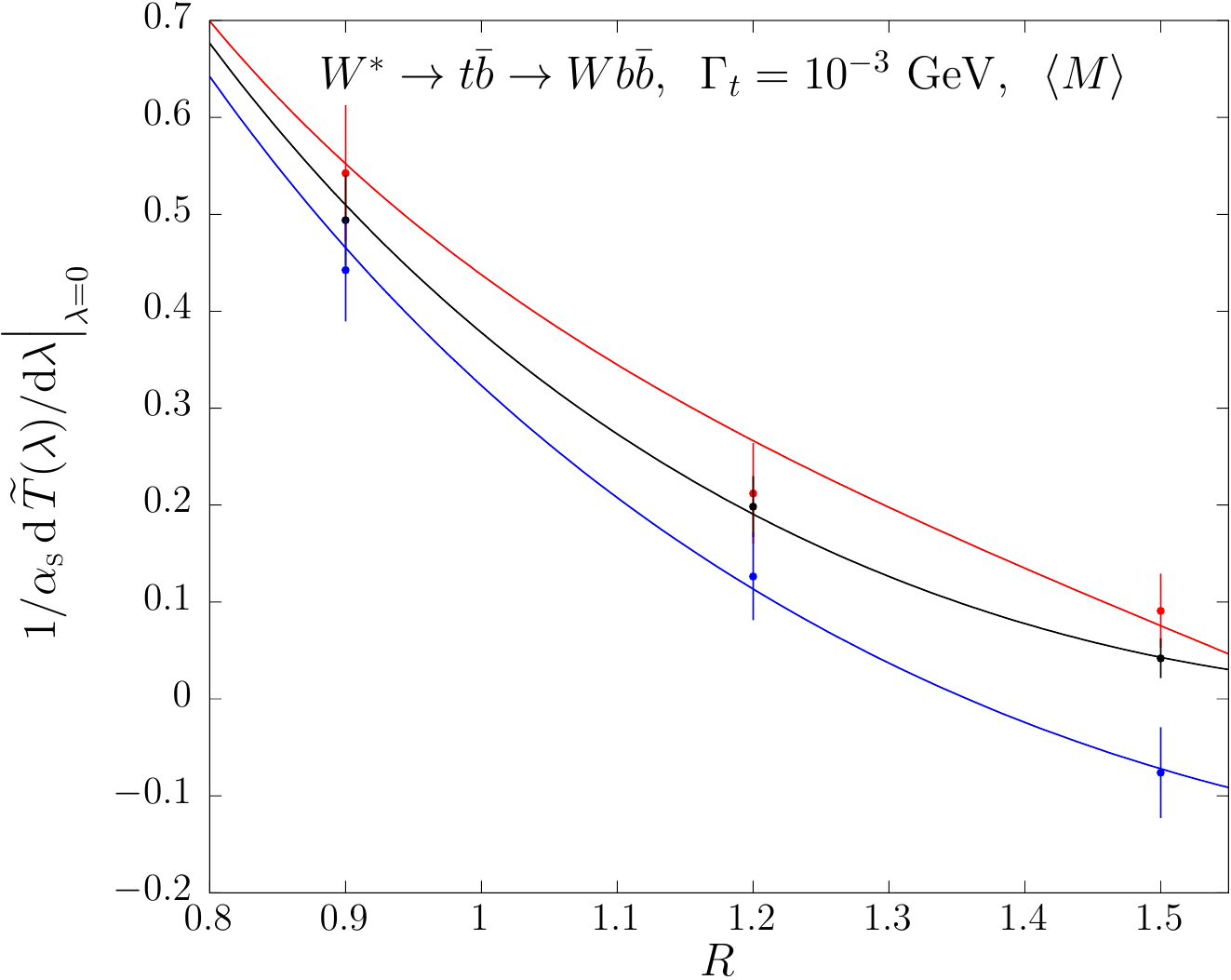}   
    \caption{$R$ dependence of the slope of $\widetilde{T}(\lambda)$ for the
      average reconstructed mass $\langle M \rangle$, at $\lambda=0$, computed with
      $\Gamma_t=10^{-3}$~GeV. The results obtained by reconstructing the $b$
      jet using only the top-decay products, without imposing any cut and
      with the cuts of Sec.~\ref{sec:cuts}, are shown in red and in black,
      respectively.  In blue, the results for a blind analysis with cuts. The
      solid lines represent fits of parametric form $a/R+b+c\,R+d\,R^2$.  The
      black and the blue curves are almost completely overlapping and are
      indistinguishable in the plot on the left. A blowup of the high-$R$
      region is illustrated in the plot on the right.  }
  \label{fig:NW-mctruth}
\end{figure}
The results of these calculations are reported in Fig.~\ref{fig:NW-mctruth},
where the slope at $\lambda=0$ of $\widetilde{T}$ for our observable is plotted as
a function of the jet radius $R$. As expected we see the shape proportional
to $1/R$ for small $R$~\cite{Dasgupta:2007wa}.

In the case of the calculation of $\langle M \rangle$ performed using only
the top decay products, and without any cuts, we expect that, for large
values of $R$, the average value of $M$ should get closer and closer to the
input top pole mass, irrespective of the value of $\lambda$. Thus, the slope of
$\widetilde{T}(\lambda)$ for $\lambda=0$ should become smaller and smaller. We find
in this case that, for the largest value of $R$ we are using ($R=1.5$), the
slope has a value around 0.09. When cuts are introduced this value becomes
even smaller, around 0.04. This curve is fairly close to the one obtained
using all final-state particles and including cuts. The large-$R$ value in
this case is $-0.08$.

\begin{figure}[tb]
  \centering
  \includegraphics[width=0.65\textwidth]{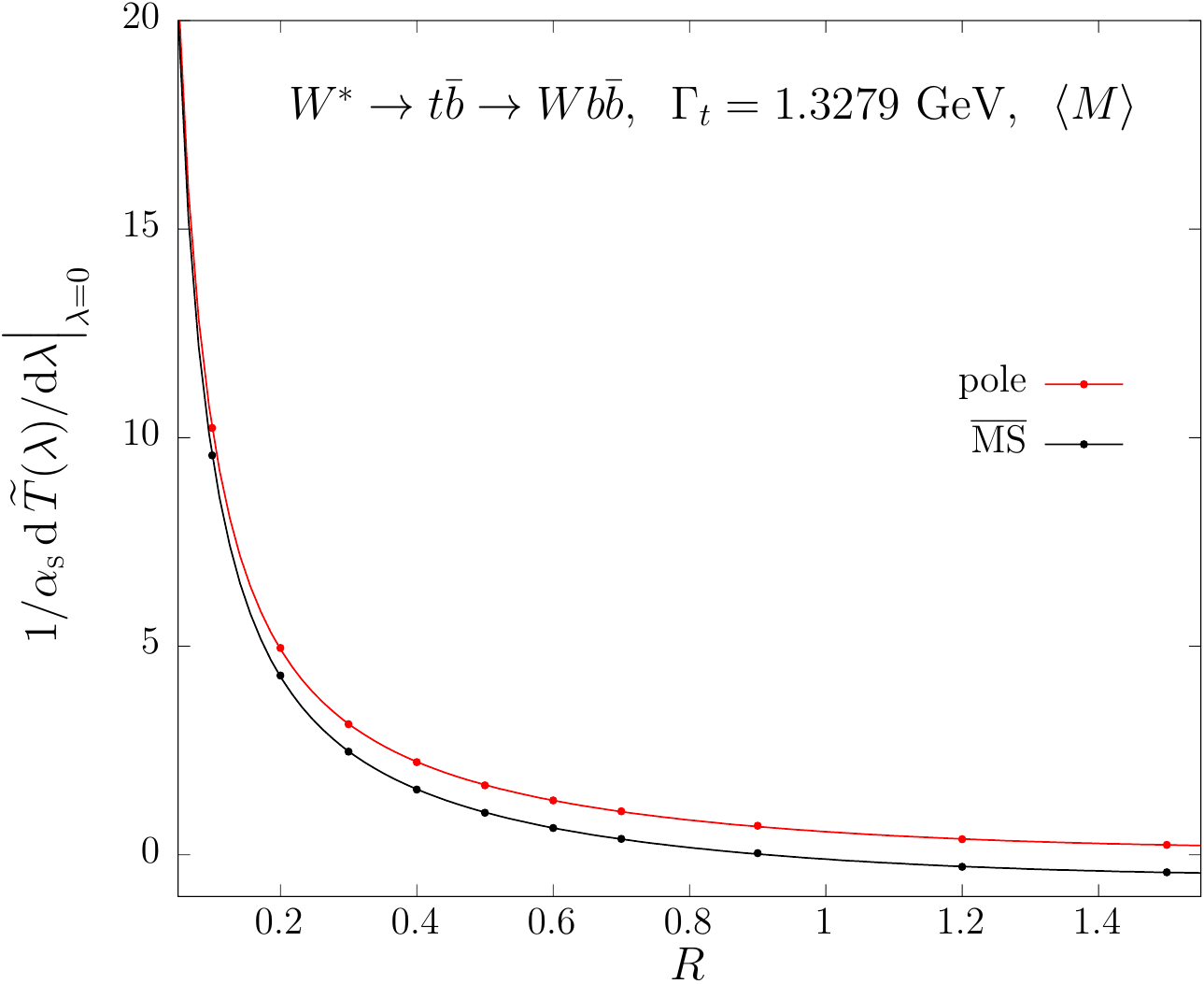}
  \caption{ $R$ dependence of the slope of $\widetilde{T}$ for the averaged
    reconstructed mass $M$.  The solid lines are the result of a fit of the
    form $a/R+b+c\,R+d\,R^2$.}
  \label{fig:Mrec_slope}
\end{figure}

If we change scheme from the pole mass to the \MSB{} one, the corresponding
change of $\widetilde{T}$ is given by eq.~(\ref{eq:Ttchangescheme}), and, for
the observable at hand, the derivative term is very near~1. The change in
slope when going to the \MSB{} scheme is roughly $-\CF/2\approx-0.67$. Thus,
if we insisted in using the \MSB{} mass for the present observable, for large
jet radii, we would get an ambiguity larger than if we used the pole mass
scheme.  The same holds even if we employ a finite top width, as shown in
Fig.~\ref{fig:Mrec_slope}, where the $R$ dependence of the
$\widetilde{T}(\lambda)$ slope for $\Gamma_t=1.3279$~GeV is plotted.

\begin{figure}[tb]
  \centering
  \includegraphics[width=0.65\textwidth]{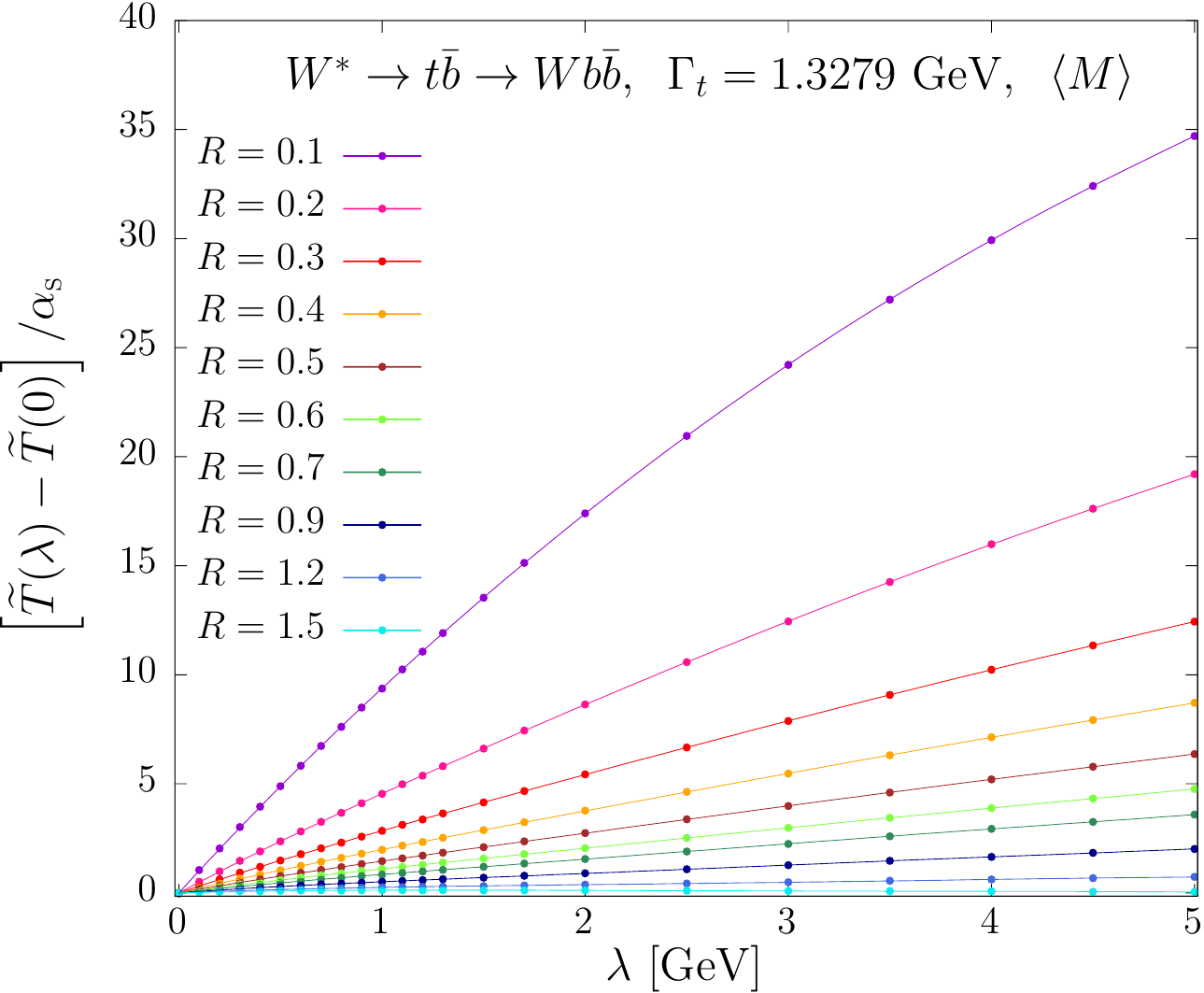}
  \caption{Small $\lambda$ behaviour of $\widetilde{T}(\lambda)$ for the averaged
    reconstructed-top mass, for several values of the jet radius $R$. The
    solid lines represent the polynomial fit of the computed points.  For $R
    \ge 1.2$ a $4^{\rm th}$ order polynomial is adopted, while for the other
    $R$ values a $5^{\rm th}$ order polynomial is employed.  }
  \label{fig:Mrec_smallk}
\end{figure}
\begin{figure}[tb]
  \centering
  \includegraphics[width=0.65\textwidth]{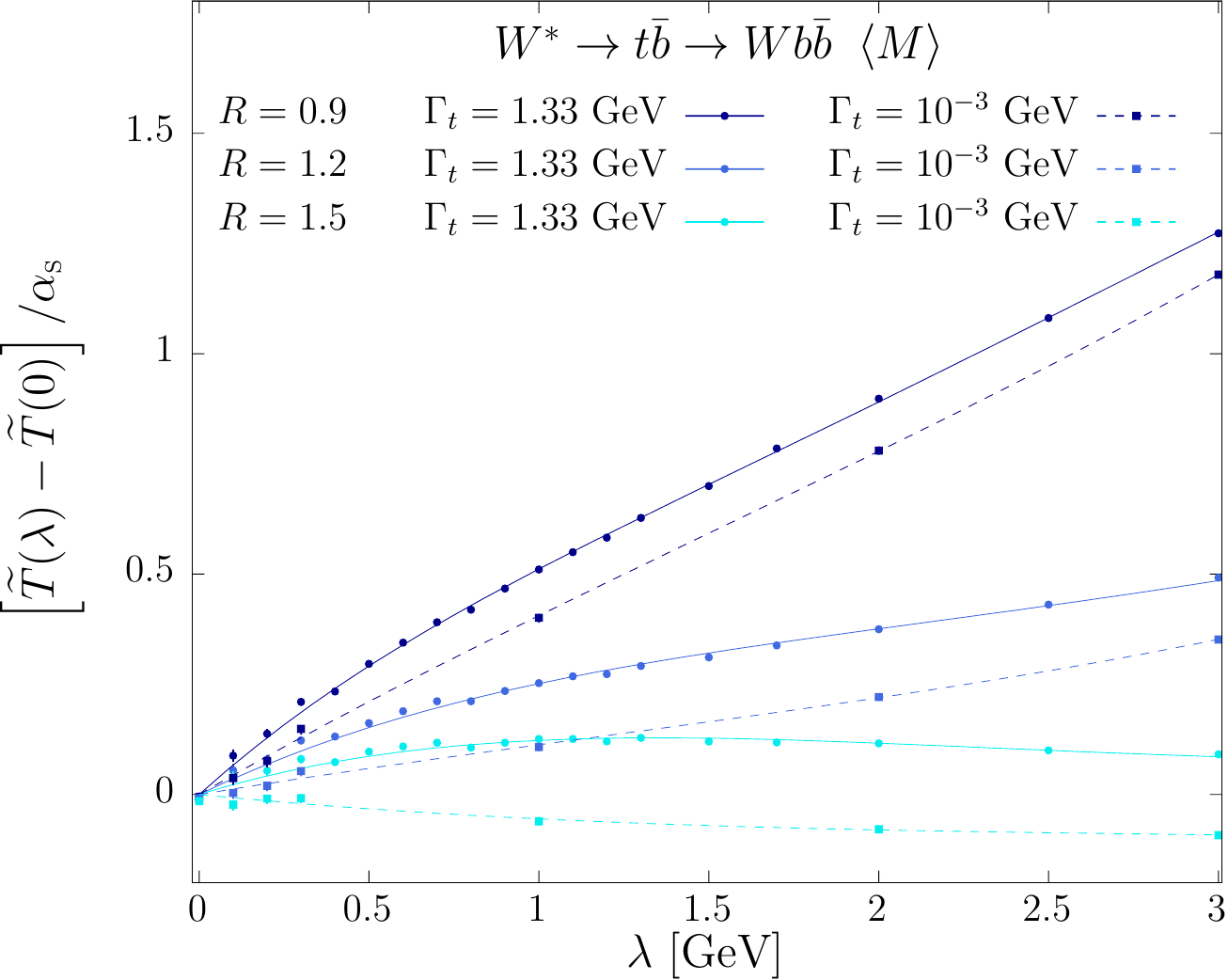}
  \caption{Small $\lambda$ behaviour of $\widetilde{T}(\lambda)$ for the averaged
    reconstructed-top mass for large values of the jet radius $R$, for two
    different values of the top decay width, $\Gamma_t=1.3279$~GeV~(solid
    lines) and $\Gamma_t=10^{-3}$~GeV~(dashed lines). The dashed lines are a
    cubic fit of the computed points. The solid lines are the same displayed
    in Fig.~\ref{fig:Mrec_smallk}.}
  \label{fig:Mrec_smallk-zoom}
\end{figure}

In Fig.~\ref{fig:Mrec_smallk} we plot the small $\lambda$ behaviour of
$\widetilde{T}(\lambda)$ for the reconstructed-top mass, computed with the
finite top width, for several values of the jet radius $R$.  It is clear that
our observable is strongly affected by the jet renormalon.  The same plot for
only the three largest values of $R$ is shown in
Fig.~\ref{fig:Mrec_smallk-zoom}.
The figure shows clearly that the slope of $\widetilde{T}(\lambda)$ for small
$\lambda$ computed with $\Gamma_t=1.3279$~GeV changes when $\lambda$ goes below $1$~GeV,
that is to say, when it goes below the top width. This behaviour is expected,
since the top width acts as a cutoff on soft radiation.  In the figure we
also report the $\lambda$ behaviour in the narrow-width approximation. It is
evident that the slopes computed in this limit are similar to the slopes with
$\Gamma_t=1.3279$~GeV, for values of $\lambda$ larger than the top width.  It is
also clear that the slopes that we find here for the largest $R$ value are
considerably smaller than the slope change induced by a change to a short
distance mass scheme, that amounts to $-0.67$. In other words, the pole mass
scheme is more appropriate for this observable, irrespective of finite width
effects.

We notice that, in the present case, for values of $R$ below 1, the \MSB{}
scheme seems to be better, because of a cancellation of the $R$ dependent
renormalon and the mass one. From our study, however, it clearly emerges that
such cancellation is accidental, and one should not rely upon it to claim an
increase in accuracy.

\section{$\boldsymbol{W}$ boson  energy}
\label{sec:Ew}
In this section we study the behaviour of the average value of the $W$
energy, $E_{\sss W}$. This observable does not depend upon the jet
definition, and can thus be considered a representative of pure ``leptonic''
observables in top-mass measurements. In this study, we do not apply any cut,
in order to avoid jet renormalons.  Our goal is to see if this observable is
free of renormalons in some mass scheme.

\begin{figure}[tb]
  \centering
  \includegraphics[width=0.65\textwidth]{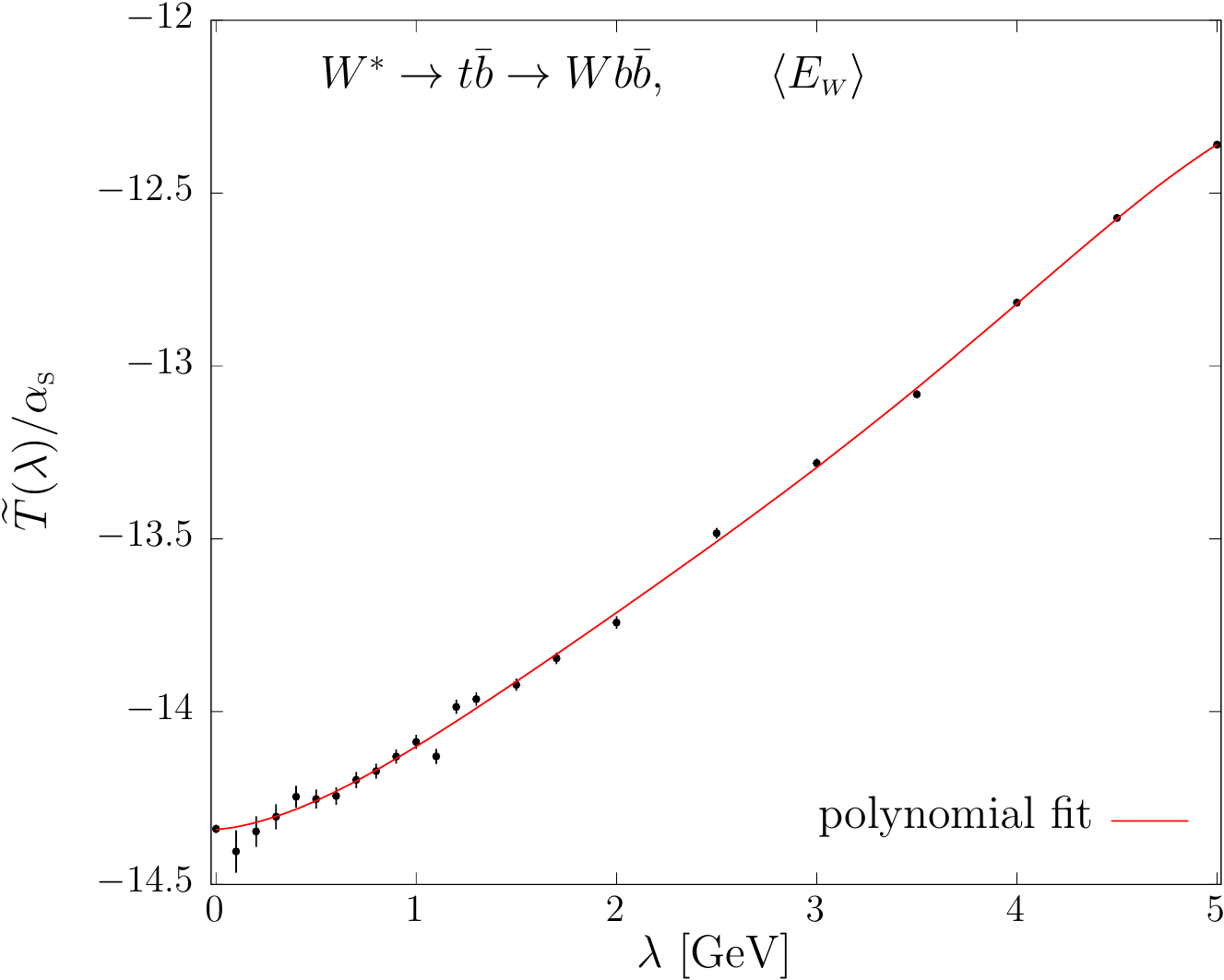}
  \caption{Small $\lambda$ behaviour of $\widetilde{T}(\lambda)$ for $\langle
    E_{\sss W} \rangle$. The solid line represents a $5^{\rm th}$ order
    polynomial fit.}
  \label{fig:Ew_smallk}
\end{figure}
In order to change scheme, according to eq.~(\ref{eq:Ttchangescheme}), we
need the derivative of the Born value of the observable with respect to the
real part of the top mass. We have computed numerically this quantity, and
found the value
\begin{equation}
  \label{eq:Ewmass-slope}
  \frac{\partial \langle E_{\sss W} \rangle_b}{\partial\, {\rm Re}(m)}=0.0980
  \,(8)  \,.
\end{equation}
The small-$\lambda$ dependence of the corresponding $\widetilde{T}$ function is
shown in Fig.~\ref{fig:Ew_smallk}. For values of $\lambda$ much larger than the
width, the slope of the curve is roughly 0.45. Thus, under these conditions,
a renormalon is clearly present whether we use the pole or the \MSB{} scheme,
since the correction in slope due to the use of the latter would be
$-0.098\times\CF/2=-0.065$.

\begin{figure}[tb]
  \centering
  \includegraphics[width=0.65\textwidth]{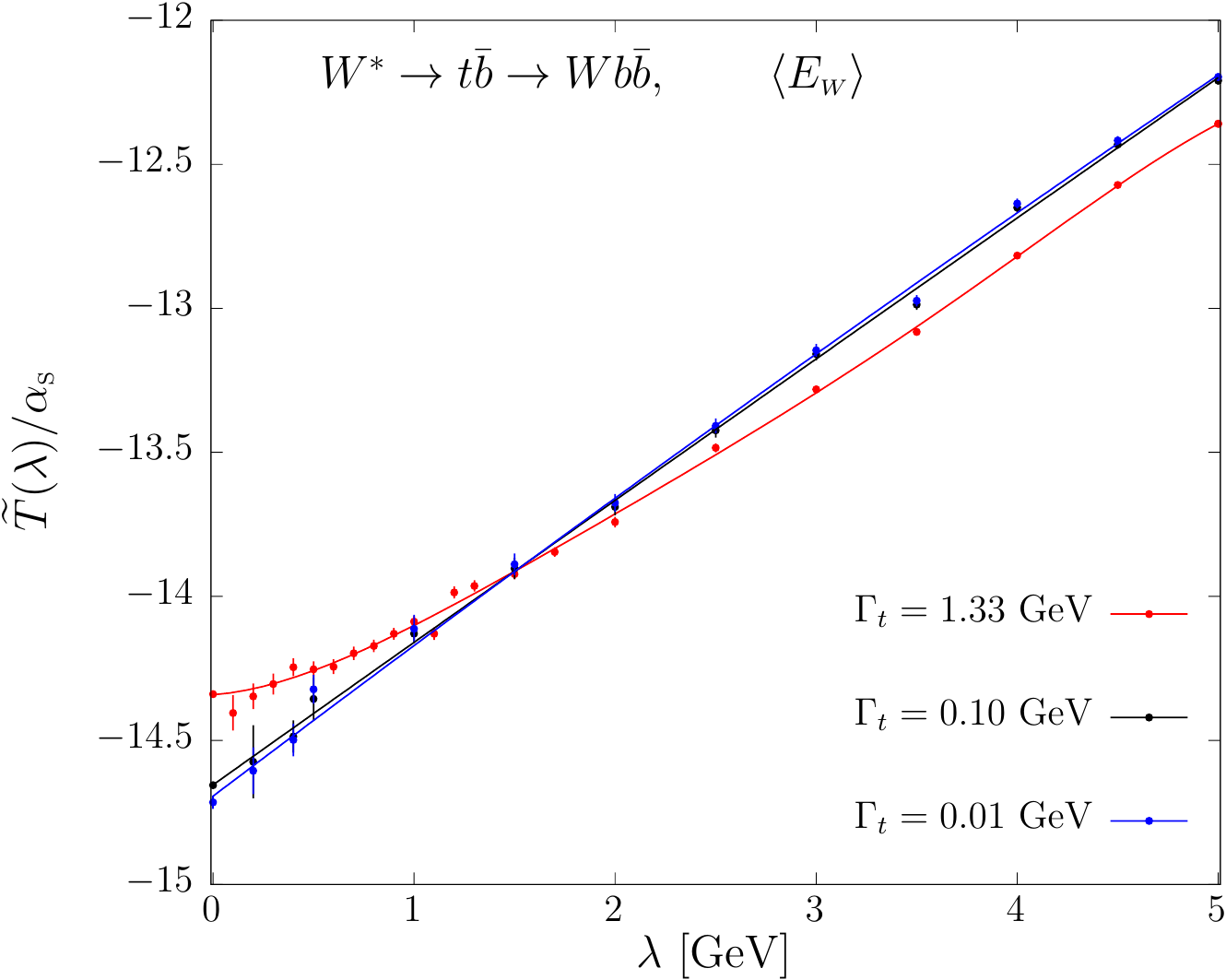}
  \caption{Small $\lambda$ behaviour of $\widetilde{T}(\lambda)$ for $\langle
    E_{\sss W} \rangle$, for increasingly smaller values of $\Gamma_t$. The
    blue and the black solid lines are a parabolic fit of the computed
    points, the red line is the same one displayed in
    Fig.~\ref{fig:Ew_smallk}. }
  \label{fig:Ew-nocuts-smallwidth}
\end{figure}

For $\lambda$ below the top width we see a reduction in slope, that is too
difficult to estimate because of the lack of statistics. In order to check
that the change in slope is related to the top finite width, we ran the
program with a reduced $\Gamma_t$, expecting to see a constant slope
extending down to smaller values of $\lambda$.  This is illustrated in
Fig.~\ref{fig:Ew-nocuts-smallwidth}.
We clearly see that, as $\Gamma_t$ becomes smaller, the slope of the $\lambda$
dependence remains constant, near the value $0.45$ found before, down to
smaller values of $\lambda$.  Since we have that
\begin{eqnarray}
  \frac{\partial \langle E_{\sss W} \rangle_b}{\partial\, {\rm
      Re}(m)} &=& 0.098\,(4)\,, \quad {\rm for }\quad \Gamma_t=0.1~{\rm
    GeV}, 
 \\
  \frac{\partial \langle E_{\sss W} \rangle_b}{\partial\, {\rm
      Re}(m)} &=& 0.10\,(3) \,, \phantom{0} \quad {\rm for }\quad
  \Gamma_t=0.01~{\rm GeV},   
\end{eqnarray}
it is clear that, for a vanishing top width, $\langle E_W \rangle$ has linear
renormalons both in the \MSB{} and pole mass scheme.

We also performed a run with $\Gamma_t=10$~GeV and $\Gamma_t=20$~GeV, in
order to estimate more accurately the value of the slope for $\lambda \ll \Gamma_t$.
\begin{figure}[tb]
  \centering
 \includegraphics[width=0.65\textwidth]{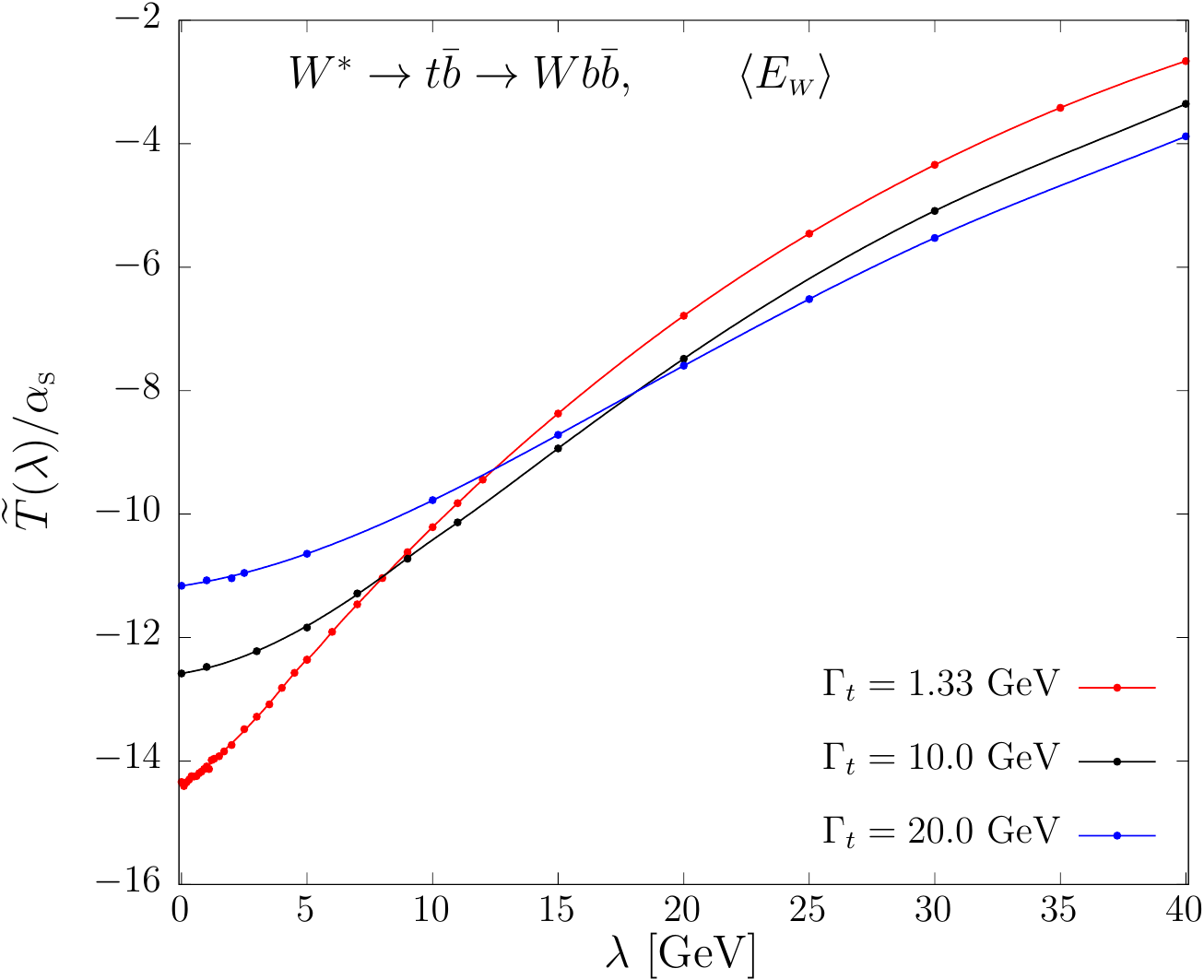}
  \caption{Results for the small $\lambda$ behaviour of $\widetilde{T}$ for
    $\langle E_{\sss W}\rangle$, at different values of $\Gamma_t$.  The
    error bar associated to each point computed at a given value of $\lambda$ is
    also plotted, but is too small to be visible on the scale of the figure.
    The red line~($\Gamma_t=1.33$~GeV) is a $5^{\rm th}$ order polynomial fit
    for $\lambda \le 5$~GeV and a spline for larger $\lambda$ values.  The blue and the
    black solid lines, that interpolates the results obtained with
    $\Gamma_t=10$~GeV and $\Gamma_t=20$~GeV respectively, are a cubic fit for
    $\lambda<\Gamma_t$ and a spline for $\lambda>\Gamma_t$.}
  \label{fig:Ew-nocuts-largewidth}
\end{figure}
The results are shown in Fig.~\ref{fig:Ew-nocuts-largewidth}.  In
Tab.~\ref{tab:slopes_Ew} we illustrate the slopes of $\widetilde{T}(\lambda)$
for small $\lambda$, obtained from the polynomial interpolation displayed in
Fig.~\ref{fig:Ew-nocuts-largewidth}, and the corresponding value in the
\MSB{} scheme, obtained by adding $ -\frac{\CF}{2}\frac{\partial \langle
  E_{\sss W} \rangle_b}{\partial\, {\rm Re}(m)} $ to the fitted slope.  This
shows that the linear sensitivity largely cancels in the \MSB{} scheme.
\begin{table}[htb]
  \centering
\begin{tabular}{|c|c|c|c|c|}
  \hline
  $\Gamma_t$ & slope (pole) & $  \displaystyle \frac{\partial \langle E_{\sss
      W}^{\phantom{\big(}}
    \rangle_b}{\partial\, {\rm Re}(m)}$ & $\displaystyle -\frac{\CF}{2}\frac{\partial
    \langle E_{\sss W} \rangle_b}{\partial\, {\rm Re}(m)}_{\phantom{\big(}}$ & slope (\MSB)\\
  \hline
  10~GeV & $0.058\,(8)$ & $0.0936\,(4)$ & $-0.0624\,(3)$ & $0.004\,(8)$\\
  \hline
  20~GeV & $0.061\,(2)$ &  $0.0901\,(2)$ & $-0.0601\,(1)$  &
  $0.001\,(2)$\\
  \hline
\end{tabular}
\caption{Slopes for $\widetilde{T}(\lambda)$ computed for $\langle E_{\sss W}
  \rangle$ in the pole-mass scheme and the derivative terms needed to change
  to the \MSB{} one, for large top widths.}
\label{tab:slopes_Ew}
\end{table}
One may now wonder if this cancellation is exact, or just accidental.  In
fact, we prove in App.~\ref{app:KLN} that the cancellation is exact.

\section{All-orders expansion in $\boldsymbol{\as}$}
\label{sec:AllOrderExp}
We consider now the all-orders expansion of various quantities, in order to
see how it is affected by the infrared renormalons, both in the pole and in
the \MSB-mass scheme.

One may think that in our framework we could, for example, compute a mass
sensitive observable, extract the mass (for a given value of the observable)
in different schemes, and finally convert all results to the pole mass
scheme, thus assessing the reliability of the methods used to estimate the
renormalon ambiguity in the pole mass~\cite{Pineda:2001zq, Ayala:2014yxa,
  Beneke:2016cbu, Hoang:2017btd}.  In fact, within the large-$n_f$
approximation, if the method adopted to resum the perturbative expansion is
linear, as is the case of the Borel transform method, we should find
identical results (always in the large-$n_f$ sense) in the \MSB{} and the
pole-mass schemes.  In fact, following eq.~(\ref{eq:changescheme}), the
relation between the pole and \MSB-mass scheme, for a generic observable, is
given by
\begin{eqnarray}
  \obsb(m,m^*)+\as \, \langle \, O \rangle ^{(1)}(m,m^*)
  &=&\obsb(\mMSB,\mMSB^*)+\left[\frac{\partial
      \obsb(\mpole,\mpole^*)}{\partial\mpole}  
    (m-\mMSB) + {\rm cc}\right] \nonumber \\
   &&{} + \as\,\langle O \rangle ^{(1)}(m,m^*) + {\cal O}\(\as^2 \(b_0\as \)^n\).
\end{eqnarray}
Neglecting subleading terms, this is an identity, since the expansion of
$\obsb$ in the mass difference stops at the first order in the large-$n_f$
limit. When performing the calculation in the pole-mass scheme, we need to
resum the expansion of $\langle O \rangle^{(1)}$, while if we perform the
calculation in the \MSB{} scheme, we are resumming the expansion of the sum
of terms in the curly bracket. If the resummation method is linear, this last
resummation can be performed on the individual terms inside the curly
bracket. This is exactly what we would do on the left-hand side if, after the
resummation, we wanted to express the same result in the \MSB{} scheme.  In
other words, if one uses the Borel method to perform the resummation, and
defines the pole mass to be the sum of the mass relation formula
eq.~(\ref{eq:msbpole}), all results obtained in the \MSB{} scheme would be
identical to those obtained in the pole mass scheme up to terms of relative
order $\as \CF$, provided the same Borel sum method is used also for the
observables.

In the following we will estimate the terms of the perturbative expansion by
extrapolating our large-$n_f$ results to the realistic QCD case, (i.e.~in the
large-$b_0$ approximation).  In order to do this, we will replace the $b_0$
of the large-$n_f$ theory with the $b_0$ of QCD, and perform other minor
adjustments, detailed later.  Needless to say, corrections to the large-$b_0$
approximation may be non-negligible.  We thus expect that, by changing
scheme, we will generate potentially important differences.  These
differences should not therefore be interpreted as due to large ambiguities
related to the choice of mass scheme, but rather to the violation of the
large-$b_0$ approximation.

The procedure we adopt in order to compute the terms of the perturbative
expansion follows from eq.~(\ref{eq:final_M_wcuts}).  We fit numerically the
$\lambda$ dependence of the appropriate $T$ or $\widetilde{T}$ function, and we
take the derivative of the fit. The arctangent factor is instead expanded
analytically, and the integration is performed numerically for each
perturbative order. The details of the procedure that we followed in order to
go from the large-$n_f$ theory to the large-$b_0$ approximation are described
at the end of App.~\ref{sec:dressed_gluon}, eqs.~(\ref{eq:Pi_Pig_expanded})
to~(\ref{eq:Cgvalue}).

\subsection{Mass-conversion formula}
The procedure for the calculation of the mass-conversion formula is described
in App.~\ref{app:PoleMSB}.  Here we switch to the realistic $b_0$ and $C$
values as discussed in the previous section.  The expansion of the mass
conversion formula reads
\begin{equation}
\label{eq:mpoleMSBnumer2}  
  \mMSB(\mu)= \mpole \(1-\sum_{i=1}^\infty  c_i \, \as^i\).
\end{equation}
and the $c_i$ coefficients are tabulated in Tab.~\ref{tab:ck}, with
$\mu^2={\rm Re}(\mpole^2)=m_0^2$, where $m_0$ is given in eq.~(\ref{eq:m0}).
\begin{table}[tb]
  \centering \input{tables-M/poleMSbar_complex_conversion.tex}
\caption{Real and imaginary parts of the coefficients $c_i$ of the mass
  relation~(\ref{eq:mpoleMSBnumer2}), up to the tenth order in the strong
  coupling constant $\as(\mu)$, with $\mu^2={\rm Re}(\mpole^2)$.}
\label{tab:ck}
\end{table}
Since we are using the complex-mass scheme, the $c_i$ coefficients are
complex, with a small imaginary part, and they have a slight dependence upon
the ratio $\Gamma_t/{\rm Re}(m)$. For small $\Gamma_t$ they become
independent on $m$ and $\Gamma_t$, and their imaginary part vanishes.

We have checked that for $\Gamma_t=0$ and in the large-$n_f$ limit,
i.e.~setting $\CA = 0$ in our numerical code used to produce the coefficients
of Tab.~\ref{tab:ck}, we obtain the same results presented in
Ref.~\cite{Ball:1995ni}.

\subsection{The inclusive cross section}
In this section we deal with the perturbative expansion of the inclusive
cross section, first without cuts, and then with cuts.

The function $T(\lambda)$ of eq.~(\ref{eq:Tdef}), needed to calculate the
integral in eq.~(\ref{eq:final_sigma_wcuts}), is obtained as an interpolation
of $T(\lambda)$ computed at several fixed values of $\lambda$. We have chosen a
polynomial fit, for values of $\lambda$ less than 5~GeV, of the form
\begin{equation}
    T(\lambda)  = p_0 \,+\,p_1\, \lambda \,+\, p_2\, \lambda^2\, +\, \ldots\,,
    \label{eq:T_polynom}
\end{equation}
and a cubic spline for larger values of $\lambda$. The two forms are required to
match in value and slope at the joining point.  The same approach is adopted
to evaluate $\widetilde{T}(\lambda)$ of eq.~(\ref{eq:Ttdef}), both for the
averaged reconstructed-top mass and for the averaged $W$-boson energy.

The fitting functions that we obtain are seen to represent sufficiently well
the numerical results for $T$, with the only caveat that, for small $\lambda$,
these have themselves non-negligible errors.  These errors strongly affect
the coefficient $p_1$, and have negligible effects on the other
coefficients. In fact, $p_0$ is obtained directly from the results computed
with a massless gluon, and has a totally negligible error. The $p_2$ and
higher-order coefficients are controlled by the larger values of $\lambda$, where
our computation has a smaller error.  We thus propagated the errors of our
numerical data to the $p_1$ coefficient only, and then to the calculation of
the coefficients of the perturbative expansion.

\subsubsection{Inclusive cross section without cuts}
As discussed in Sec.~\ref{sec:Xtot_wocuts}, $T(\lambda)$ for the inclusive
cross section does not have any term linear in $\lambda$, if expressed in terms of
the \MSB{} mass. It follows that the total cross section computed in the
\MSB{} scheme should not have any $\LambdaQCD/\mpole$ renormalon and should
display a better behaviour at large orders.

\begin{table}[tb]
  \centering
  \input{tables-M/xsec-nocuts.tex}
  \caption{Coefficients of the $\as$ expansion~(\ref{eq:sigma_expansion}) of
    the inclusive cross section to all orders, computed in the large-$b_0$
    approximation, normalized to the total Born cross section computed in the
    pole-mass scheme.  The errors reported in parenthesis are due to the
    uncertainty on the linear coefficient of the fit (i.e.~$p_1$ in
    eq.~(\ref{eq:T_polynom})).  }
  \label{tab:sigtot_expansion}
\end{table}
The coefficients $c_i$ of the expansion of eq.~(\ref{eq:final_sigma_wcuts}) in
terms of $\as$
\begin{equation}
  \label{eq:sigma_expansion}
   \sigma = \sigma_{\rm b}^{\rm nocuts}(\mpole)\( c_0 + \sum_{i=1}^{\infty}
   c_i\, \as^i \)
\end{equation}
are collected in Tab.~\ref{tab:sigtot_expansion}, in the pole~(left) and in
the \MSB{}~(right) schemes.  At large orders, the \MSB{} inclusive cross
section receives much smaller contributions than in the pole-mass scheme. On
the other hand, in the pole-mass scheme the factorial growth is already
visible at the $\rm N^3LO$ order, and the minimum of the series is reached
for $i=8$ (that corresponds to an $\mathcal{O}(\as^8)$ correction), and it is
two orders of magnitude larger than the corresponding contribution computed
in the \MSB{} scheme.  We also notice that the \MSB{} result has an NLO
correction larger than the pole mass result, an NNLO correction that is
similar, and smaller N$^3$LO and higher order corrections.

\subsubsection{Inclusive cross section with cuts}
As we have seen in Sec.~\ref{sec:xsec_cuts}, the presence of selection cuts
introduces a linear renormalon in the inclusive cross section proportional to
$1/R$.
\begin{table}[tb]
  \centering
\resizebox{\textwidth}{!}
{  \input{tables-M/xsec-cut2-R0.1.tex}
  \input{tables-M/xsec-cut2-R0.5.tex}}
  \caption{Coefficients $c_i$ of the $\as$
    expansion~(\ref{eq:sigma_expansion}) of the cross section with cuts, to
    all orders, computed in the large-$b_0$ approximation, normalized to the
    total Born cross section computed in the pole-mass scheme, for two
    different values of the jet radius ($R=0.1$ in the left pane and $R=0.5$
    in the right one).  The errors reported in parenthesis are due to the
    uncertainty on the linear coefficient of the fit (i.e.~$p_1$ in
    eq.~(\ref{eq:T_polynom})).  }
  \label{tab:sigtot_expansion_cut2}
\end{table}
In Tab.~\ref{tab:sigtot_expansion_cut2} we present the results for the
inclusive cross section, in the pole and in the \MSB-mass scheme, for a small
jet radius, $R=0.1$, and a more realistic value, $R=0.5$.  For small radii,
the perturbative expansion displays roughly the same bad behaviour, either
when we use the pole or the \MSB-mass scheme.  For larger values of $R$, the
size of the coefficients are typically smaller than the corresponding ones
with smaller values of $R$.  In particular, if we compare the coefficients
for $R=0.1$ and $R=0.5$, the second ones are one order of magnitude smaller
than the first ones.  Furthermore, for $R=0.5$, the coefficients computed in
the \MSB{}-mass scheme are roughly half of the ones computed in the pole-mass
scheme. This follows from the fact that, for large values of $R$, the cross
section with cuts approaches the total cross section, thus partially
inheriting its properties.

\subsection{Reconstructed-top mass}
In this section, we discuss the terms of the perturbative expansion for the
average reconstructed mass~$\langle M\rangle$
  \begin{equation}
    \label{eq:M_expansion}
    \langle M \rangle = \sum_{i=0}^\infty c_i\, \as^i\,,
\end{equation}
for three values of the $R$ parameter. We apply the cuts of
Sec.~\ref{sec:cuts} and the results are collected in Tab.~\ref{tab:M-coeffs}.

\begin{table}[htb]
  \resizebox{\textwidth}{!}
  {\input{tables-M/Obs-cut2-R-paper.tex}}
  \centering
  \caption{Values of the $c_i\, \as^i$ terms of the perturbative expansion
    for the average value of the reconstructed-top mass, defined in
    eq.~(\ref{eq:M_expansion}), for three different jet radii, in the
    pole-mass and \MSB{}-mass scheme.  The errors reported in parenthesis are
    due to the uncertainty on the linear coefficient of the fit (i.e.~$p_1$
    in eq.~(\ref{eq:T_polynom})).}
    \label{tab:M-coeffs}
\end{table}
From the table we can see that, for very small jet radii, the asymptotic
character of the perturbative expansion is manifest in both the pole and
\MSB{} scheme. For the realistic value $R=0.5$, the \MSB{} scheme seems to
behave slightly better. In fact, this is only a consequence of the fact that
the jet-renormalon and the mass-renormalon corrections have opposite signs,
with the mass correction in the \MSB{} scheme largely prevailing at small
orders, yielding positive effects.

As the radius becomes very large, the jet renormalon becomes less and less
pronounced, in the pole-mass scheme, leading to smaller corrections at all
orders. This is consistent with the discussion given in
Sec.~\ref{sec:rec-top-mass}, where we have seen that, for large radii, the
reconstructed mass becomes strongly related to the top pole mass, since it
approaches what one would reconstruct from the ``true'' top decay
products.\footnote{We remind here that, in the narrow width limit, and in
  perturbation theory, the concept of a ``true'' top decay final state is
  well defined.}

\subsection[${W}$ boson energy]{$\boldsymbol{W}$ boson energy}
The coefficients of the perturbative expansion of the average energy of the
$W$ boson 
  \begin{equation}
    \label{eq:Ew_expansion}
    \langle E_{\sss W}\rangle = \sum_{i=0}^\infty c_i\, \as^i \,,
  \end{equation}
in the pole and \MSB-mass schemes, are displayed in Tab.~\ref{tab:Wexp}.
\begin{table}[tb]
  \centering
  \input{tables-Ew/Obs-nocuts-R0.5.tex}
  \caption{Coefficients of the perturbative expansion~(\ref{eq:Ew_expansion})
    of the average $W$-boson energy in the pole and \MSB-mass schemes. The
    errors reported in parenthesis are due to the uncertainty on the linear
    coefficient of the fit (i.e.~$p_1$ in eq.~(\ref{eq:T_polynom})).}
  \label{tab:Wexp}
\end{table}
We notice that the perturbative expansions are similarly behaved in both
schemes up to $i\approx 6$, while, for higher orders, the \MSB{} scheme
result displays a better convergence. This supports the observation, discussed
in Sec.~\ref{sec:Ew}, that the top width screens the renormalon effects if
the \MSB{} mass is used. In fact, the $6^{\rm th}$ order renormalon
contribution is dominated by scales of order $m_t \, e^{-5}\approx 1.16$, as
illustrated in Sec.~\ref{sec:renorm_intro}, very near the top width.

\section{Conclusions}
In this work we have examined non-perturbative corrections related to
infrared renormalons relevant to typical top-quark mass measurements, in the
simplified context of a $W^* \rightarrow t\bar{b} \to W b \bar{b}$ process,
with an on-shell final-state $W$ boson and massless $b$ quarks.  As a further
simplification, we have considered only vector-current couplings.  We have
however fully taken into account top finite-width effects.

We have investigated non-perturbative corrections that arise from the
resummation of light-quark loop insertions in the gluon propagator,
corresponding to the so called large-$n_f$ limit of QCD. The large-$n_f$
limit result can be turned into the so called large-$b_0$ approximation, by
replacing the large-$n_f$ beta function coefficient with the true QCD
one. This approximation has been adopted in several contexts for the study of
non-perturbative effects (see e.g.~Refs.~\cite{Beneke:1994qe, Beneke:1994sw,
  Beneke:1994bc, Bigi:1994em, Ball:1995ni, Seymour:1994df}).

In this paper we have developed a method to compute the large-$n_f$ results
exactly, using a combination of analytic and numerical methods. The latter is
in essence the combination of four parton level generators, that allowed us
to compute kinematic observables of arbitrary complexity. We stress that,
besides being able to study the effect of the leading renormalons, we can
also compute numerically the coefficients of the perturbative expansion up
and beyond the order at which it starts to diverge.

Although our findings have all been obtained in the simplified context just
described, we can safely say that all effects that we have found are likely
to be present in the full theory, although we are not in a position to
exclude the presence of other effects related to the non-Abelian nature of
QCD, or to non-perturbative effects not related to renormalons.

Our findings can be summarized as follows:
\begin{itemize}
\item The total cross section for the process at hand is free of
  \emph{physical} linear renormalons, i.e.~its perturbative expansion in
  terms of a short distance mass is free of linear renormalons. This result
  holds both for finite top width and in the narrow-width limit.  In the
  former case, the absence of a linear renormalon is due to the screening
  effect of the top finite width, while, in the latter case, it is a
  straightforward consequence of the fact that both the top production cross
  section and the decay partial width are free of physical linear
  renormalons.

  By examining the perturbative expansion order by order, we find that,
  already at the NNLO level, the \MSB{} scheme result for the cross section
  is much more accurate than the pole-mass-scheme one.

  We stress that our choice of 300~GeV for the incoming energy corresponds to
  a momentum of 100~GeV for the top quark, that in turn roughly corresponds
  to the peak value of the transverse momentum of the top quarks produced at
  the LHC. Thus, the available phase space for soft radiation at the LHC is
  similar to the case of the process considered here, so that it is
  reasonable to assume that our result gives an indication in favour of using
  the \MSB{} scheme for the total cross section without cuts at the LHC.
\item As soon as jet requirements are imposed on the final state, corrections
  of order $\LambdaQCD$ arise. They have a leading behaviour proportional to
  $1/R$, where $R$ is the jet radius, for small
  $R$~\cite{Dasgupta:2007wa}. These corrections are present irrespective of
  the top-mass scheme being used. They are however reduced if the efficiency
  of the cuts is increased, for example by increasing the jet radius, giving
  an indication in favour of the use of the \MSB{} scheme for the total cross
  section calculation also in the presence of cuts.  It should be stressed,
  however, that with a typical jet radius of 0.5 the behaviour of the
  perturbative expansion in the \MSB{} and pole-mass schemes are very
  similar, with a rather small advantage of the first one over the latter.
\item The reconstructed-top mass, defined as the mass of the system
  comprising the $W$ and the $b$ jet, has the characteristic power correction
  due to jets, with the typical $1/R$ dependence. No benefit, i.e.~reduction
  of the power corrections, seems to be associated with the use of a
  short-distance mass. In particular, at large jet radii, when the jet
  renormalon becomes particularly small, in the pole-mass scheme the linear
  renormalon coefficient is smaller. This observation is justified if one
  considers that, in the narrow-width limit, the production and decay
  processes factorize to all orders in the perturbative expansion, yielding a
  clean separation of radiation in production and decay. In this limit, the
  system of the top-decay products is well defined, and its mass is exactly
  equal to the pole mass.  Consistently with this observation, we have shown
  that, for very large jet radii, the linear renormalon coefficient for the
  reconstructed top mass is quite small (if the observables is expressed in
  terms of the pole mass).  One may then worry that, when reconstructing the
  top mass from the full final state, renormalons associated with soft
  emissions in production from the top and from the $\bar{b}$ quark may
  affect the reconstructed mass, since these soft emissions may enter the
  $b$-jet cone. By comparing the reconstructed mass obtained using only the
  top-decay products to the one obtain using all final-state particles, we
  have shown that these effects are in fact small.

  We should also add, however, that the benefit of using very large jet radii
  cannot be exploited at hadron colliders, since we expect other renormalon
  effects, due to soft-gluon radiation in production entering the jet
  cone. This problem can in principle be investigated with our approach, by
  applying it to the process of $t\bar{t}$ production in hadronic collisions.
\item We have considered, as a prototype for a leptonic observable relevant
  for top mass measurement, the average energy of the $W$ boson. We have
  found two interesting results:
  \begin{itemize}
  \item In the narrow-width limit, this observable has a linear renormalon,
    irrespective of the mass scheme being used for the top. This finding does
    not support the frequent claim that leptonic observables should be better
    behaved as far as non-perturbative QCD corrections are concerned.  It
    also reminds us that, even if we wanted to measure the top-production
    cross section by triggering exclusively upon leptons, we may induce
    linear power corrections in the result that cannot be eliminated by going
    to the \MSB{} scheme.

    The presence of renormalons in leptonic observables seems to be in
    contrast with what is found in inclusive semileptonic decays of heavy
    flavours~\cite{Bigi:1994em, Beneke:1994bc}.  We have however verified
    that there is no contradiction with this case.  If the average value of
    the $W$ energy is computed in the top rest frame (which makes it fully
    analogous to a leptonic observable in $B$ decay) then no renormalon is
    present if the result is expressed in terms of the \MSB{} mass.
  \item For finite widths, if a short-distance mass is used, there is no
    linear renormalon. We verified this numerically, and furthermore we were
    also able to give a formal proof of this finding.  What this means in
    practice is that the perturbative expansion for this quantity will have
    factorial growth up to an order $n\approx 1+\log(m/\Gamma_t)$, that will
    stop for higher orders.  In practice, for realistic values of the width,
    this turns out to be a relatively large order. Thus, although in
    principle we cannot exclude a useful direct determination of the top
    short-distance mass from leptonic observables, it seems clear that
    finite-order calculations should be carried out at relatively high orders
    (up to the fourth or fifth order) in order to exploit it. Although it
    seems unlikely that results at these high orders may become available in
    the foreseeable future, perhaps it is not impossible to devise methods to
    estimate their leading renormalon contributions, still allowing a viable
    mass measurement.
  \end{itemize}  
\end{itemize}
In this work we have made several simplifying assumptions.  These assumptions
were motivated by the fact that the calculational technique is new, and we
wanted to make it as simple as possible.  Some of these restrictions may be
removed in future works. For example, we could consider hadronic collisions,
the full left-handed coupling for the $W$, the $W$ finite width and the
effects of a finite $b$ mass.  Although removing these limitations can lead
to interesting results, we should not forget that our calculation does not
exhaust all sources of non-perturbative effects that can affect the mass
measurement. As an obvious example, we should consider that confinement
effects are not present in our large-$b_0$ approximation, while, on the other
hand, it is not difficult to show that they may give rise to linear power
corrections. It is clear that theoretical problems of this sort should be
investigated by different means.

\section*{Acknowledgments}
P.N. wishes to thank G.~Salam for frequent discussions on this research
project; P.~Gambino and L.~Silvestrini for discussion in connections with $B$
leptonic decays; and M.~Beneke and A.~Hoang for comments on the manuscript.

\appendix

\section{The dressed gluon propagator}
\label{sec:dressed_gluon}
In this section we collect some technical details about the dressed gluon
propagator to all orders in the large-$n_f$ limit.  The insertion of an
infinite number of self-energy corrections
\begin{equation}
  \Pi^{\mu\nu}\!\(k,\mu^2\) = (-g^{\mu\nu}k^2 + k^\mu k^\nu) \, i\, \Pi\!\(k^2,\mu^2\)\,,
\end{equation}
along a gluon propagator of momentum $k$, gives rise to the dressed gluon propagator
\begin{equation}
 \frac{-i}{k^2} \, g^{\mu\nu} +  \frac{-i}{k^2} \,  \Pi^{\mu\nu}\!\(k,\mu^2\)\,
 \frac{-i}{k^2} + \ldots =   -\frac{i}{k^2} \, g^{\mu\nu} \frac{1}{1+\Pi(k^2,\mu^2)}\,,
\end{equation}
where we have dropped all the longitudinal terms.  In the limit of large
number of flavours, i.e.~considering only light-quark loops, the exact
$d$-dimensional expression of $\Pi(k^2,\mu^2)$ is given by
\begin{equation}
    \label{eq:Pi_unren_MSB}
  \Pi\(k^2,\mu^2\) = \as \frac{ \TF}{ \pi}\, e^{\epsilon\gameul}\,
  \frac{\Gamma(1+\ep)\, \Gamma^2(1-\ep)}{\Gamma(1-2\ep)} 
  \frac{1-\ep}{(3-2\ep)(1-2\ep)} \, \frac{1}{\ep} \(
  -\frac{k^2 + i\eta}{\mu^2}\)^{-\ep},
\end{equation}
where $\TF=n_l \, \TR$, $\TR =1/2$, $i\eta$ is a small imaginary part
attached to $k^2$ in order to perform the analytic continuation, $\gameul$ is
the Euler-Mascheroni constant, and where we have made the replacement $\mu^2
\to \mu^2 e^\gameul/(4\pi)$, according to the \MSB{} scheme.  In the
following, we also need an expansion in $\epsilon$ of
eq.~(\ref{eq:Pi_unren_MSB})
\begin{equation}\label{eq:Pi-expanded}
  \Pi(k^2,\mu^2)= \as\,\frac{ \TF}{3 \pi} 
\lq  \frac{1}{\ep} + \frac{5}{3} - \log \left| \frac{k^2}{\mu^2}\right|  +
i\pi\,\theta(k^2) \rq +   {\cal O}(\ep)\,,
\end{equation}
from which we can read its  counterterm in the \MSB{} scheme
\begin{equation}
  \label{eq:Pi_ct}
  \Pi_{\rm ct} = \as \, \frac{\TF}{3 \pi} \frac{1}{\epsilon}\,.
\end{equation}
The renormalized gluon propagator dressed with the sum of all quark-loop
insertions is then given by
\begin{equation}
 \label{eq:dress_ren_prop}
  -\frac{i}{k^2} \, g^{\mu\nu} \frac{1}{1+\Pi(k^2,\mu^2)- \Pi_{\rm ct}}\,.
\end{equation}
The above expressions are exact in the large-$n_f$ limit. For our
phenomenological estimate of the contribution to the vacuum polarization
coming from the insertion of gluon loops, we naively assume that this
contribution can be written as
\begin{equation}
  \Pi_g\(k^2,\mu^2\) = -\as \frac{ 11\, \CA}{12\pi}\, e^{\epsilon\gameul}\,
  \frac{\Gamma(1+\ep)\, \Gamma^2(1-\ep)}{\Gamma(1-2\ep)} 
  \left( 1+ \ep\, C_g \right) \frac{1}{\ep} \(
  -\frac{k^2 + i\eta}{\mu^2}\)^{-\ep}\,,
\end{equation}
and we add it to eq.~(\ref{eq:Pi_unren_MSB}) to get
\begin{eqnarray}
  \label{eq:Pi_Pig}
  \Pi\(k^2,\mu^2\) &=& -\as\,\lq \frac{11\, \CA}{12\pi}\(1+\ep\, C_g\)
   - \frac{\TF}{\pi}\frac{1-\ep}{(3-2\ep)(1-2\ep)} \rq
\nonumber \\
   && \times \, e^{\epsilon\gameul} \,
  \frac{\Gamma(1+\ep)\, \Gamma^2(1-\ep)}{\Gamma(1-2\ep)} 
   \frac{1}{\ep} \(
   -\frac{k^2 + i\eta}{\mu^2}\)^{-\ep}\,,
\end{eqnarray}
whose expansion in $\epsilon$ is given by
\begin{equation}
  \label{eq:Pi_Pig_expanded}
  \Pi\(k^2,\mu^2\) = -\as\,b_0 \lq \frac{1}{\epsilon} + C
  - \log \left| \frac{k^2}{\mu^2}\right|  +
i\pi\, \theta(k^2) \rq +{\cal O}(\epsilon) \,,
\end{equation}
where
\begin{equation}
   C =  \frac{1}{b_0} \( \frac{11\, \CA}{12\pi} C_g  - \frac{\TF}{3\pi} \frac{5}{3} \)\,,
\end{equation} 
and  
\begin{equation}
  \label{eq:b_0}
b_0 = \frac{11\, \CA}{12\pi}  - \frac{n_l \,\TR }{3\pi}
\end{equation}
is the first coefficient of the QCD $\beta$ function. The counterterm of
eq.~(\ref{eq:Pi_ct}) is now replaced by
\begin{equation}
  \label{eq:Pi_ct_new}
  \Pi_{\rm ct} =-\as\frac{b_0}{\ep}\,.
\end{equation}
By setting $C_g$ to the value
\begin{equation} \label{eq:Cgvalue}
C_g = \frac{67-3\pi^2}{33} \approx 1.133\,,
\end{equation}
our formula becomes appropriate to describe a QCD effective coupling, as
given in Ref.~\cite{Catani:1990rr}.

\section{Calculation of the large-$\boldsymbol{n_f}$ all-order corrections to an
  infrared-safe observable}
\label{app:details_calc}
In this section we describe the calculation in the large-$n_f$ all-order
corrections that has led to the results for a generic infrared-safe
observable $\obs$ illustrated in Sec.~\ref{sec:description_calc}. We separate
the calculation into different contributions
\begin{equation}
  \label{eq:obs_ave}
  \obs =  \obs_{\rm b} +  \obs_{\rm v}  + \obs_g +  \obs_{q \bar{q}} \,,
\end{equation}
where
\begin{eqnarray}
  \label{eq:obs_b}
  \obs_{\rm b}
  & \equiv & \int \mathd \Phi_{\rm b} \,
  \sigma_{\rm b} (\Phi_{\rm b}) \, \obs (\Phi_{\rm b})\,,
  \\
  \label{eq:obs_v}
  \obs_{\rm v}
  & \equiv &  \int \mathd \Phi_{\rm b} \,
  \sigma_{\rm v} (\Phi_{\rm b}) \, \obs (\Phi_{\rm b})\,,
  \\
    \label{eq:obs_g}
    \obs_g
    &  \equiv  & \int \mathd \Phi_g  \, \sigma_g (\Phi_g) \, \obs
  (\Phi_g) + \int \mathd
  \Phi_{q \bar{q}} \; \sigma_{q \bar{q}} (\Phi_{q \bar{q}}) \, \obs
  (\Phi_{g^{*}})\,,
  \\
    \label{eq:obs_qq}
    \obs_{q \bar{q}}
    & \equiv &\int \mathd \Phi_{q \bar{q}}\; \sigma_{q \bar{q}} (\Phi_{q \bar{q}})
  \, \lq\obs (\Phi_{q \bar{q}}) - \obs (\Phi_{g^{*}})\rq .
\end{eqnarray}

\subsection[The ${\obs_{q \bar{q}}}$ contribution]
           {The $\boldsymbol{\obs_{q \bar{q}}}$ contribution}
           
$\obs_{q \bar{q}}$ receives contributions only from the real graphs with a
final state $W b \bar{b} q \bar{q}$, where $q \bar{q}$ is a pair of light
quarks as depicted in
Fig.~\ref{fig:wbbbar}~(d). We denote with $k^2$ the invariant mass of the
$q \bar{q}$ pair.
Starting from its ${\cal O}(\as^2)$ tree-level
cross section, that we indicate with $\sigma^{(2)}_{q \bar{q}}(\Phi_{q
  \bar{q}})$, with no vacuum polarization insertions in the gluon propagator,
we obtain the differential cross section $\sigma_{q \bar{q}}(\Phi_{q
  \bar{q}})$ with the insertion of all the light-quark bubbles by simply
replacing the bare gluon propagators with the dressed one of
eq.~(\ref{eq:dress_ren_prop})
\begin{equation}
\sigma_{q \bar{q}} = \sigma^{(2)}_{q \bar{q}}(\Phi_{q \bar{q}})\left|
\frac{1}{1 + \Pi (k^2, \mu^2) - \Pi_{\rm ct} } \right|^2.
\end{equation}
From eq.~(\ref{eq:obs_qq}), we get
\begin{equation}
  \label{eq:mrecqqb_dressed}
  \obs_{q \bar{q}} = 
  \int \mathd \Phi_{q \bar{q}} \, \sigma^{(2)}_{q \bar{q}} (\Phi_{q
    \bar{q}}) \, \lq \obs (\Phi_{q \bar{q}}) - \obs (\Phi_{g^{*}})\rq \, \left|
  \frac{1}{1 + \Pi (k^2, \mu^2) - \Pi_{\rm ct} } \right|^2.
\end{equation}
We define
\begin{equation}
  \Delta(\lambda) \equiv  \frac{3\pi}{\as  \TF} \, \lambda^2 \int  \mathd
  \Phi_{q \bar{q}} \; \delta\!\(\lambda^2-k^2\) \sigma^{(2)}_{q \bar{q}} (\Phi_{q
    \bar{q}}) \, \lq \obs (\Phi_{q \bar{q}}) - \obs (\Phi_{g^{*}})\rq ,
\end{equation}
so that we can rewrite eq.~(\ref{eq:mrecqqb_dressed}) as
\begin{equation}
  \label{eq:mrecqqb_dressed_fin}
  \obs_{q \bar{q}} =  \int_0 \frac{\mathd \lambda}{\pi} \,
\frac{2\as \TF}{3}\,\frac{\Delta(\lambda)}{\lambda}
  \left|\frac{1}{1 + \Pi (\lambda^2, \mu^2) - \Pi_{\rm ct} } \right|^2.
\end{equation}
For inclusive observables and for observables that only depend upon leptonic
variables, $\Delta(\lambda)$ is obviously identically zero.  For generic
observables involving jets, we have found that $\Delta(\lambda)\propto \lambda$ for
small $\lambda$.  The following considerations hold for infrared-safe observables
that satisfy this property.

We now make use of the following replacement
\begin{eqnarray}
  \label{eq:asb_square_dress_prop}
  \left| \frac{1}{1 + \Pi (\lambda^2, \mu^2) - \Pi_{\rm ct} } \right|^2 & = & -
  \frac{1}{{\rm Im} \, \Pi (\lambda^2, \mu^2) }\, {\rm Im} \left[ \frac{1}{1 + \Pi
      (\lambda^2, \mu^2) - \Pi_{\rm ct}} \right]
  \nonumber\\
  & = & - \frac{3}{\as \TF} \,{\rm Im} \left[ \frac{1}{1 + \Pi (\lambda^2,
      \mu^2) - \Pi_{\rm ct}} \right]
  \nonumber\\
  & \Longrightarrow & - \frac{3 \,\lambda^2}{\as \TF} \,{\rm Im} \left[ \frac{1}{\lambda^2 + i\eta} \, \frac{1}{1
      + \Pi (\lambda^2, \mu^2) - \Pi_{\rm ct}} \right]
  \nonumber\\
  & = & \frac{3}{\as \TF} \, \frac{3 \pi}{\as \TF}\, \lambda^2\,
  \frac{\mathd\ }{\mathd \lambda^2}\,{\rm Im} \lg \log \lq 1 + \Pi\! \(\lambda^2, \mu^2\) -
    \Pi_{\rm ct}\rq \rg
  \nonumber\\
  & = & \frac{1}{2}\frac{3}{\as  \TF} \, \frac{3 \pi}{\as \TF} \, \lambda\,
  \frac{\mathd\ }{\mathd \lambda} \,{\rm Im} \lg \log \lq 1 + \Pi \!\(\lambda^2, \mu^2\) -
  \Pi_{\rm ct}\rq \rg.\phantom{aaa}
\end{eqnarray}
This works correctly, since the imaginary part of $1 /(\lambda^2 + i\eta)$ in the
square bracket leads, for small $\lambda$, to a contribution in eq.~(\ref{eq:mrecqqb_dressed_fin})
of the form
\begin{equation}
  \int \mathd \lambda\, \frac{\Delta(\lambda)}{\lambda}\, \frac{\lambda^2}{\log
    (\lambda^2/\mu_{\sss C}^2)}\, 
    \delta\(\lambda^2\)\,,
\end{equation}
that, under the assumption that $\Delta(\lambda)$ vanishes as $\lambda$ for small $\lambda$,
is zero.

Equation~(\ref{eq:mrecqqb_dressed_fin}) becomes
\begin{equation}
  \obs_{q \bar{q}} = \int_0 \frac{\mathd \lambda}{\pi}\,
\frac{\Delta(\lambda)}{\lambda}  
  \frac{3 \pi}{\as \TF} \, \lambda\, 
  \frac{\mathd\ }{\mathd \lambda} \,{\rm Im} \lg \log \lq 1 + \Pi (\lambda^2, \mu^2) -
  \Pi_{\rm ct}\rq \rg,
\end{equation}
and integrating by parts
\begin{equation}
  \obs_{q \bar{q}} = - \int_0 \frac{\mathd \lambda}{ \pi}\,
  \frac{\mathd  \Delta(\lambda) }{\mathd \lambda} \frac{3 \pi}{\as \TF}
  \,{\rm Im} \lg \log \lq 1 + \Pi (\lambda^2, \mu^2) - \Pi_{\rm ct}\rq \rg.
\end{equation}
The boundary terms are absent, because $\Delta(\lambda)$ vanishes for small
(by assumption) and large (for kinematic reasons) values of $\lambda$.

\subsection[The ${ \obs_g}$ contribution]
           {The $\boldsymbol{ \obs_g}$ contribution}

The $\obs_g $ term of eq.~(\ref{eq:obs_g}),
\begin{equation}
  \label{eq:obs_g_N0}
  \obs_g   =  \int \mathd \Phi_g  \, \sigma_g (\Phi_g) \, \obs
  (\Phi_g) +  \int \mathd  \Phi_{q \bar{q}}  \,\sigma_{q \bar{q}} (\Phi_{q \bar{q}}) \, \obs
  (\Phi_{g^{*}})\,,
\end{equation}
receives contributions from final states with both a single real gluon or a
$q \bar{q}$ pair.  Both these contributions have collinear divergences
related to the $q \bar{q}$ splitting, that cancel in the sum.

\subsubsection{The gluon contribution}\label{subsec:gcontr}
The first contribution of eq.~(\ref{eq:obs_g_N0}) can be computed starting
from $\sigma_g^{(1)}$, the tree-level cross section for the emission of a
single (massless) gluon. In general, this contribution will produce
soft and collinear singularities, the latter due to the fact that we
consider massless $b$ quarks. We must assume that we are using
dimensional regularization for this contribution.
In order to make the discussion more transparent,
it is convenient to introduce as
regulator a small mass $m_q$ for the quarks in the self-energy corrections,
that we denote with $\Pi (\lambda^2, m_q^2, \mu^2)$.
We then have
\begin{equation}
  \label{eq:gluoncontr}
 \int \mathd \Phi_g \, \sigma_g (\Phi_g) \, \obs (\Phi_g) = \int \mathd \Phi_g
  \, \sigma^{(1)}_g (\Phi_g)\,  \obs (\Phi_g)  \, \frac{1}{1 + \Pi (0, m_q^2, \mu^2)
    - \Pi_{\rm ct}} \,.
\end{equation}
Notice that from the integration of the cross section we may get terms of
order $1/\epsilon^2$ in $d=4-2\epsilon$ dimensions. Thus, in the denominator
of the last factor in eq.~(\ref{eq:gluoncontr}), although the $1/\epsilon$
pole of the UV divergence in $\Pi$ cancels against the one in $\Pi_{\rm ct}$
we should imagine to keep also terms up to order $\epsilon^2$ at this stage,
since they may yield finite contributions when combined with the double pole
of the integration.  We will see that, at the end, when combining all
contributions, these terms actually cancel.

\subsubsection[The  ${q \bar{q}}$ contribution]
{The  $\boldsymbol{q \bar{q}}$ contribution}
We begin by splitting the real phase space $\mathd\Phi_{q\bar{q}}$ into the product of the
phase space for the production of a gluon with virtuality $\lambda$, that we call
$\mathd \Phi_{g^*}$, and its decay into a $q \bar{q}$ pair, that we call $\mathd
\Phi_{\rm dec}$
\begin{equation}
  \label{eq:dphi_qq}
 \mathd \Phi_{q \bar{q}} = \frac{\mathd \lambda^2}{2 \pi} \, \mathd
 \Phi_{\rm dec} \,\mathd \Phi_{g^*}\,.
\end{equation}
Using the optical theorem, we easily obtain the relation
\begin{equation}
 \int \mathd \Phi_{\rm dec} \,\sigma^{(2)}_{q \bar{q}} (\Phi_{q \bar{q}}) =
 \sigma^{(1)}_{g^*} (\lambda,\Phi_{g^*}) \,
 \frac{1}{\lambda^2}\,  2\, {\rm Im}\left[ \Pi\! \(\lambda^2, m_q^2, \mu^2\)\right] \,,
\end{equation}
where $\sigma^{(2)}_{q \bar{q}} (\Phi_{q \bar{q}})$ is the tree-level cross
section for $W^*\to W b\bar{b} q\bar{q}$, and
$\sigma^{(1)}_{g^*}(\lambda,\Phi_{g^*})$ is the tree-level cross section for
the process $W^*\to W b\bar{b} g^*$, where $g^*$ is a gluon with mass
$\lambda$. We have again given a small mass $m_q$ to the light quarks, in
order to match what we did in App.~\ref{subsec:gcontr}.  Thus the second term
on the right-hand side of eq.~(\ref{eq:obs_g_N0}) can be written as
\begin{eqnarray}
  \label{eq:qqcontr}
  &&\int \mathd
  \Phi_{q \bar{q}} \,\sigma_{q \bar{q}} (\Phi_{q \bar{q}})\, \obs (\Phi_{g^*})
  \nonumber\\
  && \hspace{1cm}  =
   \int \frac{\mathd \lambda^2}{2 \pi} \, \mathd \Phi_{g^*} \,
  \sigma_{g^*}^{(1)} (\lambda,\Phi_{g^*})\, \obs (\Phi_{g^*})\,  \frac{2\,
     {\rm Im}\left[ \Pi (\lambda^2, m_q^2, \mu^2)\right]
    }{\lambda^2\,\left| 1
  + \Pi (\lambda^2, m_q^2, \mu^2) 
  - \Pi_{\rm ct} \right|^2}\,,\phantom{aaaaaaa} 
\end{eqnarray}
where we have inserted the dressed gluon propagators.
As long as $m_q>0$, no divergences arise from this contribution,
since the imaginary part of $\Pi$ vanishes for $\lambda<2m_q$.

\subsection[Combination of the gluon and $q\bar{q}$ contributions]
           {Combination of the gluon and $\boldsymbol{q\bar{q}}$ contributions}
Defining
\begin{equation}
  R(\lambda)=\int \mathd \Phi_{g^*}\, \sigma_{g^*}^{(1)}(\lambda,\Phi_{g^*} )\,O(\Phi_{g^*})\,,
\end{equation}
we can combine eq.~(\ref{eq:gluoncontr}) and~(\ref{eq:qqcontr}) and get
\begin{equation}\label{eq:finalOg}
  O_g=R^{(\epsilon)}(0)\frac{1}{1+\Pi(0,m_q^2,\mu^2)-\Pi_{\rm ct}}-\frac{1}{\pi}\int
  \frac{\mathd \lambda^2}{\lambda^2} R(\lambda)\, {\rm Im}\frac{1}{1+\Pi(\lambda^2,m_q^2,\mu^2)-\Pi_{\rm ct}}\,.
\end{equation}
With the notation $R^{(\epsilon)}(0)$ we remind that for $\lambda=0$ there are
infrared divergences in $R$ that are regulated in dimensional regularization.

\begin{figure}[tb]
  \centering
  \includegraphics[width=0.5\textwidth]{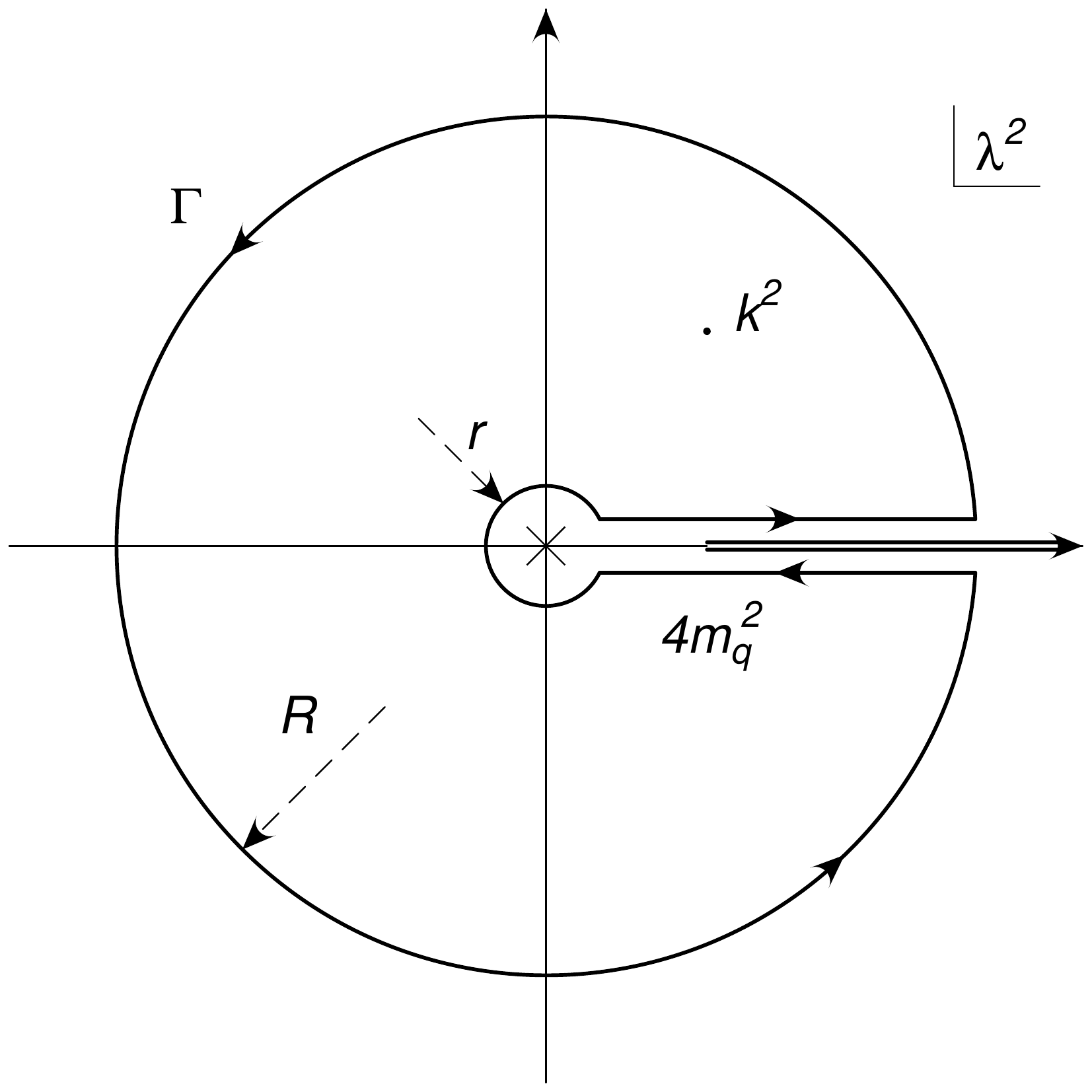}
  \caption{Integration contour in the complex $\lambda^2$ plane of the
    integral in eq.~(\ref{eq:count_integ_mq}). The cut of $\Pi (\lambda^2,
    m_q^2, \mu^2)$, starting at $4 m_q^2$, is also shown.}
  \label{fig:integr_path}
\end{figure}

\subsection[The ${\obs_{\rm v}}$ contribution]
           {The $\boldsymbol{\obs_{\rm v}}$ contribution}
\label{sec:virtual}           
The virtual contribution with all
polarization insertions can be obtained by performing the replacement
\begin{equation}
\frac{1}{k^2+i\eta} \to \frac{1}{k^2+i\eta} \,
\frac{1}{1+\Pi(k^2,m_q^2,\mu^2)-\Pi_{\rm ct}}
\end{equation}
in the computation of the NLO virtual corrections for $W^*\to W b\bar{b}$,
where $k$ is the momentum flowing in the virtual gluon propagator.
Ultraviolet divergences arise in individual diagrams, and soft and collinear
singularities also arise, so that the calculation must be performed in
$d=4-2\epsilon$ dimensions.  For a generic complex $k^2$, using the residue
theorem, we can write\footnote{A similar procedure is suggested in
  Ref.~\cite{Beneke:1994qe}. The form we have adopted here, that combines the
  cuts at $k^2=0$ and $k^2>4m_q^2$, has the advantage that it does not
  require subtractions.}
\begin{eqnarray}
  \label{eq:count_integ_mq}
&& \frac{1}{k^2} \,  \frac{1}{1 + \Pi
    (k^2, m_q^2, \mu^2) - \Pi_{\rm ct}} = \frac{1}{2\pi i}
    \oint_\Gamma d\lambda^2  \frac{1}{ \lambda^2-k^2}  \, \frac{1}{\lambda^2}\,\frac{1}{1 + \Pi
    (\lambda^2, m_q^2, \mu^2) 
    - \Pi_{\rm ct}}
\\ \label{eq:virt_trick}
&& =   \frac{1}{k^2}\frac{1}{1 + \Pi 
   (0, m_q^2, \mu^2)  - \Pi_{\rm ct}}    -\frac{1}{\pi} \int_{4m_q^2}^{+\infty}\!\!  \frac{d\lambda^2}{\lambda^2}
 \, \frac{1}{ k^2 - \lambda^2}\, {\rm Im}
 \,\frac{1}{1 + \Pi 
   (\lambda^2, m_q^2, \mu^2)  - \Pi_{\rm ct}}, \phantom{aaaaaa}
\end{eqnarray}
where $\Gamma$ is the contour depicted in Fig.~\ref{fig:integr_path}.
Notice that when we write $\Pi(\lambda^2,m_q^2,\mu^2)$ for real $\lambda^2$
we imply, consistently with eq.~(\ref{eq:Pi_unren_MSB}),
that a positive tiny imaginary part should be added to $\lambda^2$.

From eq.~(\ref{eq:virt_trick}), we see that we get two contributions
to the virtual corrections, corresponding to its two terms
\begin{equation}
  \label{eq:virt_contrib}
\obs_{\rm v}=  \frac{V^{(\epsilon)}(0)}{1 + \Pi 
  (0, m_q^2, \mu^2)  - \Pi_{\rm ct}}
- \frac{1}{\pi} \int_{0}^{\infty} \frac{\mathd \lambda^2}{\lambda^2}
\, V (\lambda)\,{\rm Im}\,
\frac{1}{1 + \Pi 
       (\lambda^2, m_q^2, \mu^2)  - \Pi_{\rm ct}},
\end{equation}
where we have set the lower integration limit to 0, since the
imaginary part is zero for $\lambda^2 < 4 m_q^2$, and 
where $V(\lambda)$ stands for the virtual contribution to our
observable computed with the substitution
\begin{equation}
 \frac{1}{k^2+i\eta}\ \rightarrow \ \frac{1}{k^2 - \lambda^2+i\eta}\,, 
\end{equation}
in all the NLO virtual diagrams. For $\lambda>0$, this corresponds to replace
the massless gluon propagator with the propagator of a gluon with mass
$\lambda$, while for $\lambda=0$ nothing is changed. As before we have added
the superscript $(\epsilon)$ to $V(0)$, to remind us that this quantity
contains poles in $\epsilon$ due to collinear and soft singularities. As
before, in the denominator of the factor multiplying $V^{(\epsilon)}(0)$ we
should keep terms up to order $\epsilon^2$. On the other hand, for
$\lambda>0$, $V(\lambda)$ is finite if a mass counterterm has been included
in the calculation, and the appropriate wave-function renormalization of the
external legs has been carried out.  We also notice that
eq.~(\ref{eq:virt_contrib}) is meaningful only if $V(\lambda)$ vanishes as
$\lambda$ goes to infinity. This turns out to be the case, provided that mass
renormalization is carried out in the pole mass scheme.

\subsection[Combination of the gluon, $q\bar{q}$ and virtual contributions]
           {Combination of the gluon, $\boldsymbol{q\bar{q}}$ and virtual
             contributions}
           
Defining
\begin{equation}
    S(\lambda)\equiv R(\lambda)+V(\lambda)
\end{equation}
and adding up eqs.~(\ref{eq:finalOg}) and~(\ref{eq:virt_contrib})
we get
\begin{equation}\label{eq:g_plus+_v1}
  \obs_g + \obs_{\rm v}=  S(0)\frac{1}{1 + \Pi 
  (0, m_q^2, \mu^2)  - \Pi_{\rm ct}}
- \frac{1}{\pi} \int_{0}^{\infty} \frac{\mathd \lambda^2}{\lambda^2}
\, S(\lambda)\,
{\rm Im}\frac{1}{1 + \Pi 
       (\lambda^2, m_q^2, \mu^2)  - \Pi_{\rm ct}}.
\end{equation}
Notice that we have written $S(0)=R^{(\epsilon)}(0)+V^{(\epsilon)}(0)$, since
the $\epsilon$ infrared poles cancel in the sum, provided the observable
we are considering is IR safe. Furthermore, for the same reason, $S(\lambda)$
has a well defined limit for $\lambda \to 0$, that is equal to $S(0)$.
In addition, since $S(0)$ is finite, we can neglect terms of order $\epsilon$ in the
denominator of the factor that multiplies it, so that all $\epsilon$
dependences cancel in eq.~(\ref{eq:g_plus+_v1}).

\begin{figure}[tb]
  \centering
  \includegraphics[width=0.5\textwidth]{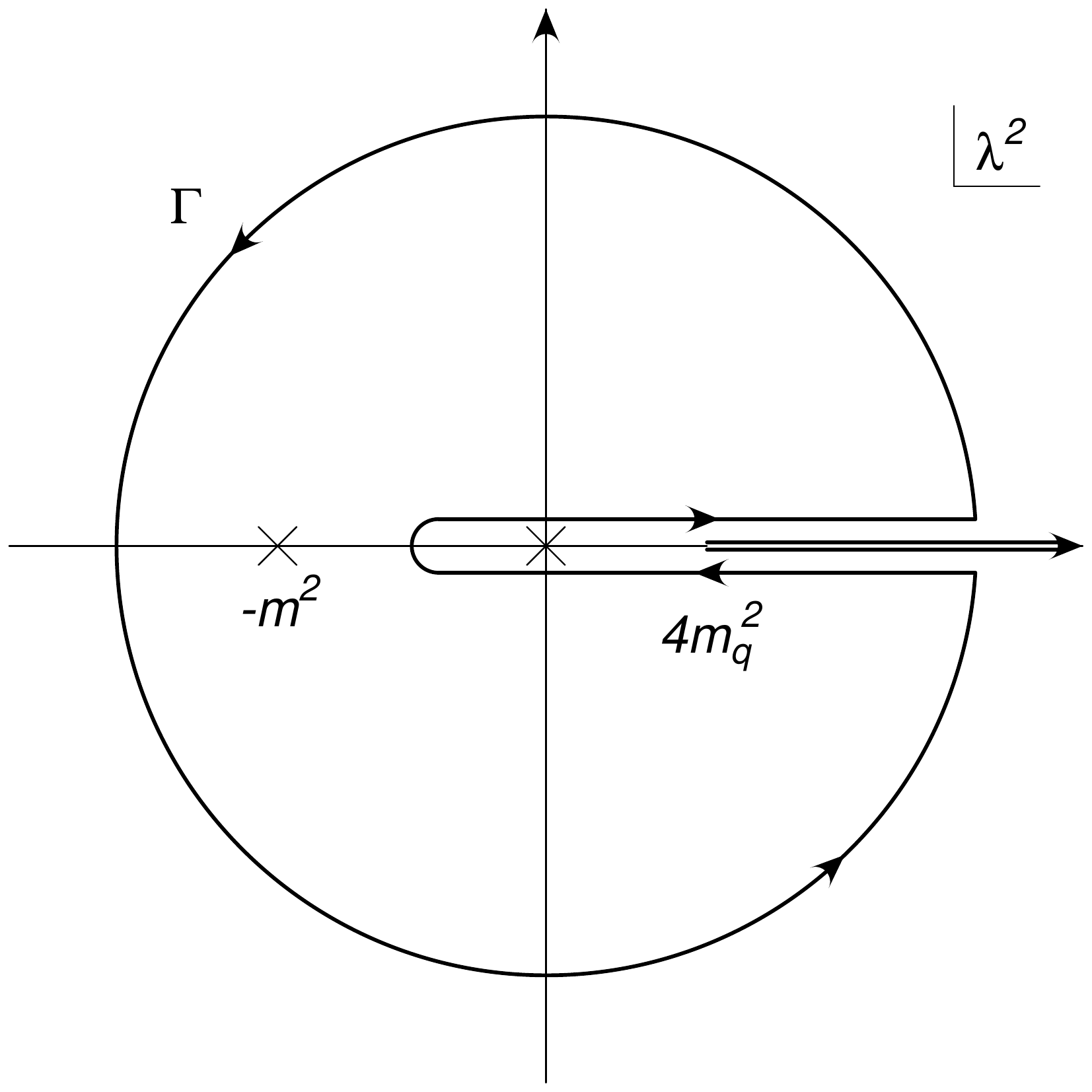}
  \caption{Integration contour in the complex $\lambda^2$ plane, used to perform
    the integral in eq.~(\ref{eq:g_plus_v3}).}
  \label{fig:integr_path1}
\end{figure}

We would like now to take the limit $m_q\to 0$ in eq.~(\ref{eq:g_plus+_v1}).
In doing so, we must be careful to handle properly the singularities at $\lambda=0$.
We thus split eq.~(\ref{eq:g_plus+_v1}), by adding and subtracting the
same quantity, as follows
\begin{eqnarray}
 \obs_g + \obs_{\rm v}&=&  S(0)\frac{1}{1 + \Pi 
  (0, m_q^2, \mu^2)  - \Pi_{\rm ct}} 
- \frac{1}{\pi} \int_{0}^{\infty} \frac{\mathd \lambda^2}{\lambda^2}
\, \frac{S(0)}{\frac{\lambda^2}{m^2}+1}\,
{\rm Im}\,\frac{1}{1 + \Pi(\lambda^2, m_q^2, \mu^2)  - \Pi_{\rm ct}} \nonumber \\
  && +\frac{1}{\pi} \int_{0}^{\infty} \frac{\mathd \lambda^2}{\lambda^2}
\, \left\{\frac{S(0)}{\frac{\lambda^2}{m^2}+1}-S(\lambda)\right\}\,
{\rm Im}\frac{1}{1 + \Pi 
     (\lambda^2, m_q^2, \mu^2)  - \Pi_{\rm ct}} \label{eq:g_plus_v2} \\
  &=&-  \int_{0^-}^\infty
  \frac{\mathd \lambda^2}{\pi} \frac{S(0)}{\frac{\lambda^2}{m^2}+1} {\rm Im}\left[ 
    \frac{1}{\lambda^2+i\eta}  \,
  \frac{1}{1 + \Pi (\lambda^2, m_q^2, \mu^2) - \Pi_{\rm ct}}\right] \nonumber \\
  & +&\int_{0}^{\infty} \frac{\mathd \lambda^2}{\pi}
\, \left\{\frac{S(0)}{\frac{\lambda^2}{m^2}+1}-S(\lambda)\right\}\,
 {\rm Im}\left[ 
    \frac{1}{\lambda^2+i\eta}  \,
    \frac{1}{1 + \Pi (\lambda^2, m_q^2, \mu^2) - \Pi_{\rm ct}}\right],\phantom{aaa}
 \label{eq:g_plus_v3}
\end{eqnarray}
where the first two terms in eq.~(\ref{eq:g_plus_v2}) have been merged in the
first term of eq.~(\ref{eq:g_plus_v3}), and the last term in
eq.~(\ref{eq:g_plus_v2}) can be turn into the last term of
eq.~(\ref{eq:g_plus_v3}) because the imaginary part of $1/(\lambda^2+i\eta)$
is a $\delta(\lambda^2)$ function, that yields a zero when multiplied by the
expression in the curly brackets.  The notation $0^-$ for the lower bound of
the first integral in eq.~(\ref{eq:g_plus_v3}) simply means that the
integration range should start slightly below 0, so that the
$\delta(\lambda^2)$ arising from the imaginary part acts in a well defined
way.  Notice also that the separation of terms in eq.~(\ref{eq:g_plus_v2})
does not spoil the convergence at large $\lambda$. We can then take safely
the limit $m_q\to 0$ in both terms of eq.~(\ref{eq:g_plus_v3}). In fact, the
first term can be expressed as an integral along the contour $\Gamma$ of
Fig.~\ref{fig:integr_path1},
that in turn can be transformed in the residue at $\lambda^2=-m^2$, that has
a well defined limit for $m_q\to 0$, and the second term is not singular in
the $\lambda \to 0$ region.

After having taken the limit $m_q\to 0$, we make use of the identity (see
eqs.~(\ref{eq:Pi-expanded}) and~(\ref{eq:Pi_ct}))
\begin{equation}
  \frac{1}{\lambda^2+i\eta}  \,
  \frac{1}{1 + \Pi (\lambda^2, \mu^2) - \Pi_{\rm ct}}= -\frac{3\pi}{\as \TF}
   \frac{\mathd}{\mathd \lambda^2} \log\left[1 + \Pi\!\(\lambda^2, \mu^2\) - \Pi_{\rm ct}\right]\,,
 \end{equation}
and rewrite (\ref{eq:g_plus_v3}) as
\begin{eqnarray}
 \obs_g + \obs_{\rm v}&=& - \int_{0^-}^\infty
  \frac{\mathd \lambda^2}{\pi} \frac{S(0)}{\frac{\lambda^2}{m^2}+1} \(-\frac{3\pi}{\as \TF}\) {\rm Im}\lg
   \frac{\mathd}{\mathd \lambda^2} \log\left[1 + \Pi\!\(\lambda^2, \mu^2\) -
     \Pi_{\rm ct}\right] \rg
\nonumber \\
  & +&\int_{0}^{\infty} \frac{\mathd \lambda^2}{\pi}
\, \left\{\frac{S(0)}{\frac{\lambda^2}{m^2}+1}-S(\lambda)\right\}\,
 \(-\frac{3\pi}{\as \TF}\) {\rm Im}\lg 
   \frac{\mathd}{\mathd \lambda^2} \log\left[1 + \Pi\!\(\lambda^2, \mu^2\) -
     \Pi_{\rm ct}\right] \rg\!.
 \nonumber  \\
   \label{eq:g_plus_v4} 
\end{eqnarray}
We can now integrate by parts in $\lambda^2$. The boundary term at
$\lambda=0^-$ in the first integral vanishes, since the imaginary part
vanishes for $\lambda<0$, while in the second integral it vanishes because
the expression in the curly bracket vanishes. We are left with
\begin{eqnarray}
 \obs_g + \obs_{\rm v}&=& - \int_{0^-}^\infty
                          \frac{\mathd \lambda^2}{\pi} \left[\frac{\mathd}{\mathd \lambda^2} \frac{S(0)}{\frac{\lambda^2}{m^2}+1}\right]
                          \frac{3\pi}{\as \TF} {\rm Im} \lg
   \log\left[1 + \Pi\!\(\lambda^2, \mu^2\) - \Pi_{\rm ct}\right]\rg  \nonumber \\
  & & {}+\int_{0}^{\infty} \frac{\mathd \lambda^2}{\pi}
\, \left[\frac{\mathd}{\mathd \lambda^2}\left\{\frac{S(0)}{\frac{\lambda^2}{m^2}+1}-S(\lambda)\right\}\right]\,
 \frac{3\pi}{\as \TF} {\rm Im}\lg
       \log\left[1 + \Pi\!\(\lambda^2, \mu^2\) - \Pi_{\rm ct}\right] \rg\nonumber \\
  &=& - \int_{0}^{\infty} \frac{\mathd \lambda^2}{\pi}
\, \frac{\mathd S(\lambda)}{\mathd \lambda^2}\,
 \frac{3\pi}{\as \TF} \,{\rm Im}\lg
       \log\left[1 + \Pi\!\(\lambda^2, \mu^2\) - \Pi_{\rm ct}\right] \rg \!. \label{eq:g_plus_v5}
\end{eqnarray}

\subsection{Summary}
\label{app:summary}
Using eqs.~(\ref{eq:Pi_Pig_expanded}) and~(\ref{eq:Pi_ct_new}), we can write
the renormalized polarization contribution $\Pi(\lambda^2,\mu^2)$, for
$\lambda^2>0$, in the form
\begin{equation}
  \label{eq:Piren_pheno}
  \Pi(\lambda^2,\mu^2)-\Pi_{\rm ct}=-\as \, b_0 \left[C-\log
    \left(\frac{\lambda^2}{\mu^2}\right)+i\pi\right]=\as \, b_0 \left[\log
    \frac{\lambda^2}{\mu_{\sss C}^2} - i\pi\right] ,
\end{equation}
where we have defined
\begin{equation}
  \label{eq:mu_C}
  \mu_{\sss C}=\mu \, \exp\(\frac{C}{2}\).
\end{equation}
Notice that, in the large-$n_f$ limit, $C=5/3$ and $b_0=-\TF/(3\pi)$. We can
then write
\begin{equation}
  {\rm Im} \lg \log \lq 1 + \Pi\(\lambda^2, \mu^2\) -
  \Pi_{\rm ct} \rq \rg=-\atanpilam . 
\end{equation}
Summarizing our findings, we have
\begin{eqnarray}
\label{eq:final_obs}
  \obs  &=&  \obs_{\rm b} + \obs_{\rm v}  +  \obs_{g} +  \obs_{q \bar{q}}
  \nonumber \\
  &=& \obs_{\rm b} - \int_0^{\infty} \frac{\mathd \lambda}{\pi} \, \frac{\mathd\ }{\mathd \lambda}
  \!\lq V(\lambda) + R(\lambda) +   \Delta(\lambda) \rq
   \frac{1}{b_0\,\as}\,
  \atanpilam, \phantom{aaaa}
\end{eqnarray}
where
\begin{eqnarray}
   \obs_{\rm b} &=&  \int \mathd \Phi_{\rm b} \, \sigma_{\rm b}
 (\Phi_{\rm b})\, \obs (\Phi_{\rm b}), \\
 V(\lambda) &=&  \int \mathd \Phi_{\rm b} \, \sigma_{\rm v}^{(1)}
 (\lambda,\Phi_{\rm b})\, \obs (\Phi_{\rm b}),
 \\
 R(\lambda) &=&  \int \mathd \Phi_{g^*}\, \sigma^{(1)}_{g^{*} } (\lambda,
 \Phi_{g^*}) \, \obs (\Phi_{g^*})\,,
 \\
  \Delta(\lambda) &=& \frac{3\pi}{\as  \TF} \,  \lambda^2 \int
   \mathd \Phi_{q \bar{q}} \; \delta\!\(\lambda^2-k^2\) \sigma^{(2)}_{q \bar{q}} (\Phi_{q
    \bar{q}})  \lq \obs (\Phi_{q \bar{q}}) - \obs (\Phi_{g^*})\rq.
\end{eqnarray}
In case one is interested in a normalized observable $\langle O \rangle$,
where the normalization factor $N_\Theta$ is the inverse of the total cross
section as given in eq.~(\ref{eq:def_N}), the resulting final expressions are
given in eqs.~(\ref{eq:final_M_wcuts})--(\ref{eq:final_Deltaqqtilde}).

\section{Pole\,-\,$\boldsymbol{\MSB}$ mass conversion with a fully dressed gluon
  propagator}

In order to extract the pole-\MSB{} mass relation at all orders in the
large-$n_f$ limit, we follow a strategy similar to the one described in
App.~\ref{sec:virtual}, where the virtual contribution is expressed in terms
of the NLO correction computed keeping a finite gluon mass $\lambda$.  We
thus begin by calculating the one-loop top-quark self energy with a massive
gluon in App.~\ref{sec:PoleMasslambda}.  This result is then used in
App.~\ref{app:PoleMSB} to derive the mass conversion formula.

\subsection{Pole mass with a massive gluon}
\label{sec:PoleMasslambda}
\begin{figure}[tb]
  \centering
  \includegraphics[width=0.5\textwidth]{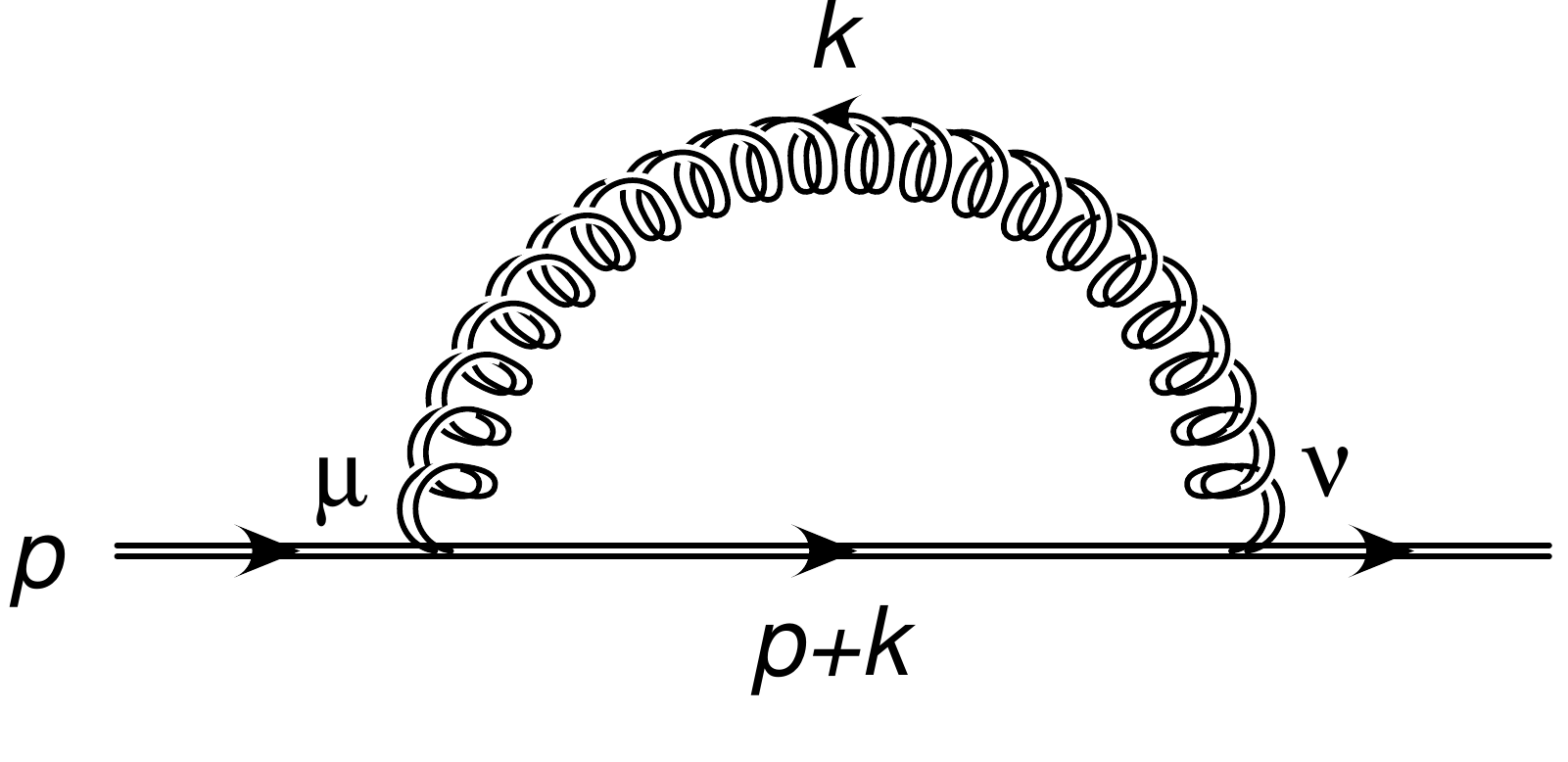}
  \caption{One-loop self-energy Feynman diagram for the propagation of a
    quark of momentum $p$ and bare mass $\mb$, and a gluon of mass
    $\lambda$.}
  \label{fig:quark_self_energy}
\end{figure}
The one-loop self energy, depicted in Fig.~\ref{fig:quark_self_energy}, for a
quark with bare mass $m_b$, due to the exchange of a gluon of mass $\lambda$,
computed in $d=4-2\ep$ dimensions, with $\mu$ rescaled according to the \MSB{}
prescription,  is
\begin{eqnarray}
  \label{eq:Sigmalambda}
  \Sigma^{(1)}_{\lambda} (\slashed{p}, m_b,\mu,\as) & = & 4\pi\as\!
\left( \frac{\mu^2 e^{\gamma_E}}{4 \pi}
  \right)^{\epsilon}\!\!\!  \int \!\frac{\mathd^d k}{(2 \pi)^d} (- i \gamma^{\nu} t^a) 
  \frac{i}{\slashed{p} + \slashed{k} - \mb+ i \eta} (- i \gamma^{\mu} t^a)
  \frac{- i g_{\mu\nu}}{k^2-\lambda^2+ i \eta}
  \nonumber\\[2mm]
  & \equiv & 4\pi\as\, \CF \,i \left[ \mathcal{A}_{\lambda} \!\(\mb,\mu\) \slashed{p} - 
    \mathcal{B}_{\lambda} \!\(\mb,\mu\)   m_b  \right] ,  
\end{eqnarray}
where we assume $p^2=m_b^2$, and we have defined
\begin{eqnarray}
  \mathcal{A}_{\lambda}\! \(\mb,\mu\) & = & - \frac{(2 - d)}{2 i m_b^2} \left(
  \frac{\mu^2 e^{\gamma_E}}{4 \pi} \right)^{\epsilon}  \left\{ \int \frac{\mathd^d
  k}{(2 \pi)^d}  \frac{p^2 - \lambda^2 + m_b^2}{\left[(p + k)^2 - m_b^2 + i
      \eta \right]\left[k^2 - \lambda^2 + i \eta\right]}  \right.
  \nonumber\\
  &  &  +\left. \int \frac{\mathd^d k}{(2 \pi)^d} \frac{1}{k^2 - \lambda^2 + i
  \eta} - \int \frac{\mathd^d k}{(2 \pi)^d} \frac{1}{k^2 - m_b^2 + i
    \eta} \right\},
  \\
  \mathcal{B}_{\lambda} \!\(\mb,\mu\) & = & \frac{d}{i}  \left(
  \frac{\mu^2 e^{\gamma_E}}{4 \pi} \right)^{\epsilon}
  \int \frac{\mathd^d k}{(2 \pi)^d}  \frac{1}{\left[(p + k)^2 -
  m_b^2 + i \eta\right] \left[ k^2 - \lambda^2 + i \eta\right]}\,.
\end{eqnarray} 
The dressed quark propagator at one loop then reads 
\begin{eqnarray}
&&\frac{i}{\slashed{p}-\mb} + \frac{i}{\slashed{p}-\mb} \Sigma^{(1)}_{k}
(\slashed{p}, m_b,\mu,\as)   \frac{i}{\slashed{p}-\mb} 
\nonumber\\
  \displaystyle
&& \hspace{2cm}=\frac{i}{\slashed{p}-\mb} \left\{ 1 - 4\pi\as \CF \left[\mathcal{A}_{\lambda}
    \(\mb, \mu\) \slashed{p} -  \mathcal{B}_{\lambda} \(\mb, \mu\) 
    \, m_b \right] \frac{1}{\slashed{p}-\mb} \right\}
  \nonumber \\
&& \hspace{2cm}=\frac{i\left[ 1+ 4\pi\as\CF \, \mathcal{A}_{\lambda} \(\mb,
      \mu\)\right]^{-1}}{\slashed{p}
    - m_b \lg 1+ +4\pi\as\CF \lq  \mathcal{B}_{\lambda} \(\mb, \mu\) -\mathcal{A}_{\lambda} \(
    \mb, \mu\) \rq \rg} +\mathcal{O}\(\as^2\). \phantom{aaa}
\end{eqnarray}
The position of the pole in the propagator defines the pole mass
\begin{equation}
  m \equiv m_b \lg 1 + 4\pi\as \CF \lq \mathcal{B}_{\lambda}\(\mb, \mu\)
  -\mathcal{A}_{\lambda}\(\mb, \mu\) \rq \rg.
\end{equation}
Neglecting terms of the order $\as^2$, we can write
\begin{equation}
  \mpole = \mb \lq 1+ \as \,r_{\lambda} (m, \mu) \right]\,,
\end{equation}
where we have defined
\begin{eqnarray}\label{eq:rlambda}
  r_{\lambda} (m, \mu) & \equiv & 4 \pi \,\CF \lq \mathcal{B}_{\lambda} \(
  \mpole, \mu\)  - \mathcal{A}_{\lambda} \(\mpole, \mu\) \rq.
\end{eqnarray}
Furthermore, separating the finite and divergent part of $r_{\lambda}$ according to the
\MSB{} prescription
\begin{equation}
  r_{\lambda} (m, \mu) = \frac{1}{\epsilon} r^{\rm (d)}_{\lambda} (m, \mu)
  + r^{\rm (f)}_{\lambda} (m, \mu) + {\cal O}(\epsilon),
\end{equation}
we have that the \MSB{} mass is given by
\begin{equation}
  \overline{m}(\mu)= \mb \left[ 1+ \as \frac{1}{\epsilon} r^{\rm (d)}_{\lambda}
      \(\overline{m}(\mu), \mu\)\right],
\end{equation}
so that the relation between the pole mass and the \MSB{} mass\footnote{At
  the order that we consider, the mass appearing in the term of order $\as$
  can be either the pole or the \MSB{} one.} is
\begin{equation}
  \overline{m}(\mu)=  m\left[1- \as\, r^{(\rm f)}_{\lambda}
    \(m, \mu\) \right].
\label{eq:mMSB}
\end{equation}
For a generic $\lambda$ value we have
\begin{eqnarray}
  \mathcal{A}_{\lambda} (m,\mu) 
  & = & \frac{1}{(4 \pi)^2} \left\{ \frac{1}{\epsilon} + 2 - \log
  \frac{m^2}{\mu^2} - x (1 + \log x) +_{} (2 - x) H (x) \right\} +{\cal O}(\ep)\,, \\
  \mathcal{B}_{\lambda} (m,\mu) 
  & = & \frac{1}{(4 \pi)^2} \left\{ \frac{4}{\epsilon} + 6 - 4 \log
  \frac{m^2}{\mu^2} + 4 H (x) \right\}+{\cal O}(\ep)\, ,
\end{eqnarray}
where
\begin{equation}
  x = \frac{\lambda^2}{m^2}
\end{equation}
and
\begin{equation}
\label{eq:H}
H(x) =
\left\{
\begin{array}{ll}
 \displaystyle{  -\frac{x}{2} \log{x} - \sqrt{x(4-x)}
   \,\arctan\sqrt{\frac{4-x}{x}}} & x < 4\,,
 \\[2mm]
  \displaystyle{   -\frac{x}{2} \,\log{x} + \frac{1}{2}  \sqrt{x(x-4)} \log \frac{\sqrt{x} +
      \sqrt{x-4}}{\sqrt{x} - \sqrt{x-4}}} \quad \quad& x>4   \,.
\end{array}
\right.
\end{equation}
This leads to
\begin{eqnarray}
  \label{eq:r_k}  
 r_{\lambda} (m, \mu) =\frac{\CF}{4 \pi} \left\{ \frac{3}{\epsilon} + 4 - 3 \log
  \frac{m^2}{\mu^2} + x (1 + \log x) + (2 + x) H (x) \right\} . 
\end{eqnarray}
Our result is consistent with Ref.~\cite{Ball:1995ni}.
Notice that, for large $\lambda$, eq.~(\ref{eq:r_k}) becomes
\begin{equation}
  r_{\lambda} (m, \mu) = \frac{\CF}{4 \pi} \left\{ \frac{3}{\epsilon}
  + \frac{5}{2} - 3 \log \frac{\lambda^2}{\mu^2} + \ldots \right\} ,
\end{equation}
and for small $\lambda$
\begin{equation}
  r_{\lambda} (m, \mu) = \frac{\CF}{4 \pi} \left\{ \frac{3}{\epsilon}
  + 4 - 3 \log \frac{m^2}{\mu^2} -2\pi\frac{\lambda}{m} + \ldots \right\} .
\end{equation}
We also need the exact $d$-dimensional expression of $r_{\lambda}(m, \mu)$,
for $\lambda = 0$ and $\lambda \gg m$.  For $\lambda=0$ we have
\begin{eqnarray}
  \mathcal{A}_\lambda^0 (m,\mu) 
  & = & \frac{1}{(4 \pi)^2} \,e^{\ep \gameul}\,\frac{\Gamma(1+\ep)}{\ep}\, \left(\frac{\mu^2}{m^2}
  \right)^\ep \frac{1}{1-2\ep}\,, \\
  \mathcal{B}_\lambda^0 (m,\mu) 
  & = & \frac{1}{(4 \pi)^2}\, e^{\ep \gameul}\,\frac{\Gamma(1+\ep)}{\ep}\, \left(\frac{\mu^2}{m^2}
  \right)^\ep\frac{4-2\ep}{1-2\ep}\,, \\
  r_\lambda^0(m,\mu) &=& \frac{\CF}{4\pi} \, e^{\epsilon\gameul} \,\frac{\Gamma(1+\ep)}{\ep}\! \left(
  \frac{\mu^2}{m^2}\right)^\ep \frac{3-2\ep}{1-2\ep}.
  \label{eq:rlambda0}
\end{eqnarray}
For $\lambda \gg m$ (and $\mu \approx \lambda$) we have
\begin{eqnarray}
  \mathcal{A}^\infty_{\lambda} (\mu) &\equiv& \lim_{m \to 0} \mathcal{A}_{\lambda} (m,\mu) 
  = \frac{1}{(4 \pi)^2}\, e^{\ep \gameul}\,\frac{\Gamma(1+\ep)}{\ep}\, \left(\frac{\mu^2}{\lambda^2}
  \right)^\ep \frac{2-2\ep}{2-\ep}\,, \\
  \mathcal{B}^\infty_{\lambda} (\mu) &\equiv& \lim_{m \to 0} \mathcal{B}_{\lambda} (m,\mu) 
  = \frac{1}{(4 \pi)^2}\, e^{\ep \gameul}\,\frac{\Gamma(1+\ep)}{\ep}\, \left(\frac{\mu^2}{\lambda^2}
  \right)^\ep\frac{4-2\ep}{1-\ep}\,, \\
  r_{\lambda}^{\infty}(\mu) &\equiv&  \lim_{m \to 0}
  r_{\lambda} (m, \mu) = \frac{\CF}{4 \pi}\, e^{\ep \gameul}\,\frac{\Gamma(1+\ep)}{\ep}\, \left(\frac{\mu^2}{\lambda^2}
  \right)^\ep \frac{2(3-2\ep)}{(1-\ep)(2-\ep)}.
    \label{eq:rlambdainf}
\end{eqnarray}

\subsection{All-orders result}
\label{app:PoleMSB}
In this section we deal with the computation of the on-shell top self-energy
in $d = 4 - 2\epsilon$ dimensions, $\Sigma(\slashed{p},\mb,\mu,\as)$, with
the insertion of an infinite number of light-quark loops in the gluon line.
The one-loop self energy with a massless gluon is obtained by setting
$\lambda=0$ in eq.~(\ref{eq:Sigmalambda}).  According to
App.~\ref{sec:dressed_gluon}, once the gluon line is dressed with all
possible light-quark loop insertions, the expression for the self-energy
becomes
\begin{eqnarray}
  \label{eq:Sigma_bubbles}
  \Sigma (\slashed{p}, \mb,\mu,\as) & = & 4\,\pi\,\as
                                             \left( \frac{\mu^2 e^{\gamma_E}}{4 \pi}
  \right)^{\epsilon}  \int \frac{\mathd^d k}{(2 \pi)^d} (- i \gamma^{\nu} t^a) 
  \frac{i}{\slashed{p} + \slashed{k} - \mb} (- i \gamma^{\mu} t^a)
  \nonumber\\
  && \hspace{3cm}\times  \frac{- ig_{\mu \nu}}{(k^2 + i \eta)  [1 + \Pi (k^2,
      \mu^2) - \Pi_{\rm ct} ]}
  \nonumber\\[2mm]
  & \equiv & 4\pi\as\,\CF\,i  \left[  \mathcal{A} \!\(\mb, \mu\) \slashed{p}
    -  \mathcal{B} \!\( \mb,\mu\) \mb  \right] , 
\end{eqnarray}
where we assume $p^2=\mb^2$ and the \MSB{} scheme expressions of $\Pi$ and
$\Pi_{\rm ct}$ are given by eqs.~(\ref{eq:Pi_unren_MSB})
and~(\ref{eq:Pi_ct}), respectively. Using eq.~(\ref{eq:virt_trick}) we can
rewrite eq.~(\ref{eq:Sigma_bubbles}) as
\begin{equation}
{\Sigma}(\slashed{p},\mb,\mu,\as)   = - \frac{1}{\pi}
  \int_{0 -}^{\infty} \mathd \lambda^2\, \,\Sigma_\lambda^{(1)}\!\(\slashed{p},\mb,\mu,\as\)\,
{\rm Im} \lq 
 \frac{1}{\lambda^2+i\eta}\,\frac{1}{1 + \Pi 
   (\lambda^2,\mu^2)  - \Pi_{\rm ct}} \rq,
\end{equation}
where $\Sigma_\lambda^{(1)}$ is defined in eq.~(\ref{eq:Sigmalambda}).
Defining as before the function $r$ such that
\begin{equation}
  \mpole = m_b \left[1+ \as \, r (\mpole, \mu, \as) \right]
  +\mathcal{O}\!\(\as^2\(\as b_0\)^n\)
\end{equation}
is the pole mass position, we get
\begin{eqnarray}
  r (\mpole, \mu, \as) & = & - \frac{1}{\pi} \int_{0 -}^{\infty} \mathd
  \lambda^2 \hspace{0.17em} r_{\lambda} (\mpole, \mu) \,\mathrm{{\rm Im}}
  \left[  \frac{1}{\lambda^2 + i \eta}  
  \hspace{0.17em} \frac{1}{1 + \Pi (\lambda^2 , \mu^2) -
                             \Pi_{\rm ct}}  \right] .
\label{eq:r_allorders}  
\end{eqnarray}
In analogy with eq.~(\ref{eq:mMSB}), we can write 
  \begin{equation}
    \label{eq:mMSB_allorders}
  \mMSB(\mu) = \mpole \left[ 1 -  \as \,r^{\rm (f)} (\mpole, \mu, \as)
    \right] + \mathcal{O}(\as^2(\as b_0)^n), 
\end{equation}
where $r^{\rm (f)}(\mpole, \mu, \as)$ denotes the finite part (according to
the \MSB{} scheme) of $r (\mpole, \mu, \as)$.

In order to compute eq.~(\ref{eq:r_allorders}), since $r_\lambda$ contains a
single pole in $\epsilon$ and does not go to zero for large $\lambda$,
besides its value given in eq.~(\ref{eq:rlambda}), we need its value for
$\lambda=0$ and its asymptotic value for large $\lambda$ in $d=4-2\ep$
dimensions at all orders in $\epsilon$.  Their expressions are given in
eqs.~(\ref{eq:rlambda0}) and~(\ref{eq:rlambdainf}) respectively.  We also
express $r_\lambda$ as the sum of following two terms
\begin{eqnarray}
\label{eq:r_d}
 r_{\lambda,d}\!\(\mpole,\mu\)  &=& \frac{\mu^2}{\mu^2+\lambda^2} \, r_\lambda^0(\mpole,\mu)
                  + \frac{\lambda^2}{\mu^2+\lambda^2} \, r_\lambda^\infty\!\(\mu\),
  \\[2mm]
\label{eq:r_f}
  r_{\lambda,f}\!\(\mpole,\mu\)  &=&  r_\lambda\!\(\mpole,\mu\) - r_{\lambda,d}\!\(\mpole,\mu\)
\end{eqnarray}
and  we write 
\begin{eqnarray}
   \label{eq:rtilde_split}
 r(\mpole,\mu,\as) &=& r_f(\mpole,\mu,\as) + r_d(\mpole,\mu,\as)\,,
  \\
  \label{eq:rtilde_f}  
  r_f(\mpole,\mu,\as) &\equiv& - \frac{1}{\pi}
\int_{0 -}^{\infty} \mathd \lambda^2  \, r_{\lambda,f}\!\(\mpole,\mu\)
    {\rm Im} \lq  \frac{1}{\lambda^2+i\eta}\,\frac{1}{1 + \Pi 
      (\lambda^2,\mu^2)  - \Pi_{\rm ct}} \rq ,
 \\
\label{eq:rtilde_d}    
 r_d(\mpole,\mu,\as) &\equiv&    - \frac{1}{\pi}
\int_{0 -}^{\infty} \mathd \lambda^2  \, r_{\lambda,d}\!\(\mpole,\mu\)
    {\rm Im} \lq  \frac{1}{\lambda^2+i\eta}\,\frac{1}{1 + \Pi 
      (\lambda^2,\mu^2)  - \Pi_{\rm ct}} \rq.
\end{eqnarray}
The function $r_{\lambda,f}(\mpole,\mu)$ vanishes for $\lambda^2\to 0$ and
for $\lambda^2\to \infty$. In addition, it has a finite limit for
$\epsilon\to 0$, so that we can write
\begin{eqnarray}
  r_{\lambda,f}\!\(\mpole,\mu\)  &=&
   \frac{\CF}{4\pi}   \lg
 -3\log\! \(\frac{\mpole^2}{\mu^2}\)   + \frac{\lambda^2}{\mpole^2}\(1+\log
 \frac{\lambda^2}{\mpole^2}\) + 4 + \(2+\frac{\lambda^2}{\mpole^2}\) 
 H\(\frac{\lambda^2}{\mpole^2}\)\right.
 \nonumber \\
 &-& \left. \!\frac{\mu^2}{\mu^2+\lambda^2} \, \lq - 3 \log \!\(\frac{\mpole^2}{\mu^2}\) + 4 \rq
- \frac{\lambda^2}{\mu^2+\lambda^2} \,  \lq - 3 \log
\! \(\frac{\lambda^2}{\mu^2}\) + \frac{5}{2} \rq \rg + {\cal O}(\epsilon)\,.\phantom{aaaaaa}
\end{eqnarray}
For these reasons, we can manipulate $r_f(\mpole,\mu,\as)$ according to the
same procedure used in App.~\ref{app:details_calc}, to get
\begin{eqnarray}
 \label{eq:final_tilderf}
  r_f(\mpole,\mu,\as) & = &- \frac{3\pi}{\as\TF}
\int_{0 }^{\infty}
\frac{\mathd \lambda}{\pi} \, \frac{\mathd\ }{\mathd \lambda}
\lq r_{\lambda,f}\!\(\mpole,\mu\) \rq
{\rm Im} \lg \log \lq 1 + \Pi\(\lambda^2, \mu^2\) -
\Pi_{\rm ct} \rq \rg, \nonumber \\
& = & -\frac{1}{b_0\, \as} \int_{0 }^{\infty}
\frac{\mathd \lambda}{\pi} \, \frac{\mathd\ }{\mathd \lambda}
\lq r_{\lambda,f}\!\(\mpole,\mu\) \rq \atanpilam\,,
\end{eqnarray}
that can be evaluated numerically. We notice that $ r_f(\mpole,\mu,\as) $
contains a linear infrared renormalon, since the behaviour of
$r_{\lambda,f}\!\(\mpole,\mu\)$ for small $\lambda$ is
\begin{equation}
  \label{eq:rf_small_lambda}
  r_{\lambda,f}\!\(\mpole,\mu\) \approx -\frac{\CF}{2} \frac{\lambda}{m}\,.
\end{equation}
As far as the integral in eq.~(\ref{eq:rtilde_d}) is concerned, we can split
it into two terms, according to eq.~(\ref{eq:r_d}),
\begin{eqnarray}
 r_d(\mpole,\mu,\as)  &=&r_d^0(\mpole,\mu,\as)+r_d^\infty(\mpole,\mu,\as)\,,
  \\[1.5mm]
 r_d^0(\mpole,\mu,\as)   &\equiv&    - \frac{1}{\pi}
 \int_{0 -}^{\infty} \!\mathd \lambda^2  \,
 \frac{\mu^2}{\mu^2+\lambda^2} \, r_\lambda^0(\mpole,\mu)
   \,  {\rm Im} \lq  \frac{1}{\lambda^2+i\eta}\,\frac{1}{1 + \Pi 
      (\lambda^2,\mu^2)  - \Pi_{\rm ct}} \rq\!, \phantom{aaa}
  \\[1.5mm]
 r_d^\infty(\mpole,\mu,\as)   &\equiv&    - \frac{1}{\pi}
 \int_{0 -}^{\infty} \!\mathd \lambda^2  \,
\frac{\lambda^2}{\mu^2+\lambda^2} \, r^\infty_\lambda(\mu)
   \,  {\rm Im} \lq  \frac{1}{\lambda^2+i\eta}\,\frac{1}{1 + \Pi 
      (\lambda^2,\mu^2)  - \Pi_{\rm ct}} \rq\!.\phantom{aaaaa}
   \label{eq:final_tilderd_zero}  
  \end{eqnarray}
Using eq.~(\ref{eq:virt_trick}), we can write
\begin{equation}
 r_d^0(\mpole,\mu,\as)=  r_\lambda^0(\mpole,\mu)\, \frac{1}{1 + \Pi
   (-\mu^2,\mu^2) -  \Pi_{\rm ct}}. 
\end{equation}
In order to deal with the integral in $ r_d^\infty(\mpole,\mu,\as) $, we need
to expose the $\lambda$ dependence of the integrand.  From
eq.~(\ref{eq:rlambdainf}), we can write
\begin{equation}
 r^\infty_\lambda(\mu)   = \(\frac{\lambda^2}{\mu^2}\)^{-\ep} R^\infty, 
\end{equation}
where $R^\infty $ depends only on $\epsilon$ and no longer on
$\lambda$. Similarly, using eq.~(\ref{eq:Pi_unren_MSB}), we have
\begin{eqnarray}
   \Pi\(\lambda^2,\mu^2\) &=& \frac{\as \TF}{ \pi} e^{\epsilon\gameul}
  \frac{\Gamma(1+\ep)\, \Gamma^2(1-\ep)}{\Gamma(1-2\ep)} 
  \frac{1-\ep}{(3-2\ep)(1-2\ep)} \, \frac{1}{\ep} \(
  \frac{\lambda^2 }{\mu^2}\)^{-\ep} e^{i\epsilon\pi}
  \nonumber\\
  &=&  \Pi\(-\mu^2,\mu^2\)  \(\frac{\lambda^2 }{\mu^2}\)^{-\ep} e^{i\epsilon\pi},
\end{eqnarray}
and we can write
\begin{eqnarray}
r_d^\infty(\mpole,\mu,\as)  &=&    - \frac{R^\infty }{\pi}
 \int_{0 }^{\infty} \mathd \lambda^2  \,
\frac{1}{\mu^2+\lambda^2}\(\frac{\lambda^2}{\mu^2}\)^{-\epsilon}
 \,   {\rm Im} \lq  \frac{1}{1 + \Pi 
      (\lambda^2,\mu^2)  - \Pi_{\rm ct}} \rq
\nonumber \\
&=& - \frac{R^\infty }{\pi} \, \sum_{n = 0}^{\infty}\, (-1)^n \int_{0
}^{\infty} \mathd \lambda^2  \, 
\frac{1}{\mu^2+\lambda^2}\(\frac{\lambda^2}{\mu^2}\)^{-\epsilon}
\nonumber\\
&& \hspace{3cm} \times \, {\rm Im} \lq \Pi\(-\mu^2,\mu^2\)  \(\frac{\lambda^2 }{\mu^2}\)^{-\ep}
 e^{i\epsilon\pi} - \Pi_{\rm ct}  \rq^n 
 \nonumber \\
&=&- \frac{R^\infty }{\pi} \, \sum_{n = 0}^{\infty}\, (-1)^n \int_{0
 }^{\infty} \mathd z \,
 \frac{z^{-\epsilon}}{1+z}
 \,   {\rm Im} \lq \Pi\(-\mu^2,\mu^2\)  z^{-\ep}
 e^{i\epsilon\pi} - \Pi_{\rm ct}  \rq^n, \phantom{aaa}
 \label{eq:final_tilderd_inf}  
\end{eqnarray}
where we have performed a Taylor expansion in the second line.  By computing
the imaginary part of the $n$-th power of the term in the square brackets, we
are lead to evaluate integrals of the form
\begin{equation}
\int_0^{\infty} \mathd z\, \frac{z^{- h}}{1+z} =
 \Gamma (1 - h) \, \Gamma (h),
\end{equation}
where $h$ is a real number, so that $r_d^\infty(\mpole,\mu,\as)$ can be
straightforwardly evaluated by computer algebraic means at any fixed order in
$\as$.

We emphasize that $r_d(\mpole,\mu,\as)$ has no linear renormalon. Indeed if
we perform an $\epsilon$ expansion and we consider the small $\lambda$
contribution, by writing $\mathd\lambda^2= 2 \lambda \mathd \lambda$, we
notice that the integrand behaves as $\lambda \log^n(\lambda)$. This signals
the absence of linear renormalons, that come from terms of the type
$\log^n(\lambda)$, without any power of $\lambda$ in front.

From eq.~(\ref{eq:mMSB_allorders}) we get
\begin{equation}
\mMSB(\mu)= \mpole \left\{ 1-\as\left[ r_f(\mpole,\mu,\as)+ r_d^{\rm
      (f)}(\mpole,\mu,\as) \right] \right\} + \mathcal{O}\!\(\as^2(\as b_0)^n\),
\label{eq:mMSB_rf_rd}
\end{equation}
where $r_d^{\rm (f)}$ is the finite part (according to the \MSB{} scheme) of
$r_d$.  If we expand $\left( r_f+ r_d^{\rm (f)} \right)$ is series of $\as$
\begin{equation}
r^{\rm (f)}(\mpole,\mu,\as) = r_f(\mpole,\mu,\as)+ r_d^{\rm (f)}(\mpole,\mu,\as) \equiv
\sum_{i=0}^{\infty} c_{i+1}(\mpole,\mu) \, \as^{i},
\end{equation}
we obtain
\begin{equation}
  \label{eq:mpoleMSBnumer}
\mMSB(\mu)= \mpole \left[ 1- \, \,\sum_{i=1}^{\infty} c_i(\mpole, \mu)
  \,\as^i\right] + \mathcal{O}\!\(\as^2\(\as b_0\)^n\). 
\end{equation}


\section{Cancellation of the linear sensitivity in the total cross section
  and in ``leptonic'' observables}
\label{app:KLN}
In order to discuss the issue of the linear sensitivity cancellation in
the total cross section and in $E_{\sss W}$, it is convenient to use the
old-fashioned perturbation theory.  One writes the propagators as the sum of
an advanced and retarded part
\begin{eqnarray}
  \frac{i}{k^2 - m^2 + i \eta} & = & \frac{i}{2 E_{k, m}} \left[
    \frac{1}{k^0 - E_{k, m} + i \eta} + \frac{1}{- k^0 - E_{k, m} + i
      \eta} \right],
  \\ E_{k, m} & = & \sqrt{\vec{k}^2 + m^2},
\end{eqnarray}
while, for unstable particles, we have
\begin{eqnarray} 
  \frac{i}{k^2 - m^2 + i m \Gamma} & = & \frac{i}{2 E_{k, m, \Gamma}}  \left[
  \frac{1}{k^0 - E_{k, m, \Gamma}} + \frac{1}{- k^0 - E_{k, m, \Gamma}}
  \right],\\
  E_{k, m, \Gamma} & = & \sqrt{\underline{k}^2 + m^2 - i m \Gamma} \,.
\end{eqnarray}
In this way, each Feynman graph is separated into contributions where the
vertexes have all possible time orderings.  Each line joining two vertexes
has an energy set to its on-shell value, with an extra negative sign when
considering a retarded propagator.  For each time ordering the integration of
all the $k^0$ components yields a product of old-fashioned perturbation
theory denominators
\begin{equation}
  D_i = \frac{1}{E - E_i+i\eta}\,,
\end{equation}
where $E$ is the total energy and $E_i$ is the energy of the state $i$, given
by the sum of the energies flowing in the $i^{\rm th}$ cut of the amplitude,
times an overall delta function of energy conservation.
\begin{figure}[tb]
  \centering
  \includegraphics[width=0.5\textwidth]{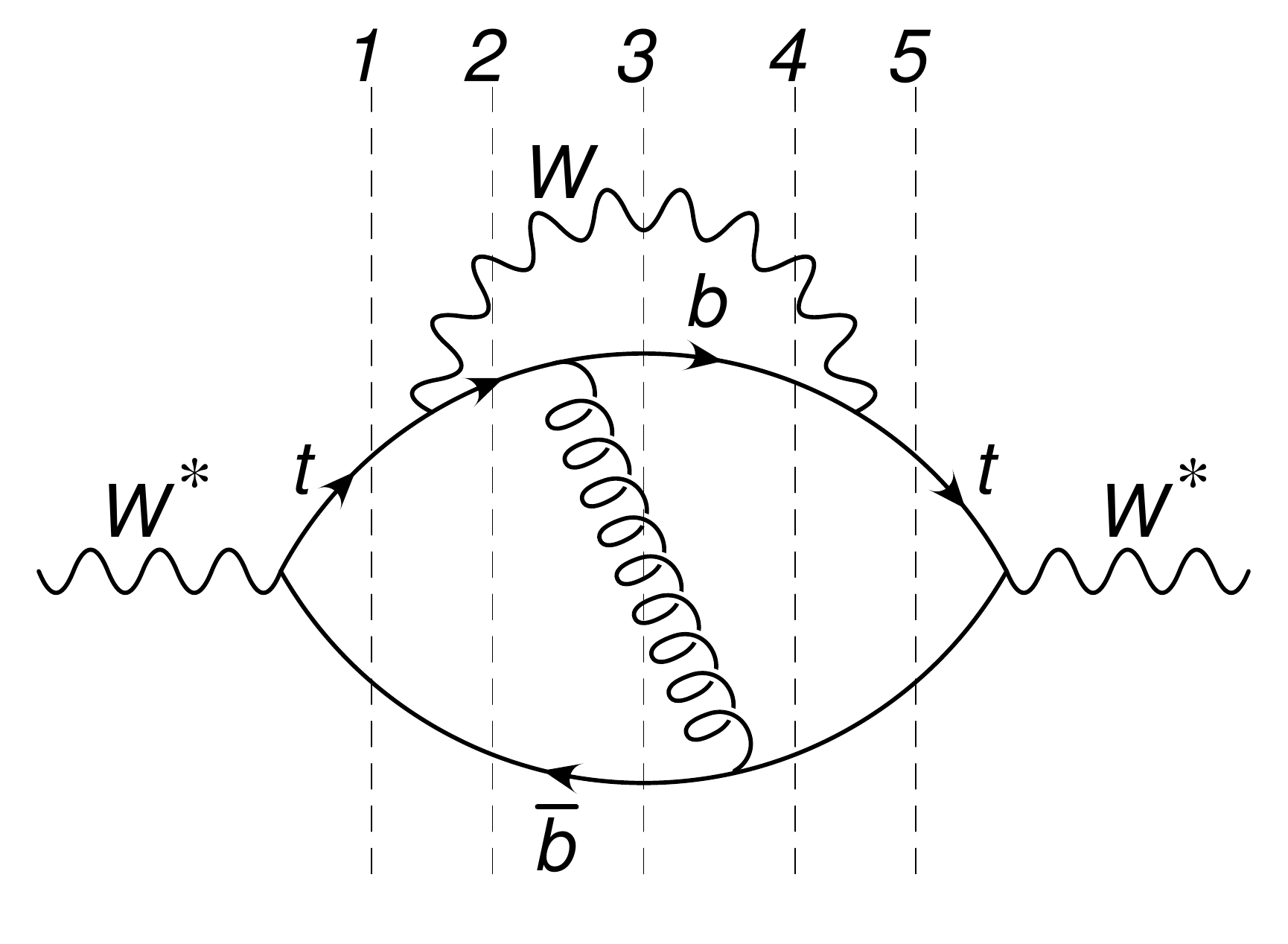}
  \caption{\label{fig:wstart} One time-ordered graph contributing to the
     $W^* \rightarrow t\bar{b} \to W b \bar{b}$ cross section. }
\end{figure}
In Fig.~\ref{fig:wstart} we show a possible time ordering for one graph
contributing to the $W^* \rightarrow t\bar{b} \to W b \bar{b}$ cross
section. The corresponding contribution to the cross section is obtained by
setting either one of the 2, 3, 4 intermediate states on the energy shell,
and changing the sign of the $i \eta$ in the denominators to the right of the
cut. We then define
\begin{eqnarray}
  D_1 & = & \frac{1}{E - E_{t, 1} - E_{\bar{b}, 1}}\,,
  \\
  D_2 & = & \frac{1}{E - E_{\sss W} - E_{b, 2} - E_{\bar{b}, 1} + i \eta}\,,
  \\
  D_3 & = & \frac{1}{E - E_{\sss W}  -
    E_{b, 3} - E_{\bar{b}, 1} - E_{g, 3} + i \eta}\,,
  \\
  D_4 & = &  \frac{1}{E - E_{\sss W} - E_{b, 3} - E_{\bar{b}, 4} + i \eta}\,,
  \\
  D_5 & = & \frac{1}{E - E_{t, 5} - E_{\bar{b}, 5}}\,,
\end{eqnarray}
where
\begin{eqnarray}
  E_{t, i} & = & \sqrt{\vec{k}_{t, i}^2 + m^2 - i m \Gamma_t}\,,
  \\
  E_{l, i} & = & \sqrt{\vec{k}_{l, i}^2}\,, \qquad {\rm for }\qquad l = b,\,
  \bar{b},\, g, 
  \,
  \\
  E_{\sss W} & = & \sqrt{{\vec{k}^2_{\sss W}} + m^2}\, .
\end{eqnarray}
Notice that the top energy has an imaginary part, so that no $i \eta$ is
needed in the denominators containing it.  We never include the corresponding
cuts since the top width prevents this particle from being on-shell. Thus,
the only intermediate states contributing to cuts will be the ones that do
not include the top.  Then, in the integrand for the cross section, we have
the sum
\begin{equation}
D_1 \lq {\rm Im}\(D_2\) D_3^{*}\, D_4^{*} + D_2\, {\rm Im}\(D_3\) D_4^{*}
  + D_2\, D_3\, {\rm Im}\(D_4\) \rq  D_5^{*}\, ,
\end{equation}
that is algebraically equal to $D_1\, {\rm Im} [D_2 D_3 D_4] D_5^{*}$. In fact
\begin{equation}
  D_1 \, {\rm Im} \lq D_2 \,D_3\, D_4\rq D_5^{*} = \frac{1}{2 i} D_1 \lq  D_2
  \, D_3 \,D_4 - \(D_2\,  D_3\,  D_4 \)^{*}\rq  D_5^{*} .
\end{equation}
When performing the 3-momentum integral for the loops not including the $W$
line, one can approach the singularity in the denominator. However, if there
is a direction in the 9-dimensional integration space (corresponding to the
three 3-momenta flowing in the loops) such that, integrating along it, it
leaves the singularities of $D_2$, $D_3$ and $D_4$ on the same side of the
complex plane, the integration contour can be deformed away from the
singularities, so that the denominators cannot contribute to mass
singularities. The singularity for small gluon mass $m_g$ is thus determined
only by the remaining factor
\begin{equation}
  \frac{\mathd^3 k_g}{\sqrt{\vec{k}_g^2 + m_g^2}}\,,
\end{equation}
that gives a quadratic sensitivity to the gluon mass. The only cases when an
appropriate deformation of the contour does not exist correspond to Landau
singularities~\cite{Landau:1959fi}. These are characterised by the presence
of configurations with intermediate on-shell particles corresponding to
classical propagation over large distances.  It can be proven that the Landau
singularities arise when several denominators go simultaneously on-shell and
the momenta of the particles are such that they meet again after having come
apart. If this is not possible, the singularity is an avoidable one.

\begin{figure}[htb]
  \centering
  \includegraphics[width=0.5\textwidth]{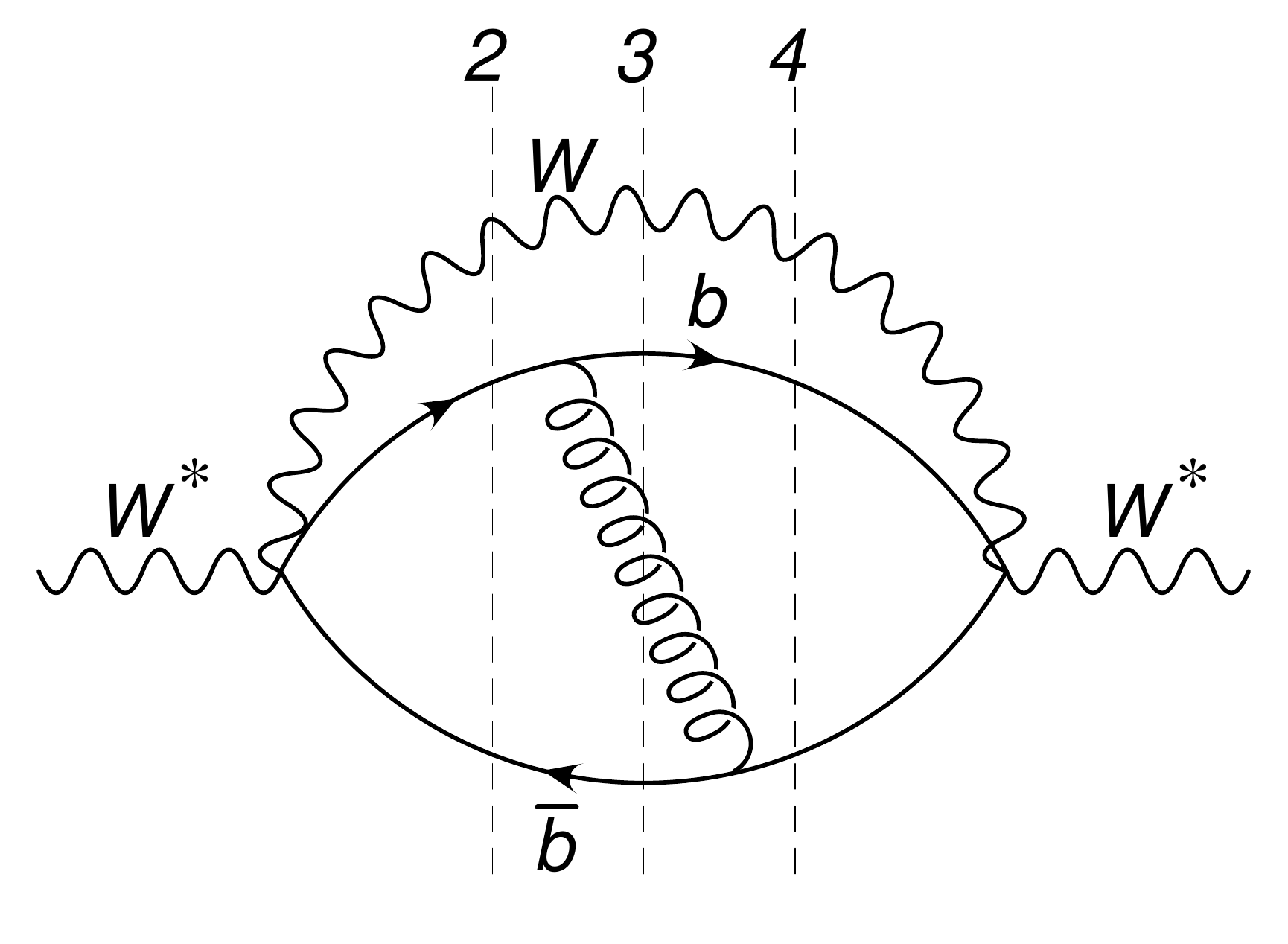}
  \caption{\label{fig:wstart-landau1} The reduced graph to look for Landau
    singularities in the graph of Fig.~\ref{fig:wstart}.}
\end{figure}

In order to explore the possible Landau configurations, one can start with
the graph of Fig.~\ref{fig:wstart} with the top lines shrunk to a point. In
fact, the top is always off-shell, and it cannot propagate over large
distances. The remaining configuration is shown in
Fig.~\ref{fig:wstart-landau1}.
In order for the 2, 3 and 4 intermediate states to be on-shell at the same
time, either both the $b$ and the $\bar{b}$ quarks should be collinear to the
gluon, or the gluon should be soft. In the first case, also the $b$, the
$\bar{b}$ and the gluon are collinear, and are all travelling in the opposite
direction with respect to the $W$. Thus, they cannot meet at the same point
on the last vertex to the right. On the other hand, if the gluon is soft, the
$b$, the $\bar{b}$ and the $W$ produced at the primary vertex have momenta
that sum to zero, so, again their velocities will make them diverge. One can
try to shrink other propagators to a point.  Shrinking one fermion propagator
to a point, for example the $b$ one at the intermediate state 2, forces the
gluon to be either collinear to the $\bar{b}$ or soft, and again one would
end up with a $W$, a $b$ and a collinear $b g$ system produced at the vertex
on the left and meeting at the vertex on the right, which is
impossible. Shrinking the $W$, the two $b$ or the two $\bar{b}$ lines to a
point shrinks the whole graph to a point, leading to nothing. Shrinking a $b$
and a $\bar{b}$ line to a point leads again to configuration with two
massless system (either $b$ quarks of collinear $b g$ systems) and a $W$,
that again cannot meet at the same point. Thus, no Landau configuration can
exist, so one infers that the $m_g$ sensitivity of the total cross section is
at least quadratic.

We can repeat the same reasoning including a factor $E_{\sss W}$ in our
Feynman graph. The argument runs as before, and so, even for the average
energy of the $W$ boson, one expects that the sensitivity to the gluon mass
is at least quadratic. Notice that, in order for this to work, one needs that
the $E_{\sss W}$ factor is the same for all cuts, which is in fact the case.

The argument fails if the top width is sent to zero. In fact, even at the
Born level (i.e.~removing the gluon line) the first and last intermediate
states have equal energy, but their $i \eta$ have opposite signs. Under these
condition, the pinch is clearly unavoidable.

As a last point, we recall that the total cross section is free of linear
$m_g$ sensitivity also in the limit of zero width. This happens because, in
the zero-width limit, the cross section factorizes into a production cross
section times a decay width, and both of them are free of linear sensitivity
to $m_g$ if the mass is in a short-distance scheme. The same, however, does
not hold for the average $E_{\sss W}$. In fact, the cancellation of mass
singularities in $\Gamma_t$ cannot be proven in the same fashion adopted
here, since logarithmic divergences are also present in the wave-function
renormalization, and cannot be treated in a straightforward way in the
old-fashioned perturbation theory.


\providecommand{\href}[2]{#2}\begingroup\raggedright\endgroup

\end{document}

%% file: texmacros.tex
\newcommand\obs{{ O}}
\newcommand\obsb{{ \langle O \rangle_{\rm b}}}

\newcommand\LambdaQCD{\Lambda_{\sss\rm QCD}}

\newcommand\RES{{\tt POWHEG~\!BOX~\!RES}}

\newcommand\Collier{{\tt COLLIER}\xspace}

\newcommand{\mathd}{\mathrm{d}}

\newcommand\sss{\mathchoice%
{\displaystyle}%
{\scriptstyle}%
{\scriptscriptstyle}%
{\scriptscriptstyle}%
}

\def\beq{\begin{equation}}
\def\beqn{\begin{eqnarray}}
\def\eeq{\end{equation}}
\def\eeqn{\end{eqnarray}}

\def\lq{\left[} 
\def\rq{\right]} 
\def\rg{\right\}} 
\def\lg{\left\{} 
\def\({\left(} 
\def\){\right)}

\newcommand\as{\alpha_{\sss\rm S}}

\newcommand\CF{C_{\sss\rm F}}
\newcommand\CA{C_{\sss\rm A}}
\newcommand\TF{T_{\sss\rm F}}
\newcommand\TR{T_{\sss\rm R}}

\newcommand\mb{m_{b}}

\newcount\minutes 
\newcount\scratch 
\def\timestamp{%
\scratch=\time 
\divide\scratch by 60 
\edef\hours{\the\scratch} 
\multiply\scratch by 60 
\minutes=\time 
\advance\minutes by -\scratch 
---$\,$\hours:\null 
\ifnum\minutes< 10 0\fi 
\the\minutes}

\definecolor{mygray}{gray}{0.5}

\newcommand\ep{\epsilon}
\newcommand{\mpole}{\ensuremath{m}}

\newcommand{\mMSB}{\ensuremath{\overline{m}}}
\newcommand{\MSB}{\ensuremath{\overline{\rm MS}}}

\newcommand{\gameul}{{\gamma_{\sss E}}}

\newcommand\atanpilam{\arctan \frac{\pi \,b_0\, \as}{1+b_0\,\as
    \log\displaystyle{\frac{\lambda^2}{\mu_{\sss C}^2}}}}

%% file: tables-M/poleMSbar_complex_conversion.tex
 \begin{tabular}{|c|c|c|c|c|}
  \cline{1-5}
 \multicolumn{5}{|c|}{ $\displaystyle \phantom{\Big|} \mpole - \mMSB(\mu)  \phantom{\Big|} $}
 \\ \cline{1-5}
  $ \phantom{\Big|}i\phantom{\Big|}$ & ${\rm Re}\(c_i\)$   & ${\rm Im}\(c_i\)$   & ${\rm Re}\(\mpole\,c_i\,\as^i\)$ & ${\rm Im}\(\mpole\,c_i\,\as^i\)$
 \\ \cline{1-5}
 \phantom{\Big|}           1 \phantom{\Big|} & $4.244\times 10^{   -1}                                                                              $ & $2.450\times 10^{   -3}                                                                              $ & $7.919\times 10^{+    0}                                                                             $ & $+1.524\times 10^{   -2}                                                                             $
 \\ \cline{1-5}
 \phantom{\Big|}           2 \phantom{\Big|} & $6.437\times 10^{   -1}                                                                              $ & $2.094\times 10^{   -3}                                                                              $ & $1.299\times 10^{+    0}                                                                             $ & $-7.729\times 10^{   -4}                                                                             $
 \\ \cline{1-5}
 \phantom{\Big|}           3 \phantom{\Big|} & $1.968\times 10^{+    0}                                                                             $ & $8.019\times 10^{   -3}                                                                              $ & $4.297\times 10^{   -1}                                                                              $ & $+9.665\times 10^{   -5}                                                                             $
 \\ \cline{1-5}
 \phantom{\Big|}           4 \phantom{\Big|} & $7.231\times 10^{+    0}                                                                             $ & $2.567\times 10^{   -2}                                                                              $ & $1.707\times 10^{   -1}                                                                              $ & $-5.110\times 10^{   -5}                                                                             $
 \\ \cline{1-5}
 \phantom{\Big|}           5 \phantom{\Big|} & $3.497\times 10^{+    1}                                                                             $ & $1.394\times 10^{   -1}                                                                              $ & $8.930\times 10^{   -2}                                                                              $ & $+1.240\times 10^{   -5}                                                                             $
 \\ \cline{1-5}
 \phantom{\Big|}           6 \phantom{\Big|} & $2.174\times 10^{+    2}                                                                             $ & $8.164\times 10^{   -1}                                                                              $ & $6.005\times 10^{   -2}                                                                              $ & $-5.616\times 10^{   -6}                                                                             $
 \\ \cline{1-5}
 \phantom{\Big|}           7 \phantom{\Big|} & $1.576\times 10^{+    3}                                                                             $ & $6.133\times 10^{+    0}                                                                             $ & $4.709\times 10^{   -2}                                                                              $ & $+2.009\times 10^{   -6}                                                                             $
 \\ \cline{1-5}
 \phantom{\Big|}           8 \phantom{\Big|} & $1.354\times 10^{+    4}                                                                             $ & $5.180\times 10^{+    1}                                                                             $ & $4.376\times 10^{   -2}                                                                              $ & $-1.031\times 10^{   -6}                                                                             $
 \\ \cline{1-5}
 \phantom{\Big|}           9 \phantom{\Big|} & $1.318\times 10^{+    5}                                                                             $ & $5.087\times 10^{+    2}                                                                             $ & $4.608\times 10^{   -2}                                                                              $ & $+4.961\times 10^{   -7}                                                                             $
 \\ \cline{1-5}
 \phantom{\Big|}          10 \phantom{\Big|} & $1.450\times 10^{+    6}                                                                             $ & $5.572\times 10^{+    3}                                                                             $ & $5.481\times 10^{   -2}                                                                              $ & $-2.909\times 10^{   -7}                                                                             $
 \\ \cline{1-5}
 \end{tabular}

%% file: tables-M/xsec-nocuts.tex
 \begin{tabular}{c|c|c|c|c|}
  \cline{2-5}
  & \multicolumn{4}{|c|}{$\sigma/\sigma^{\rm nocuts}_{\rm b}(\mpole{}) $}
 \\ \cline{2-5}
  & \multicolumn{2}{|c|}{ \phantom{\Big|}pole scheme \phantom{\Big|}}& \multicolumn{2}{|c|}{\MSB{} scheme}
 \\ \cline{1-5}
 \multicolumn{1}{|c|}{$\phantom{\Big|} i \phantom{\Big|}$}  & $c_i $ & $ c_i\, \as^i$  & $c_i $ & $ c_i\, \as^i$ 
 \\ \cline{1-5}
 \multicolumn{1}{|c|}{$\phantom{\Big|}$ 0 $\phantom{\Big|}$}& 1.00000000 &1.0000000& 0.86841331 &0.8684133
 \\ \cline{1-5}
 \multicolumn{1}{|c|}{ $\phantom{\Big|}$           1 $\phantom{\Big|}$}& $5.003 \, (0) \times 10^{   -1}$ & $5.411 \, (0) \times 10^{   -2}$& $1.480 \, (0) \times 10^{    0}$ & $1.601 \, (0) \times 10^{   -1}$
 \\ \cline{1-5}
 \multicolumn{1}{|c|}{ $\phantom{\Big|}$           2 $\phantom{\Big|}$}& $-6.20 \, (2) \times 10^{   -1}$ & $-7.25 \, (2) \times 10^{   -3}$& $4.42 \, (2) \times 10^{   -1}$ & $5.17 \, (2) \times 10^{   -3}$
 \\ \cline{1-5}
 \multicolumn{1}{|c|}{ $\phantom{\Big|}$           3 $\phantom{\Big|}$}& $-3.03 \, (2) \times 10^{    0}$ & $-3.83 \, (3) \times 10^{   -3}$& $6.4 \, (2) \times 10^{   -1}$ & $8.1 \, (3) \times 10^{   -4}$
 \\ \cline{1-5}
 \multicolumn{1}{|c|}{ $\phantom{\Big|}$           4 $\phantom{\Big|}$}& $-1.25 \, (2) \times 10^{    1}$ & $-1.70 \, (3) \times 10^{   -3}$& $0 \, (2) \times 10^{   -2}$ & $0 \, (3) \times 10^{   -6}$
 \\ \cline{1-5}
 \multicolumn{1}{|c|}{ $\phantom{\Big|}$           5 $\phantom{\Big|}$}& $-6.4 \, (2) \times 10^{    1}$ & $-9.4 \, (3) \times 10^{   -4}$& $1 \, (2) \times 10^{   -1}$ & $1 \, (3) \times 10^{   -5}$
 \\ \cline{1-5}
 \multicolumn{1}{|c|}{ $\phantom{\Big|}$           6 $\phantom{\Big|}$}& $-3.9 \, (1) \times 10^{    2}$ & $-6.2 \, (2) \times 10^{   -4}$& $0 \, (1) \times 10^{    0}$ & $0 \, (2) \times 10^{   -6}$
 \\ \cline{1-5}
 \multicolumn{1}{|c|}{ $\phantom{\Big|}$           7 $\phantom{\Big|}$}& $-2.9 \, (1) \times 10^{    3}$ & $-5.0 \, (2) \times 10^{   -4}$& $0 \, (1) \times 10^{    1}$ & $0 \, (2) \times 10^{   -6}$
 \\ \cline{1-5}
 \multicolumn{1}{|c|}{ $\phantom{\Big|}$           8 $\phantom{\Big|}$}& $-2.5 \, (1) \times 10^{    4}$ & $-4.6 \, (2) \times 10^{   -4}$& $0 \, (1) \times 10^{    2}$ & $0 \, (2) \times 10^{   -6}$
 \\ \cline{1-5}
 \multicolumn{1}{|c|}{ $\phantom{\Big|}$           9 $\phantom{\Big|}$}& $-2.4 \, (1) \times 10^{    5}$ & $-4.9 \, (2) \times 10^{   -4}$& $0 \, (1) \times 10^{    3}$ & $0 \, (2) \times 10^{   -6}$
 \\ \cline{1-5}
 \multicolumn{1}{|c|}{ $\phantom{\Big|}$          10 $\phantom{\Big|}$}& $-2.6 \, (1) \times 10^{    6}$ & $-5.8 \, (2) \times 10^{   -4}$& $0 \, (1) \times 10^{    4}$ & $-1 \, (2) \times 10^{   -6}$
 \\ \cline{1-5}
 \end{tabular}

%% file: tables-M/xsec-cut2-R0.1.tex
 \begin{tabular}{c|c|c|}
  \cline{2-3}
  & \multicolumn{2}{|c|}{$\sigma/\sigma^{\rm nocuts}_{\rm b}(\mpole{}) \;\; R=0.1$}
 \\ \cline{2-3}
  & \multicolumn{1}{|c|}{ \phantom{\Big|}pole scheme \phantom{\Big|}}& \multicolumn{1}{|c|}{\MSB{} scheme}
 \\ \cline{1-3}
 \multicolumn{1}{|c|}{$\phantom{\Big|} i \phantom{\Big|}$} & $ c_i\, \as^i$ & $ c_i\, \as^i$ 
 \\ \cline{1-3}
 \multicolumn{1}{|c|}{$\phantom{\Big|}$ 0 $\phantom{\Big|}$} &0.9985836&0.8666708
 \\ \cline{1-3}
 \multicolumn{1}{|c|}{ $\phantom{\Big|}$           1 $\phantom{\Big|}$}& $-7.953 \, (0) \times 10^{   -2}$& $2.650 \, (0) \times 10^{   -2}$
 \\ \cline{1-3}
 \multicolumn{1}{|c|}{ $\phantom{\Big|}$           2 $\phantom{\Big|}$}& $-7.22 \, (2) \times 10^{   -2}$& $-5.98 \, (2) \times 10^{   -2}$
 \\ \cline{1-3}
 \multicolumn{1}{|c|}{ $\phantom{\Big|}$           3 $\phantom{\Big|}$}& $-3.71 \, (2) \times 10^{   -2}$& $-3.24 \, (2) \times 10^{   -2}$
 \\ \cline{1-3}
 \multicolumn{1}{|c|}{ $\phantom{\Big|}$           4 $\phantom{\Big|}$}& $-1.97 \, (2) \times 10^{   -2}$& $-1.80 \, (2) \times 10^{   -2}$
 \\ \cline{1-3}
 \multicolumn{1}{|c|}{ $\phantom{\Big|}$           5 $\phantom{\Big|}$}& $-1.13 \, (2) \times 10^{   -2}$& $-1.04 \, (2) \times 10^{   -2}$
 \\ \cline{1-3}
 \multicolumn{1}{|c|}{ $\phantom{\Big|}$           6 $\phantom{\Big|}$}& $-7.0 \, (2) \times 10^{   -3}$& $-6.4 \, (2) \times 10^{   -3}$
 \\ \cline{1-3}
 \multicolumn{1}{|c|}{ $\phantom{\Big|}$           7 $\phantom{\Big|}$}& $-4.8 \, (1) \times 10^{   -3}$& $-4.3 \, (1) \times 10^{   -3}$
 \\ \cline{1-3}
 \multicolumn{1}{|c|}{ $\phantom{\Big|}$           8 $\phantom{\Big|}$}& $-3.6 \, (1) \times 10^{   -3}$& $-3.1 \, (1) \times 10^{   -3}$
 \\ \cline{1-3}
 \multicolumn{1}{|c|}{ $\phantom{\Big|}$           9 $\phantom{\Big|}$}& $-3.1 \, (1) \times 10^{   -3}$& $-2.7 \, (1) \times 10^{   -3}$
 \\ \cline{1-3}
 \multicolumn{1}{|c|}{ $\phantom{\Big|}$          10 $\phantom{\Big|}$}& $-3.2 \, (2) \times 10^{   -3}$& $-2.6 \, (2) \times 10^{   -3}$
 \\ \cline{1-3}
 \end{tabular}

%% file: tables-M/xsec-cut2-R0.5.tex
 \begin{tabular}{c|c|c|}
  \cline{2-3}
  & \multicolumn{2}{|c|}{$\sigma/\sigma^{\rm nocuts}_{\rm b}(\mpole{}) \;\; R=0.5$}
 \\ \cline{2-3}
  & \multicolumn{1}{|c|}{ \phantom{\Big|}pole scheme \phantom{\Big|}}& \multicolumn{1}{|c|}{\MSB{} scheme}
 \\ \cline{1-3}
 \multicolumn{1}{|c|}{$\phantom{\Big|} i \phantom{\Big|}$} & $ c_i\, \as^i$ & $ c_i\, \as^i$ 
 \\ \cline{1-3}
 \multicolumn{1}{|c|}{$\phantom{\Big|}$ 0 $\phantom{\Big|}$} &0.9783310&0.8511828
 \\ \cline{1-3}
 \multicolumn{1}{|c|}{ $\phantom{\Big|}$           1 $\phantom{\Big|}$}& $-4.992 \, (0) \times 10^{   -3}$& $9.705 \, (0) \times 10^{   -2}$
 \\ \cline{1-3}
 \multicolumn{1}{|c|}{ $\phantom{\Big|}$           2 $\phantom{\Big|}$}& $-2.966 \, (5) \times 10^{   -2}$& $-1.779 \, (5) \times 10^{   -2}$
 \\ \cline{1-3}
 \multicolumn{1}{|c|}{ $\phantom{\Big|}$           3 $\phantom{\Big|}$}& $-1.267 \, (6) \times 10^{   -2}$& $-8.22 \, (6) \times 10^{   -3}$
 \\ \cline{1-3}
 \multicolumn{1}{|c|}{ $\phantom{\Big|}$           4 $\phantom{\Big|}$}& $-5.37 \, (6) \times 10^{   -3}$& $-3.73 \, (6) \times 10^{   -3}$
 \\ \cline{1-3}
 \multicolumn{1}{|c|}{ $\phantom{\Big|}$           5 $\phantom{\Big|}$}& $-2.58 \, (5) \times 10^{   -3}$& $-1.66 \, (5) \times 10^{   -3}$
 \\ \cline{1-3}
 \multicolumn{1}{|c|}{ $\phantom{\Big|}$           6 $\phantom{\Big|}$}& $-1.44 \, (4) \times 10^{   -3}$& $-8.5 \, (4) \times 10^{   -4}$
 \\ \cline{1-3}
 \multicolumn{1}{|c|}{ $\phantom{\Big|}$           7 $\phantom{\Big|}$}& $-9.8 \, (4) \times 10^{   -4}$& $-5.0 \, (4) \times 10^{   -4}$
 \\ \cline{1-3}
 \multicolumn{1}{|c|}{ $\phantom{\Big|}$           8 $\phantom{\Big|}$}& $-8.1 \, (4) \times 10^{   -4}$& $-3.7 \, (4) \times 10^{   -4}$
 \\ \cline{1-3}
 \multicolumn{1}{|c|}{ $\phantom{\Big|}$           9 $\phantom{\Big|}$}& $-8.0 \, (4) \times 10^{   -4}$& $-3.4 \, (4) \times 10^{   -4}$
 \\ \cline{1-3}
 \multicolumn{1}{|c|}{ $\phantom{\Big|}$          10 $\phantom{\Big|}$}& $-9.2 \, (5) \times 10^{   -4}$& $-3.7 \, (5) \times 10^{   -4}$
 \\ \cline{1-3}
 \end{tabular}

%% file: tables-M/Obs-cut2-R-paper.tex
 \begin{tabular}{c|c|c|c|c|c|c|c|}
  \cline{2-7}
  & \multicolumn{2}{|c|}{$R=0.1$ }
  & \multicolumn{2}{|c|}{$R=0.5$ }
  & \multicolumn{2}{|c|}{$R=1.5$ }
 \\ \cline{1-7}
 \multicolumn{1}{|c|}{$i$} & \multicolumn{1}{|c|}{ \phantom{\Big|} pole  \phantom{\Big|}}& \multicolumn{1}{|c|}{\MSB{}} & \multicolumn{1}{|c|}{ \phantom{\Big|} pole\phantom{\Big|}}& \multicolumn{1}{|c|}{\MSB{}} & \multicolumn{1}{|c|}{ \phantom{\Big|} pole \phantom{\Big|}}& \multicolumn{1}{|c|}{\MSB{} }
 \\ \cline{1-7}
 \multicolumn{1}{|c|}{  $\phantom{\Big|}$           0  $\phantom{\Big|}$}
 & $172.8280                                                                                            $& $163.0146                                                                                            $
 & $172.8201                                                                                            $& $163.0040                                                                                            $
 & $172.7533                                                                                            $& $162.9244                                                                                            $
 \\ \cline{1-7}
 \multicolumn{1}{|c|}{  $\phantom{\Big|}$           1  $\phantom{\Big|}$}
 & $-7.597 \, (0) \times 10^{    0}                                                                     $& $2.163 \, (0) \times 10^{   -1}                                                                      $
 & $-2.785 \, (0) \times 10^{    0}                                                                     $& $5.030 \, (0) \times 10^{    0}                                                                      $
 & $4.446 \, (0) \times 10^{   -1}                                                                      $& $8.268 \, (0) \times 10^{    0}                                                                      $
 \\ \cline{1-7}
 \multicolumn{1}{|c|}{  $\phantom{\Big|}$           2  $\phantom{\Big|}$}
 & $-4.136 \, (2) \times 10^{    0}                                                                     $& $-2.852 \, (2) \times 10^{    0}                                                                     $
 & $-1.255 \, (1) \times 10^{    0}                                                                     $& $2.9 \, (1) \times 10^{   -2}                                                                        $
 & $1.029 \, (8) \times 10^{   -1}                                                                      $& $1.387 \, (1) \times 10^{    0}                                                                      $
 \\ \cline{1-7}
 \multicolumn{1}{|c|}{  $\phantom{\Big|}$           3  $\phantom{\Big|}$}
 & $-2.397 \, (2) \times 10^{    0}                                                                     $& $-1.973 \, (2) \times 10^{    0}                                                                     $
 & $-5.96 \, (2) \times 10^{   -1}                                                                      $& $-1.72 \, (2) \times 10^{   -1}                                                                      $
 & $1.4 \, (1) \times 10^{   -2}                                                                        $& $4.38 \, (1) \times 10^{   -1}                                                                       $
 \\ \cline{1-7}
 \multicolumn{1}{|c|}{  $\phantom{\Big|}$           4  $\phantom{\Big|}$}
 & $-1.505 \, (2) \times 10^{    0}                                                                     $& $-1.337 \, (2) \times 10^{    0}                                                                     $
 & $-3.13 \, (2) \times 10^{   -1}                                                                      $& $-1.44 \, (2) \times 10^{   -1}                                                                      $
 & $-6 \, (1) \times 10^{   -3}                                                                         $& $1.63 \, (1) \times 10^{   -1}                                                                       $
 \\ \cline{1-7}
 \multicolumn{1}{|c|}{  $\phantom{\Big|}$           5  $\phantom{\Big|}$}
 & $-1.038 \, (2) \times 10^{    0}                                                                     $& $-9.50 \, (2) \times 10^{   -1}                                                                      $
 & $-1.88 \, (2) \times 10^{   -1}                                                                      $& $-1.00 \, (2) \times 10^{   -2}                                                                      $
 & $-9.7 \, (9) \times 10^{   -3}                                                                       $& $7.86 \, (9) \times 10^{   -2}                                                                       $
 \\ \cline{1-7}
 \multicolumn{1}{|c|}{  $\phantom{\Big|}$           6  $\phantom{\Big|}$}
 & $-7.94 \, (2) \times 10^{   -1}                                                                      $& $-7.35 \, (2) \times 10^{   -1}                                                                      $
 & $-1.33 \, (1) \times 10^{   -1}                                                                      $& $-7.3 \, (1) \times 10^{   -2}                                                                       $
 & $-1.05 \, (8) \times 10^{   -2}                                                                      $& $4.89 \, (8) \times 10^{   -2}                                                                       $
 \\ \cline{1-7}
 \multicolumn{1}{|c|}{  $\phantom{\Big|}$           7  $\phantom{\Big|}$}
 & $-6.79 \, (2) \times 10^{   -1}                                                                      $& $-6.33 \, (2) \times 10^{   -1}                                                                      $
 & $-1.09 \, (1) \times 10^{   -1}                                                                      $& $-6.3 \, (1) \times 10^{   -2}                                                                       $
 & $-1.12 \, (7) \times 10^{   -2}                                                                      $& $3.53 \, (7) \times 10^{   -2}                                                                       $
 \\ \cline{1-7}
 \multicolumn{1}{|c|}{  $\phantom{\Big|}$           8  $\phantom{\Big|}$}
 & $-6.51 \, (2) \times 10^{   -1}                                                                      $& $-6.08 \, (2) \times 10^{   -1}                                                                      $
 & $-1.04 \, (1) \times 10^{   -1}                                                                      $& $-6.1 \, (1) \times 10^{   -2}                                                                       $
 & $-1.25 \, (7) \times 10^{   -2}                                                                      $& $3.08 \, (7) \times 10^{   -2}                                                                       $
 \\ \cline{1-7}
 \multicolumn{1}{|c|}{  $\phantom{\Big|}$           9  $\phantom{\Big|}$}
 & $-6.99 \, (2) \times 10^{   -1}                                                                      $& $-6.54 \, (2) \times 10^{   -1}                                                                      $
 & $-1.12 \, (1) \times 10^{   -1}                                                                      $& $-6.7 \, (1) \times 10^{   -2}                                                                       $
 & $-1.47 \, (7) \times 10^{   -2}                                                                      $& $3.09 \, (7) \times 10^{   -2}                                                                       $
 \\ \cline{1-7}
 \multicolumn{1}{|c|}{  $\phantom{\Big|}$          10  $\phantom{\Big|}$}
 & $-8.37 \, (2) \times 10^{   -1}                                                                      $& $-7.83 \, (2) \times 10^{   -1}                                                                      $
 & $-1.35 \, (1) \times 10^{   -1}                                                                      $& $-8.1 \, (1) \times 10^{   -2}                                                                       $
 & $-1.85 \, (9) \times 10^{   -2}                                                                      $& $3.57 \, (9) \times 10^{   -2}                                                                       $
 \\ \cline{1-7}
 \end{tabular}

%% file: tables-Ew/Obs-nocuts-R0.5.tex
 \begin{tabular}{c|c|c|c|c|}
  \cline{2-5}
  & \multicolumn{4}{|c|}{ $\displaystyle \phantom{\big|}\langle E_W \rangle\phantom{\big|}  $}
 \\ \cline{2-5}
  & \multicolumn{2}{|c|}{ \phantom{\Big|}pole scheme \phantom{\Big|}}& \multicolumn{2}{|c|}{\MSB{} scheme}
 \\ \cline{1-5}
 \multicolumn{1}{|c|}{$\phantom{\Big|} i \phantom{\Big|}$}  & $c_i $ & $ c_i\,\as^i$  & $c_i $ & $ c_i \, \as^i$ 
 \\ \cline{1-5}
 \multicolumn{1}{|c|}{ $\phantom{\Big|}$ 0 $\phantom{\Big|}$}& $121.5818$ & $121.5818$& $120.8654$ & $120.8654$
 \\ \cline{1-5}
 \multicolumn{1}{|c|}{$\phantom{\Big|}$           1 $\phantom{\Big|}$}& $-1.435 \, (0) \times 10^{    1}$ & $-1.552 \, (0) \times 10^{    0}$ & $-7.192 \, (0) \times 10^{    0}$ & $-7.779 \, (0) \times 10^{   -1}$ 
 \\ \cline{1-5}
 \multicolumn{1}{|c|}{$\phantom{\Big|}$           2 $\phantom{\Big|}$}& $-4.97 \, (4) \times 10^{    1}$ & $-5.82 \, (4) \times 10^{   -1}$ & $-3.88 \, (4) \times 10^{    1}$ & $-4.54 \, (4) \times 10^{   -1}$ 
 \\ \cline{1-5}
 \multicolumn{1}{|c|}{$\phantom{\Big|}$           3 $\phantom{\Big|}$}& $-1.79 \, (5) \times 10^{    2}$ & $-2.26 \, (6) \times 10^{   -1}$ & $-1.45 \, (5) \times 10^{    2}$ & $-1.84 \, (6) \times 10^{   -1}$ 
 \\ \cline{1-5}
 \multicolumn{1}{|c|}{$\phantom{\Big|}$           4 $\phantom{\Big|}$}& $-6.9 \, (4) \times 10^{    2}$ & $-9.4 \, (6) \times 10^{   -2}$ & $-5.7 \, (4) \times 10^{    2}$ & $-7.8 \, (6) \times 10^{   -2}$ 
 \\ \cline{1-5}
 \multicolumn{1}{|c|}{$\phantom{\Big|}$           5 $\phantom{\Big|}$}& $-2.9 \, (3) \times 10^{    3}$ & $-4.4 \, (5) \times 10^{   -2}$ & $-2.4 \, (3) \times 10^{    3}$ & $-3.5 \, (5) \times 10^{   -2}$ 
 \\ \cline{1-5}
 \multicolumn{1}{|c|}{$\phantom{\Big|}$           6 $\phantom{\Big|}$}& $-1.4 \, (3) \times 10^{    4}$ & $-2.2 \, (4) \times 10^{   -2}$ & $-1.0 \, (3) \times 10^{    4}$ & $-1.7 \, (4) \times 10^{   -2}$ 
 \\ \cline{1-5}
 \multicolumn{1}{|c|}{$\phantom{\Big|}$           7 $\phantom{\Big|}$}& $-8 \, (2) \times 10^{    4}$ & $-1.3 \, (4) \times 10^{   -2}$ & $-5 \, (2) \times 10^{    4}$ & $-8 \, (4) \times 10^{   -3}$ 
 \\ \cline{1-5}
 \multicolumn{1}{|c|}{$\phantom{\Big|}$           8 $\phantom{\Big|}$}& $-5 \, (2) \times 10^{    5}$ & $-9 \, (4) \times 10^{   -3}$ & $-2 \, (2) \times 10^{    5}$ & $-4 \, (4) \times 10^{   -3}$ 
 \\ \cline{1-5}
 \multicolumn{1}{|c|}{$\phantom{\Big|}$           9 $\phantom{\Big|}$}& $-3 \, (2) \times 10^{    6}$ & $-7 \, (4) \times 10^{   -3}$ & $-1 \, (2) \times 10^{    6}$ & $-2 \, (4) \times 10^{   -3}$ 
 \\ \cline{1-5}
 \multicolumn{1}{|c|}{$\phantom{\Big|}$          10 $\phantom{\Big|}$}& $-3 \, (2) \times 10^{    7}$ & $-6 \, (5) \times 10^{   -3}$ & $0 \, (2) \times 10^{    6}$ & $-1 \, (5) \times 10^{   -4}$ 
 \\ \cline{1-5}
 \multicolumn{1}{|c|}{$\phantom{\Big|}$          11 $\phantom{\Big|}$}& $-3 \, (3) \times 10^{    8}$ & $-7 \, (6) \times 10^{   -3}$ & $0 \, (3) \times 10^{    6}$ & $0 \, (6) \times 10^{   -5}$ 
 \\ \cline{1-5}
 \multicolumn{1}{|c|}{$\phantom{\Big|}$          12 $\phantom{\Big|}$}& $-4 \, (3) \times 10^{    9}$ & $-9 \, (9) \times 10^{   -3}$ & $0 \, (3) \times 10^{    8}$ & $1 \, (9) \times 10^{   -3}$ 
 \\ \cline{1-5}
 \end{tabular}